\documentclass[10pt,fleqn]{book} 

\usepackage[top=2cm,bottom=2.5cm,left=2.5cm,right=2cm,headsep=10pt,a4paper]{geometry} 
\usepackage{graphicx} 
\graphicspath{{Images/}} 
\usepackage{lipsum} 
\usepackage{tikz} 
\usepackage[english]{babel} 
\usepackage{enumitem} 
\setlist{nolistsep} 
\usepackage{pifont} 
\usepackage{booktabs} 
\usepackage{xcolor} 
\definecolor{purple}{RGB}{127,63,152} 

\usepackage{fancybox, graphicx} 

\usepackage{times} 
\usepackage{mathptmx} 

\usepackage{microtype} 
\usepackage[utf8]{inputenc} 
\usepackage[T1]{fontenc} 

\usepackage{multicol} 
\usepackage{etoolbox}
\usepackage{relsize}

\usepackage{calc} 
\usepackage{makeidx} 
\makeindex 

\usepackage{titletoc} 

\contentsmargin{0cm} 

\titlecontents{part}[0cm]
{\addvspace{16pt}\centering\large\bfseries}
{}
{}
{}

\titlecontents{chapter}[1.25cm] 
{\addvspace{12pt}\large\sffamily\bfseries} 
{\color{purple!60}\contentslabel[\Large\thecontentslabel]{1.25cm}\color{purple}} 
{\color{purple}}  
{\color{purple!60}\normalsize\;\titlerule*[.5pc]{.}\;\thecontentspage} 

\titlecontents{section}[1.25cm] 
{\addvspace{3pt}\sffamily\bfseries} 
{\color{purple!60}\contentslabel[\thecontentslabel]{1.25cm}} 
{}
{\color{purple}\ \titlerule*[.5pc]{.}\;\thecontentspage} 
[]

\titlecontents{subsection}[1.25cm] 
{\addvspace{1pt}\sffamily\small} 
{\color{purple!60}\contentslabel[\thecontentslabel]{1.25cm}} 
{}
{\color{purple}\ \titlerule*[.5pc]{.}\;\thecontentspage} 
[]

\titlecontents{figure}[0em]
{\addvspace{-5pt}\sffamily}
{\thecontentslabel\hspace*{1em}}
{}
{\ \titlerule*[.5pc]{.}\;\thecontentspage}
[]

\titlecontents{table}[0em]
{\addvspace{-5pt}\sffamily}
{\thecontentslabel\hspace*{1em}}
{}
{\ \titlerule*[.5pc]{.}\;\thecontentspage}
[]

\titlecontents{lchapter}[0em] 
{\addvspace{15pt}\large\sffamily\bfseries} 
{\color{purple}\contentslabel[\Large\thecontentslabel]{1.25cm}\color{purple}} 
{}  
{\color{purple}\normalsize\sffamily\bfseries\;\titlerule*[.5pc]{.}\;\thecontentspage} 

\titlecontents{lsection}[0em] 
{\sffamily\small} 
{\contentslabel[\thecontentslabel]{1.25cm}} 
{}
{}

\titlecontents{lsubsection}[.5em] 
{\normalfont\footnotesize\sffamily} 
{}
{}
{}

\usepackage{fancyhdr} 

\fancypagestyle{plain}{\fancyhead{}} 

\usepackage{amsmath,amsfonts,amssymb,amsthm} 

\RequirePackage[framemethod=default]{mdframed} 

\newmdenv[skipabove=7pt,
skipbelow=7pt,
rightline=false,
leftline=true,
topline=false,
bottomline=false,
linecolor=purple,
innerleftmargin=5pt,
innerrightmargin=5pt,
innertopmargin=0pt,
leftmargin=0cm,
rightmargin=0cm,
linewidth=4pt,
innerbottommargin=0pt]{dBox}

\makeatletter
\renewcommand{\@seccntformat}[1]{\llap{\textcolor{purple}{\csname the#1\endcsname}\hspace{1em}}}                    
\renewcommand{\section}{\@startsection{section}{1}{\z@}
{-4ex \@plus -1ex \@minus -.4ex}
{1ex \@plus.2ex }
{\normalfont\large\sffamily\bfseries\textcolor{purple}}}
\renewcommand{\subsection}{\@startsection {subsection}{2}{\z@}
{-3ex \@plus -0.1ex \@minus -.4ex}
{0.5ex \@plus.2ex }
{\normalfont\sffamily\bfseries\textcolor{purple}}}
\renewcommand{\subsubsection}{\@startsection {subsubsection}{3}{\z@}
{-2ex \@plus -0.1ex \@minus -.2ex}
{.2ex \@plus.2ex }
{\normalfont\small\sffamily\bfseries\textcolor{purple}}}                        
\renewcommand\paragraph{\@startsection{paragraph}{4}{\z@}
{-2ex \@plus-.2ex \@minus .2ex}
{.1ex}
{\normalfont\small\sffamily\bfseries\textcolor{purple}}}

\newcommand{\@mypartnumtocformat}[2]{%
\setlength\fboxsep{0pt}%
\noindent\colorbox{purple!20}{\strut\parbox[c][.7cm]{\ecart}{\color{purple!70}\Large\sffamily\bfseries\centering#1}}\hskip\esp\colorbox{purple!40}{\strut\parbox[c][.7cm]{\linewidth-\ecart-\esp}{\Large\sffamily\centering#2}}}%
\newcommand{\@myparttocformat}[1]{%
\setlength\fboxsep{0pt}%
\noindent\colorbox{purple!40}{\strut\parbox[c][.7cm]{\linewidth}{\Large\sffamily\centering#1}}}%
\newlength\esp
\newlength\ecart
\def\@part[#1]#2{%
\ifnum \c@secnumdepth >-2\relax%
\refstepcounter{part}%
\addcontentsline{toc}{part}{\texorpdfstring{\protect\@mypartnumtocformat{\thepart}{#1}}{\partname~\thepart\ ---\ #1}}
\else%
\addcontentsline{toc}{part}{\texorpdfstring{\protect\@myparttocformat{#1}}{#1}}%
\fi%
\startcontents%
\markboth{}{}%
{\thispagestyle{empty}%
\begin{tikzpicture}[remember picture,overlay]%
\node at (current page.north west){\begin{tikzpicture}[remember picture,overlay]%
\fill[purple!20](0cm,0cm) rectangle (\paperwidth,-\paperheight);
\node[anchor=north] at (4cm,-3.25cm){\color{purple!40}\fontsize{220}{100}\sffamily\bfseries\thepart}; 
\node[anchor=south east] at (\paperwidth-1cm,-\paperheight+1cm){\parbox[t][][t]{8.5cm}{
\printcontents{l}{0}{\setcounter{tocdepth}{1}}%
}};
\node[anchor=north east] at (\paperwidth-1.5cm,-3.25cm){\parbox[t][][t]{15cm}{\strut\raggedleft\color{white}\fontsize{30}{30}\sffamily\bfseries#2}};
\end{tikzpicture}};
\end{tikzpicture}}%
\@endpart}
\def\@spart#1{%
\startcontents%
\phantomsection
{\thispagestyle{empty}%
\begin{tikzpicture}[remember picture,overlay]%
\node at (current page.north west){\begin{tikzpicture}[remember picture,overlay]%
\fill[purple!20](0cm,0cm) rectangle (\paperwidth,-\paperheight);
\node[anchor=north east] at (\paperwidth-1.5cm,-3.25cm){\parbox[t][][t]{15cm}{\strut\raggedleft\color{white}\fontsize{30}{30}\sffamily\bfseries#1}};
\end{tikzpicture}};
\end{tikzpicture}}
\addcontentsline{toc}{part}{\texorpdfstring{%
\setlength\fboxsep{0pt}%
\noindent\protect\colorbox{purple!40}{\strut\protect\parbox[c][.7cm]{\linewidth}{\Large\sffamily\protect\centering #1\quad\mbox{}}}}{#1}}%
\@endpart}
\def\@endpart{\vfil\newpage
\if@twoside
\if@openright
\null
\thispagestyle{empty}%
\newpage
\fi
\fi
\if@tempswa
\twocolumn
\fi}

\newif\ifusechapterimage
\usechapterimagetrue
\newcommand{\thechapterimage}{}%
\newcommand{\chapterimage}[1]{\ifusechapterimage\renewcommand{\thechapterimage}{#1}\fi}%
\newcommand{\autodot}{.}
\def\@makechapterhead#1{%
{\parindent \z@ \raggedright \normalfont
\ifnum \c@secnumdepth >\m@ne
\if@mainmatter
\begin{tikzpicture}[remember picture,overlay]
\node at (current page.north west)
{\begin{tikzpicture}[remember picture,overlay]
\node[anchor=north west,inner sep=0pt] at (0,0) {\ifusechapterimage\includegraphics[width=\paperwidth]{\thechapterimage}\fi};
\draw[anchor=west] (\Gm@lmargin,-9cm) node [line width=2pt,rounded corners=15pt,draw=purple,fill=white,fill opacity=0.7,inner sep=20pt]{\strut\makebox[22cm]{}};
\draw[anchor=west] (\Gm@lmargin+.3cm,-9cm) node {\large\sffamily\bfseries\color{purple}\thechapter\autodot~#1\strut};
\end{tikzpicture}};
\end{tikzpicture}
\else
\begin{tikzpicture}[remember picture,overlay]
\node at (current page.north west)
{\begin{tikzpicture}[remember picture,overlay]
\node[anchor=north west,inner sep=0pt] at (0,0) {\ifusechapterimage\includegraphics[width=\paperwidth]{\thechapterimage}\fi};
\draw[anchor=west] (\Gm@lmargin,-9cm) node [line width=2pt,rounded corners=15pt,draw=purple,fill=white,fill opacity=0.7,inner sep=15pt]{\strut\makebox[22cm]{}};
\draw[anchor=west] (\Gm@lmargin+.3cm,-9cm) node {\large\sffamily\bfseries\color{purple}#1\strut};
\end{tikzpicture}};
\end{tikzpicture}
\fi\fi\par\vspace*{270\p@}}}

\def\@makeschapterhead#1{%
\begin{tikzpicture}[remember picture,overlay]
\node at (current page.north west)
{\begin{tikzpicture}[remember picture,overlay]
\node[anchor=north west,inner sep=0pt] at (0,0) {\ifusechapterimage\includegraphics[width=\paperwidth]{\thechapterimage}\fi};
\draw[anchor=west] (\Gm@lmargin,-9cm) node [line width=2pt,rounded corners=15pt,draw=purple,fill=white,fill opacity=0.7,inner sep=15pt]{\strut\makebox[22cm]{}};
\draw[anchor=west] (\Gm@lmargin+.3cm,-9cm) node {\large\sffamily\bfseries\color{purple}#1\strut};
\end{tikzpicture}};
\end{tikzpicture}
\par\vspace*{270\p@}}
\makeatother

\usepackage[colorlinks = true,
            linkcolor = purple,
            urlcolor  = blue,
            citecolor = blue,
            anchorcolor = blue]{hyperref}
\usepackage{hyperref}

\usepackage{bookmark}
\bookmarksetup{
open,
numbered,
addtohook={%
\ifnum\bookmarkget{level}=0 
\bookmarksetup{bold}%
\fi
\ifnum\bookmarkget{level}=-1 
\bookmarksetup{color=purple,bold}%
\fi
}
}

\newcommand\lsim{\mathrel{\rlap{\lower4pt\hbox{\hskip1pt$\sim$}}
        \raise1pt\hbox{$<$}}}
\newcommand\gsim{\mathrel{\rlap{\lower4pt\hbox{\hskip1pt$\sim$}}
        \raise1pt\hbox{$>$}}}
\def\myputfigure#1#2#3#4#5%
{\vskip#5pt\makebox[0pt]{\hskip#2in
\includegraphics[width=#3\textwidth]{#1}}\vskip#4pt\hfill}

\usepackage[natbib=true,backend=bibtex,style=authoryear,maxcitenames=3,doi=false,isbn=false,url=false]{biblatex}
\AtEveryBibitem{\clearfield{title}}
\AtEveryBibitem{\clearfield{eprint}}
\AtEveryBibitem{\clearfield{archivePrefix}}
\usepackage{ulem}
\usepackage{unitsdef}
\def\msun{\,\rm{M_\odot}}

\usepackage[italic]{mathastext}
\usepackage{isomath}

\begin{document}


\begingroup
\thispagestyle{empty}
\begin{tikzpicture}[remember picture,overlay]
\node[inner sep=0pt] (background) at (current page.center) {\includegraphics[width=\paperwidth]{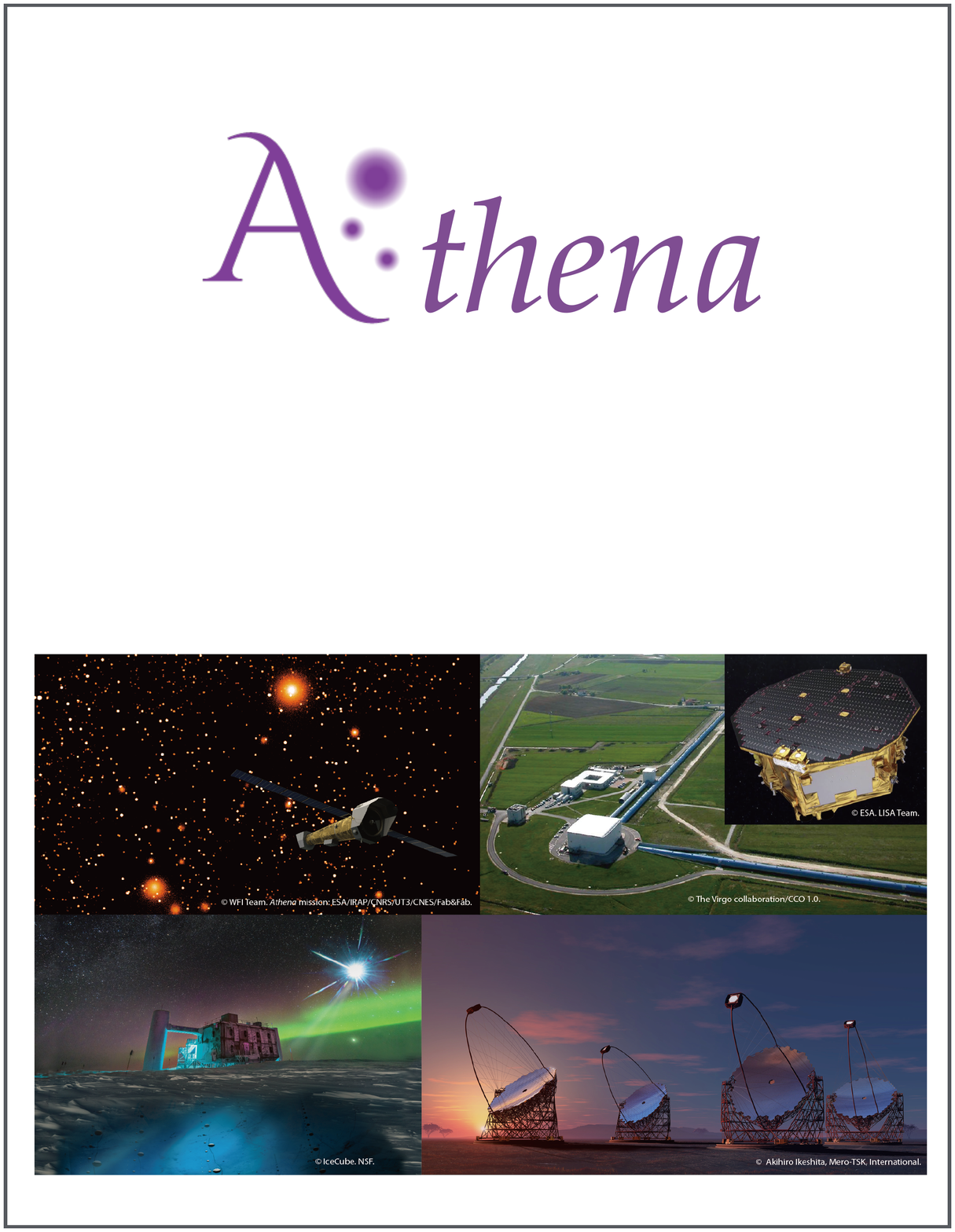}};
\draw (current page.center) (7,-10) node {\bfseries\sffamily\parbox[c][][t]{\paperwidth}{\color{purple}\centering{\Huge \hspace*{0.7cm}Multi-messenger-\textsl{Athena} Synergy White Paper}\\ 
\vspace*{0.5cm}
{\color{black}\huge Multi-messenger-\textsl{Athena} Synergy Team}}}; 
\end{tikzpicture}
\vfill
\endgroup


\newpage
~\vfill
\thispagestyle{empty}
\noindent October 2021.\\ 
Edited by: Francisco J. Carrera and Silvia Mart\'inez-N\'u\~nez on behalf of the \textsl{Athena} Community Office.\\
Revisions provided by the \textsl{Athena} Science Study Team.

\thispagestyle{empty} 
\chapterimage{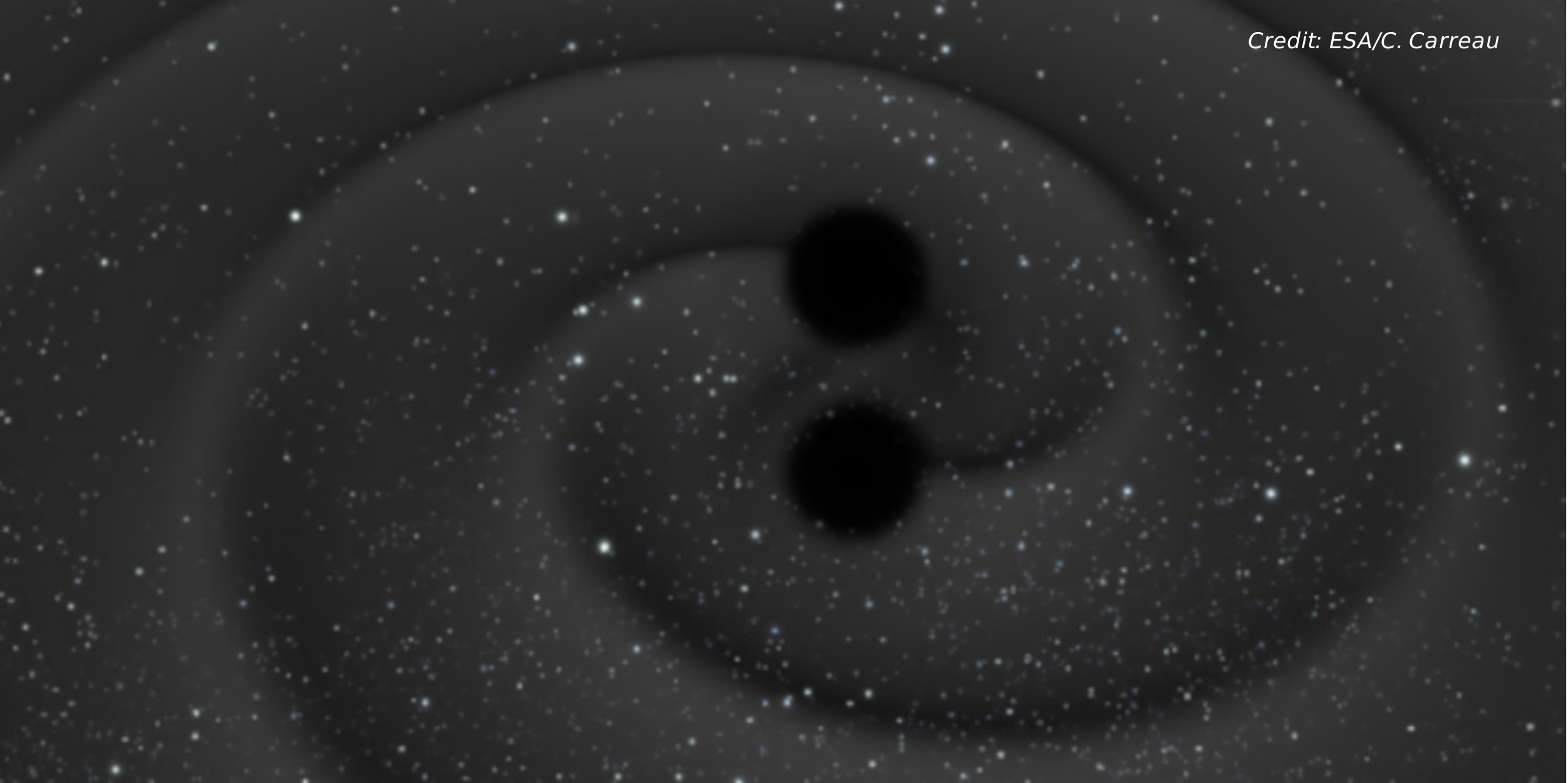} 

{\color{purple} 
\chapter*{Authorship}

\section*{Authors}}

\begin{itemize}
\small
    \item 
L. Piro, INAF-IAPS, Italy
    \item 
M. Ahlers, University of Copenhagen, Denmark
  \item 
A. Coleiro, APC-University of Paris, France
  \item 
M. Colpi, University of Milano - Bicocca, Italy
  \item 
E. de Oña Wilhelmi, DESY-Zeuthen, Germany
  \item 
M. Guainazzi, ESA
  \item 
P. G. Jonker, SRON, The Netherlands
  \item 
P. Mc Namara, ESA
  \item
D. A. Nichols, University of Virginia, USA
  \item 
P. O’ Brien, University of Leicester, UK
  \item 
E. Troja, NASA/GSFC \& University of Maryland, USA
  \item 
J. Vink, University of Amsterdam, The Netherlands
\end{itemize}

{\color{purple}
\section*{Contributors}
}
	\begin{itemize}
	\small
   \item 
J. Aird, University of Edinburgh, UK
    \item
L. Amati, INAF-OAS, Italy
  \item
S. Anand, Caltech, USA
   \item
E. Bozzo, University of Geneva, Switzerland
  \item
F. J. Carrera, Instituto de F\'\i{}sica de Cantabria (CSIC-UC), Spain
  \item
A. C. Fabian,  Institute of Astronomy, University of Cambridge
  \item
C. Fryer, Los Alamos National Laboratory
  \item
E. Hall, MIT, USA
  \item
O. Korobkin, Los Alamos National Laboratory
  \item
V. Korol, University of Birmingham, UK
   \item
 J. Osborne, University of Leicester, UK
  \item
A. Mangiagli, University of Milano - Bicocca, Italy
  \item
S. Mart\'\i{}nez-Nu\~nez, Instituto de F\'\i{}sica de Cantabria (CSIC-UC), Spain
  \item
S. Nissanke, GRAPPA centre, University of Amsterdam, The Netherlands
  \item
P. Padovani, ESO
  \item
E.M. Rossi, Leiden Observatory, The Netherlands
  \item
G. Ryan, University of Maryland, USA
  \item
A. Sesana, University of Milano - Bicocca, Italy
  \item
G. Stratta, INAF-OAS, Italy
  \item
N. Tanvir, University of Leicester, UK
  \item
H. van Eerten,  University of Bath, UK

\end{itemize}

\let\cleardoublepage\clearpage 


\thispagestyle{empty} 
\chapterimage{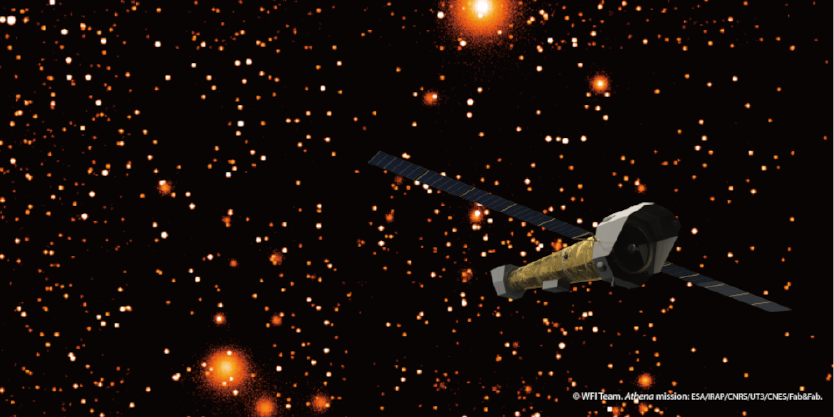} 
\tableofcontents 
\cleardoublepage 

\pagestyle{fancy} 
\fancyfoot[CE,CO]{\color{purple}Page \thepage} 

%
\chapterimage{Images/MergingBH.pdf} 
\chapter{Abstract}

In this paper we explore the  scientific synergies between \textsl{Athena} and some of the key multi-messenger facilities that should be operative concurrently with \textsl{Athena}.  These facilities include LIGO A+, Advanced Virgo+ and future detectors for ground-based observation of gravitational waves (GW), LISA for space-based observations of GW, IceCube and KM3NeT for neutrino observations, and CTA for very high energy observations. These science themes encompass  pressing issues in astrophysics, cosmology and fundamental physics such as: the central engine and jet physics in compact binary mergers, accretion processes and jet physics in Super-Massive Binary Black Holes (SMBBHs) and in compact stellar binaries, the equation of state of neutron stars, cosmic accelerators and the origin of Cosmic Rays (CRs), the origin of intermediate and high-Z elements in the Universe, the Cosmic distance scale and tests of General Relativity and the Standard Model. Observational strategies for implementing the identified science topics are also discussed.  

A significant part of the sources targeted by multi-messenger facilities is of transient nature. We have thus also discussed the synergy of \textsl{Athena} with wide-field high-energy facilities, taking THESEUS as a case study for transient discovery. This discussion covers all the \textsl{Athena} science goals that rely on follow-up observations of high-energy transients identified by external observatories, and includes also topics that are not based on multi-messenger observations, such as the search for missing baryons or the observation of early star populations and metal enrichment at the cosmic dawn with Gamma-Ray Bursts (GRBs). 


\chapterimage{Images/Athena.pdf}
\chapter{Introduction}
Recent years witnessed a blossoming of multi-messenger astrophysics, in which gravitational waves (GWs), neutrinos, and photons provide complementary views of the universe. The astounding results obtained from the joint electromagnetic-gravitational wave observations of the compact binary merger GW170817 or from the neutrino-electromagnetic (EM) observations of the blazar TXS~056+056  showed the tremendous discovery potential of this field, that will be progressively exploited throughout the next decade, as  observing facilities are deployed. A substantial step forward could be expected by early 2030s, when the second and third generations of GW and neutrino detectors will become operational.

A full exploitation of the potential of multi-messenger astronomy demands also capabilities in the X-ray band that are beyond those achievable by current and near future missions, but consistent with the performance planned for \textsl{Athena}. For example, population studies of X-ray counterparts to GW mergers at the distances probed by next generation of GW detectors require an X-ray sensitivity that only \textsl{Athena} can provide.
  A significant part of the sources targeted by multi-messenger facilities is of transient nature. We have thus also discussed the synergy with wide-field high-energy facilities. This discussion is extended to those \textsl{Athena} science goals that, while not being strictly multi-messenger, rely on follow-up observations of high-energy transients identified by external observatories. 

ESA has established the \textsl{Athena} Science Study Team (ASST) to provide guidance on all scientific aspects of the \textsl{Athena} mission. One of the ASST’s tasks is to identify and elaborate synergies with various astronomical facilities, which will be available in the 2030s time frame. The Multi-messenger-\textsl{Athena} Synergy Team has been tasked by the ASST and the facilities involved to single out the potential scientific synergies between \textsl{Athena} and some of the key multi-messenger and high-energy astronomical facilities covering GW, neutrinos, very high energy (VHE) and high energy transients that should be operative concurrently with \textsl{Athena}. These facilities include LIGO A+, Advanced VIRGO+ and future detectors for ground-based observation of GW, LISA for space-based observations of GW, IceCube and KM3NeT for neutrino observations, CTA for VHE observations and THESEUS as a case study for transient discovery in X-rays.
Although THESEUS was eventually not selected by ESA as an M5 mission, its advanced assessment allowed us to carry out a detailed study of the synergy, and provide a reference for future high energy transient missions.


In this paper we discuss the main synergy science themes, emphasize their relevance with respect to the core science of each facility, and elaborate detailed science topics, outlining the observational strategy required to their successful achievement.
 
\section{\textsl{Athena} as a multi-messenger observatory}\label{sec:athena_intro}
 
\begin{figure*}
\centering
\vspace{-1\baselineskip}
\includegraphics[width=0.4\textwidth]{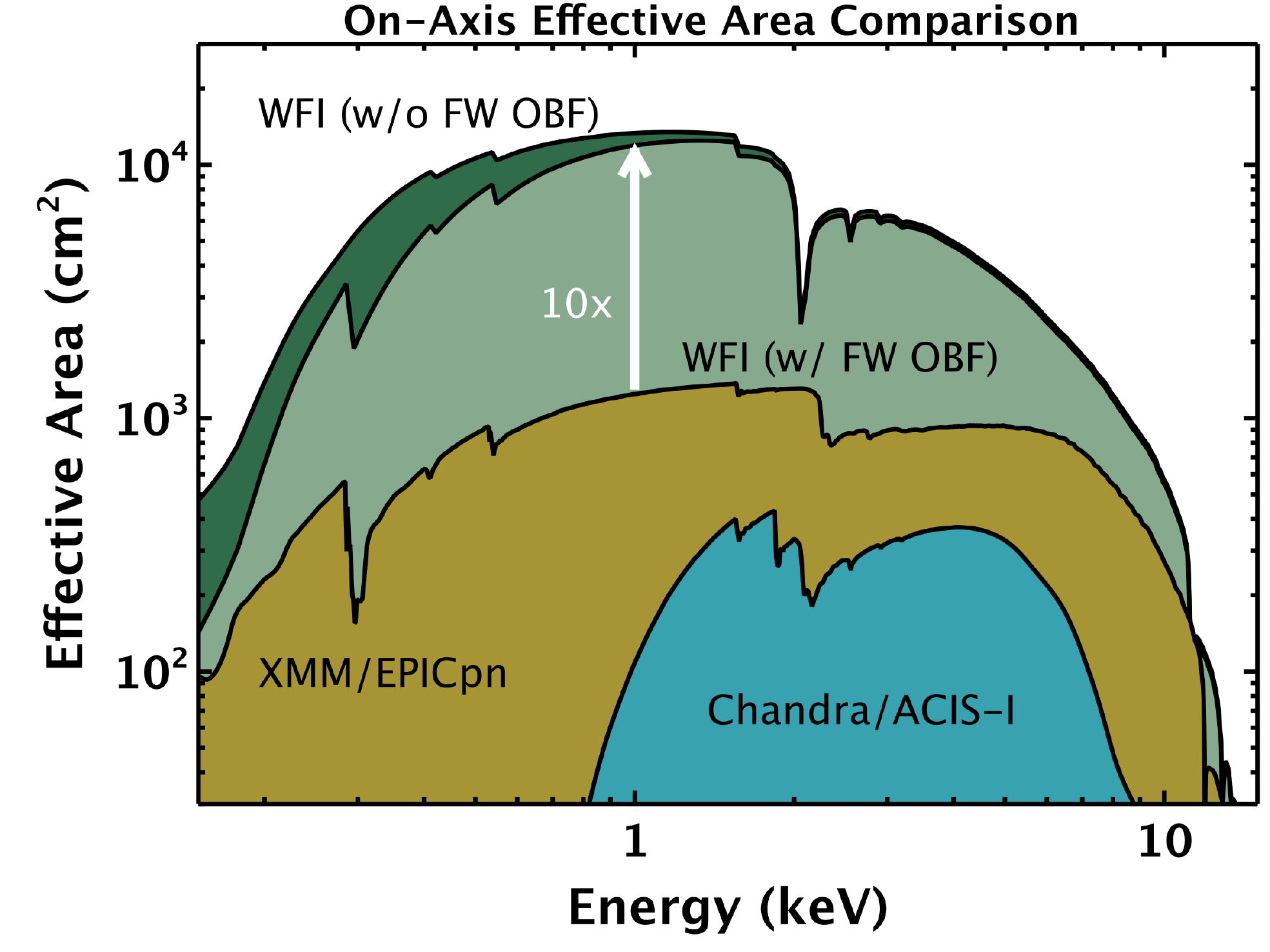}
\hspace*{1cm}
\includegraphics[width=0.4\textwidth]{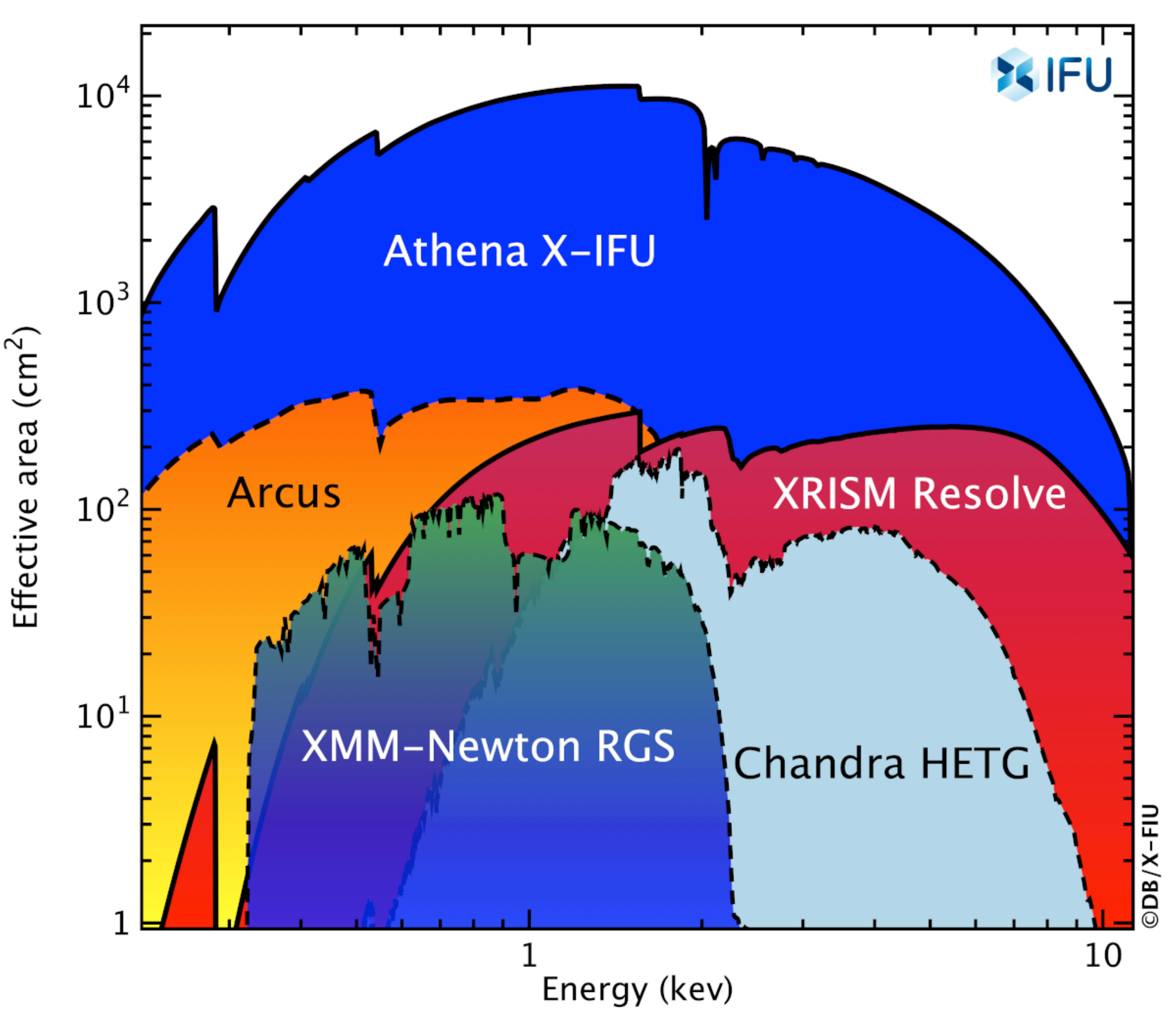}
\caption{
{\it Left:} \textsl{Athena}/WFI effective on-axis area (with and without Filter Wheel Optical Blocking Filter) compared to XMM-\textsl{Newton}/EPIC-pn (medium filter) and Chandra/ACIS-I. The \textsl{Athena}/WFI curve assumes an SPO optics with 15 mirror rows, 2.3 mm rib pitch, and Ir+B4C coating. Credits: A.Rau/WFI Team.  {\it Right}: Effective area comparison between the \textsl{Athena}/X-IFU and other currently in operation and future X-ray spectrometers. The X-IFU provides an increase of effective area by a factor of 45 and 6 at 1 and 7~keV with respect to the XRISM/{\it Resolve}, respectively. Credits: X-IFU Consortium.}\label{fig:AthenaFoM}
\end{figure*}

\textsl{Athena} (\href{http://www.the-athena-x-ray-observatory.eu/}{Advanced Telescope for High ENergy Astrophysics}) is the X-ray observatory large mission selected by the European Space Agency (ESA), within its Cosmic Vision 2015-2025 programme, to address the Hot and Energetic Universe scientific theme \parencite{ath14}, and it is provisionally due for launch in the early 2030s. \textsl{Athena} will have three key elements to its scientific payload: an X-ray telescope with a focal length of 12 m and two instruments: a Wide Field Imager (\href{https://www.mpe.mpg.de/ATHENA-WFI/}{\textsl{Athena}/WFI}) \citep{WFIpaper} for high count rate, moderate resolution spectroscopy over a large Field of View (FoV) and an X-ray Integral Field Unit (\href{http://x-ifu.irap.omp.eu/}{\textsl{Athena}/X-IFU}) 
\citep{barret:2021}for high-spectral resolution imaging. In Tab.~\ref{Athena_specs} we report the main requirements of the \textsl{Athena} observatory, as constrained by the key core science drivers of the mission and Figure~\ref{fig:AthenaFoM} shows the effective area of the two instruments.

\begin{figure*}
\centering
\vspace{-1\baselineskip}
\includegraphics[width=0.45\textwidth]{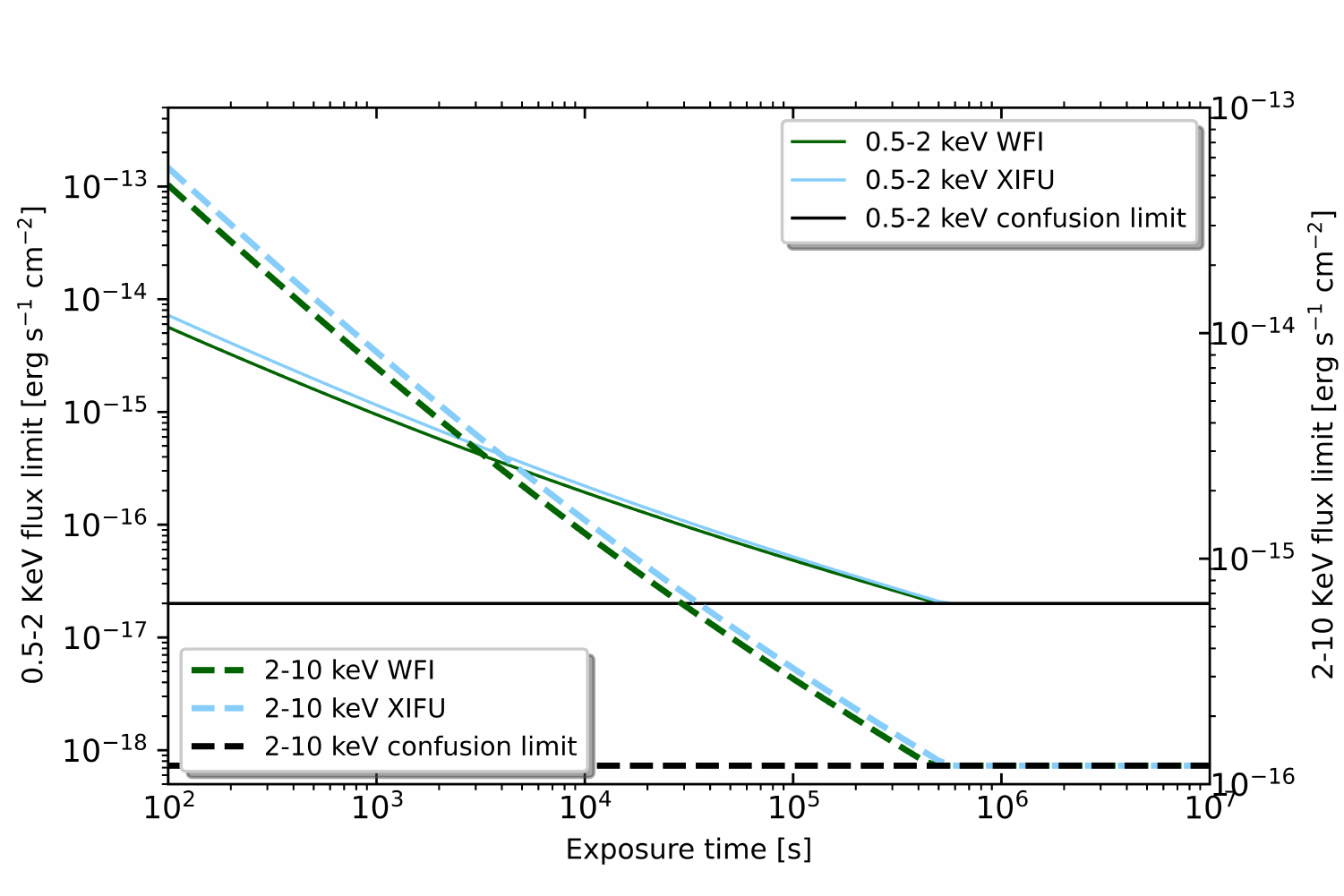}
\includegraphics[width=0.35\textwidth]{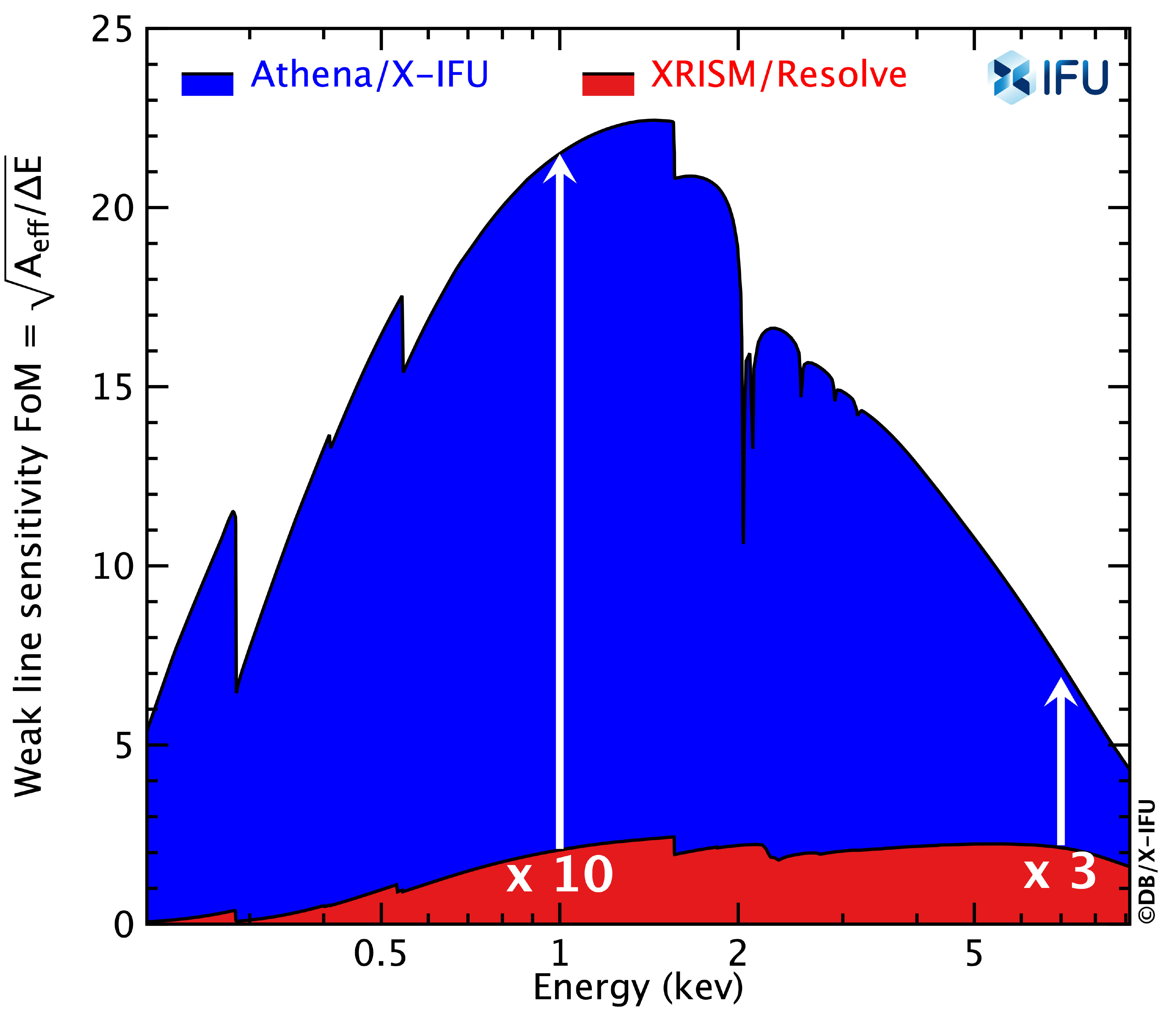}
\caption{Left: \textsl{Athena}/WFI and X-IFU sensitivity (0.5-2~keV on the left axis and 2-10~keV on the right axis) for on-axis point source as function of integration time. At around 400~ks the sensitivity reaches the confusion limit. Right: Comparison of the weak line sensitivity of \textsl{Athena}/X-IFU and XRISM/{\it Resolve} spectrometers. }\label{fig:Athenasensitivity}
\end{figure*}

\begin{figure*}
\centering
\vspace{-1\baselineskip}
\includegraphics[width=0.8\textwidth]{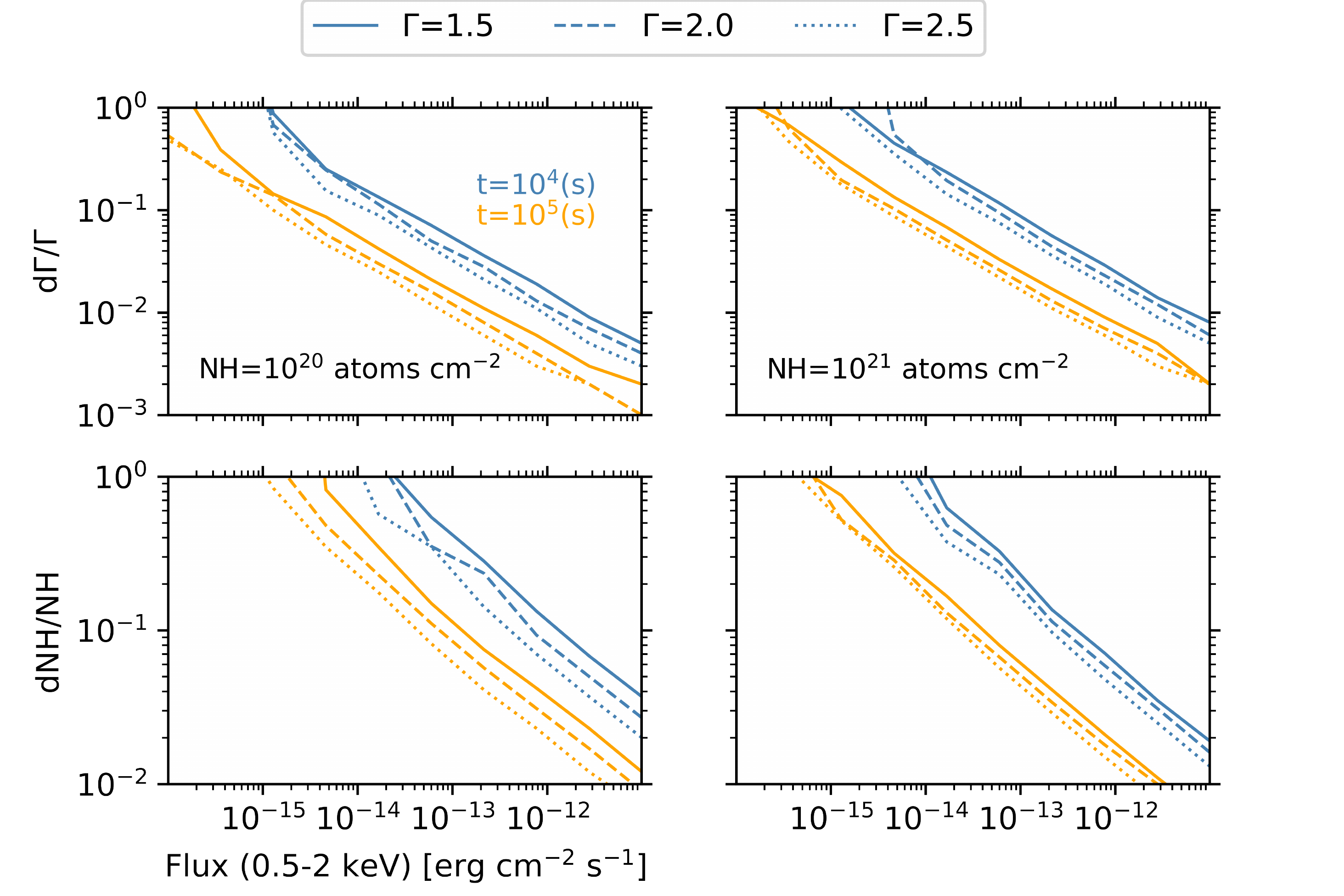}
\caption{
 \textsl{Athena} WFI/X-IFU capability to measure the continuum spectral shape (expressed as relative error on the photon index of a power-law spectrum and on the equivalent neutral Hydrogen absorption column density) as a function of the source flux for different integration times and spectral parameters.}
 \label{fig:Athenaspectrum}
\end{figure*}

\begin{figure*}
\centering
\vspace{-1\baselineskip}
\includegraphics[width=0.8\textwidth]{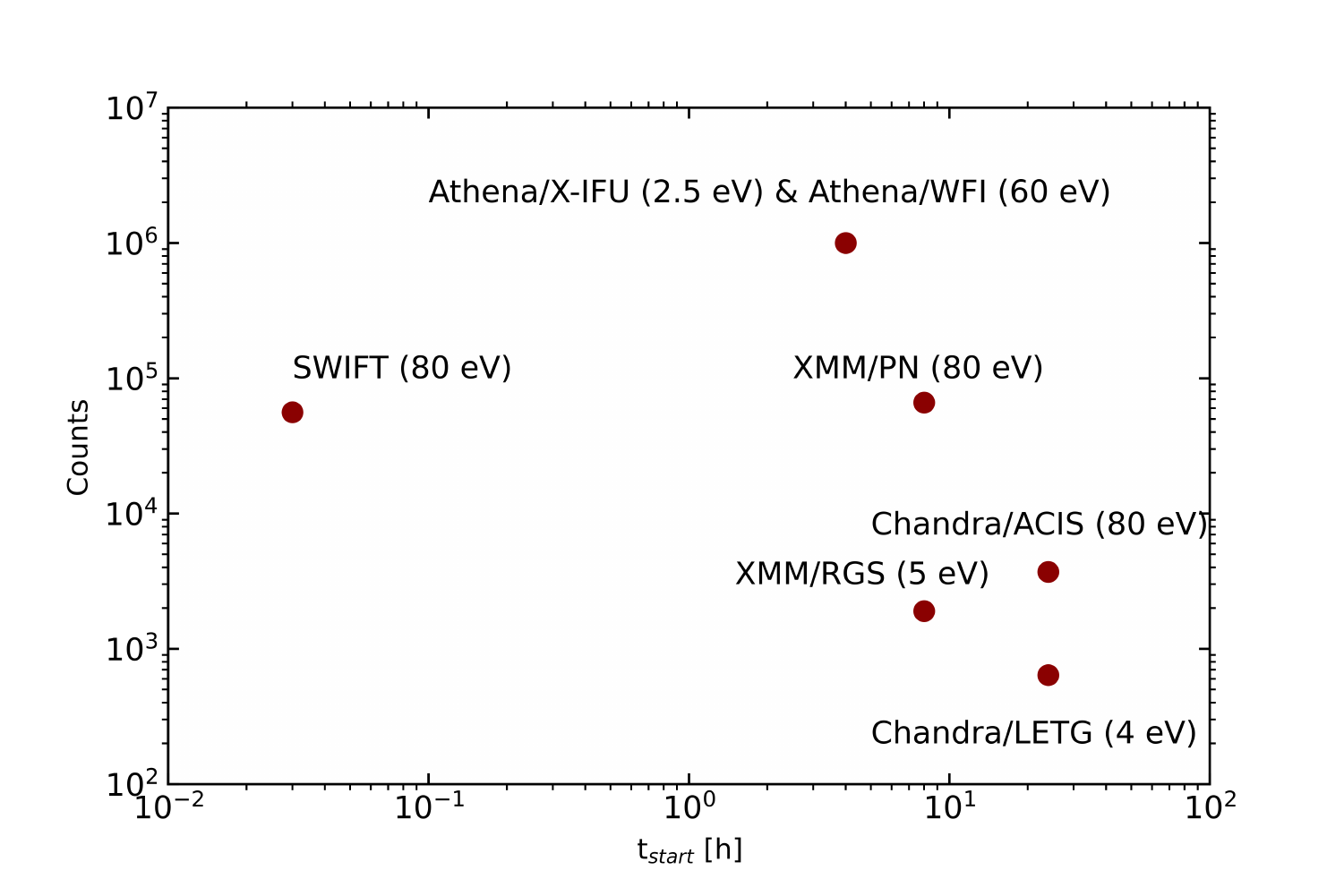}
\caption{
 \textsl{Athena} capability on Targets of Opportunity (ToO). Number of counts gathered for a bright GRB integrated for about 50~ks as a function of the typical ToO response time for various instruments (energy resolution at 1~keV in parenthesis). The one million counts observed by \textsl{Athena} enable high resolution absorption spectroscopy of extremely weak lines. }
 \label{fig:AthenaTOO}
\end{figure*}

Most of the sources targeted by multi-messenger astronomy are related to energetic phenomena, such as stellar explosions, compact objects (black hole [BH]; neutron star [NS]; white dwarf [WD]), accelerations sites at all scales, and transients. These are the main constituents of the \textsl{Athena} science themes and, as such, have driven the science performance of the mission, that are therefore already largely tuned for a multi-messenger approach.
In this regard \textsl{Athena} provides a unique combination of performances for the benefit of multi-messenger astronomy.
\begin{itemize}
    \item The large FoV (0.4 $\deg^2$) catered by the WFI, boosted by the capability of carrying out mosaic or raster scans up to 10 $\deg^2$,  allows us to cover error boxes of GW, neutrino and VHE sources down to the unprecedented sensitivity enabled by \textsl{Athena} (next bullet). This capability, coupled with the much smaller number of serendipitous sources expected in X-rays than  at lower frequencies, will help in discovering the EM counterpart.
    \item A sensitivity of $2\times 10^{-17} \rm{erg\ cm^{-2}\ s^{-1}} $ (Fig.~\ref{fig:Athenasensitivity}, left panel) enables the discovery and the characterization of the temporal evolution of the faintest X-ray counterparts of multi-messenger events, such as the X-ray kilonovae.
    \item The combination of large area and low background  allows \textsl{Athena} to characterize the spectral properties of  faint X-ray sources, important e.g. for constraining leptonic vs hadronic models in neutrino and VHE sources, or tracking the spectral evolution of afterglows of GRBs and GW mergers.
    (Fig.~\ref{fig:Athenaspectrum})
    \item A source positional accuracy of $\approx 1 \arcsec$ allows a precise location of the counterpart, enabling follow-up observations by other large EM facilities with a narrow FoV.
    \item The high spectral resolution (2.5 eV) and imaging capabilities of the X-IFU enable searches of extremely faint narrow lines from a rich variety of sources, from WHIM to radioactive decay from kilonova remnants, as well as to uniquely characterize the sites of particle acceleration in the Universe.
    \item The reaction time to Target of Opportunity (ToO) (4~h) coupled with the large fraction of the sky accessible at any time (>50\%) and the large effective area allow \textsl{Athena} to follow-up GRBs and other multi-messenger transients fast enough to a) gather an adequate number of photons to enable high resolution absorption spectroscopy with the X-IFU (Fig.~\ref{fig:AthenaTOO}); b) to detect dim and fastly decaying sources of counterparts of multi-messenger events.
\end{itemize}

\begin{table}[h]
\caption{\textsl{Athena} main scientific requirements}\label{Athena_specs}
\vspace*{0.5cm}
\footnotesize
\begin{tabular}{l|l|l}
{\bf Parameter} & {\bf Requirements} & {\bf Scientific driver} \\
\hline
Effective Area at 1keV & $\ge$1.4 \rm{m}$^2$  &  Early groups, cluster entropy \& metal evolution, WHIM, first stars with GRBs\\
Effective Area at 6keV & 0.25 \rm{m}$^2$ &  Cluster bulk motions and turbulence, AGN winds and outflows, BH spin \\
HEW (spatial resolution) & 5'' on-axis, 10'' off-axis & High-z AGN, early groups, AGN feedback in galaxy clusters \\
WFI point source sensitivity & 2.4 $\times 10^{-17}$ \rm{erg/s/cm}$^2$  & AGN evolution, early groups\\
 & (in 450 ks at 0.5-2~keV) & \\
X-IFU spectral resolution & 2.5 eV & WHIM, cluster hot gas energetics and AGN feedback, AGN outflows\\
WFI spectral resolution & 150 eV (at 6~keV) & Galactic BH spin, reverberation mapping \\
WFI FoV & 40' $\times$ 40' square &  High-z AGN, AGN census, early groups, cluster entropy evolution \\
X-IFU FoV & 5' effective diameter &  AGN feedback in clusters, intergalactic medium physics\\
ToO Trigger efficiency & 50\% & WHIM, GRBs\\
ToO Reaction time & $\le$ 4 hrs & WHIM, first stars with GRBs \\
\hline		    
\noalign{\smallskip}
\end{tabular}
\end{table}

\chapterimage{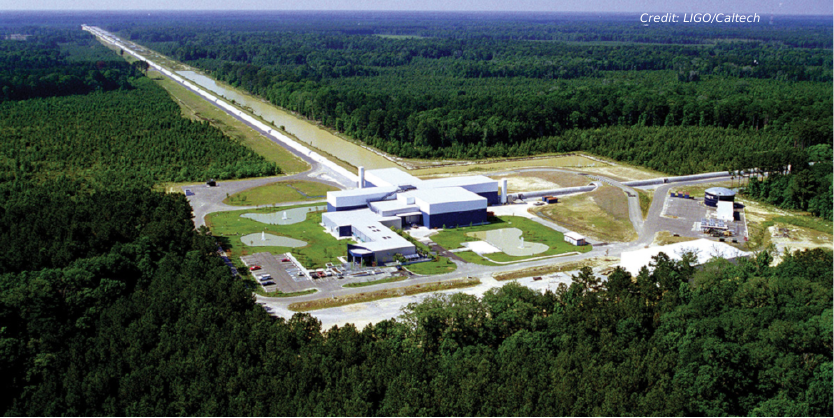} 
\chapter{ \textsl{Athena} and ground-based GW observatories: LIGO, Virgo, and future detectors}\label{chapter:GW}

\textit{\color{purple}{By E. Troja, D. A. Nichols, P. G. Jonkers, S. Anand, C. Fryer, E. Hall,  O. Korobkin, S. Nissanke, L. Piro, G. Ryan, H. van Eerten.}}

\section{Science Themes}\index{}\label{sec:GWthemes}

The observation of gravitational waves (GWs) from mergers of stellar-mass black holes (BHs) and neutron stars (NSs) opened a new vantage point on some of the most luminous and energetic events in the Universe.
A major aim of the ground-based GW observatories like \href{https://www.ligo.caltech.edu/}{LIGO}, \href{https://www.virgo-gw.eu/}{Virgo} and \href{https://gwcenter.icrr.u-tokyo.ac.jp/en/}{KAGRA} is to use these novel types of observations to explore a broad range of science themes, such as understanding the extremes of gravity and matter, studying cosmology and fundamental physics, observing the multi-messenger universe, investigating the formation and evolution of compact binaries, and searching for new classes of GW sources~\parencite{LIGOScientific:2019vkc,Sathyaprakash:2019rom,Sathyaprakash:2019yqt}.
A number of science themes can be achieved solely through GW observations, whereas others are significantly enhanced by having a multi-messenger event.
The power of multi-messenger observations is nicely illustrated by the detection of GWs from the merger of two NSs, GW170817, and the variety of EM counterparts that followed it, GRB~170817A and AT2017gfo~\parencite{Abbott-multimess}.
By combining information from the observed GWs and the associated gamma-ray burst (GRB), GRB afterglow, and kilonova, it was possible to put novel constraints on the nuclear Equation of State (EoS), to infer that NS-NS mergers are a significant source of r-process elements, to measure the value of the Hubble constant, to verify the equivalence of the speed of GWs and the speed of light, and to gain new insights into the engine that drives short GRBs~\parencite{TheLIGOScientific:2017qsa,Abbott-multimess}.

This section examines the common scientific themes and topics in compact-object astrophysics for which multi-messenger observations by ground-based GW observatories and the X-ray observatory \textsl{Athena} would be particularly fruitful.
We begin by outlining the capabilities of the current detectors LIGO, Virgo, and KAGRA for detecting compact binaries; we also give similar metrics for the next generation of detectors like LIGO Voyager, Einstein Telescope (ET), and Cosmic Explorer (CE).   
We then turn to the synergistic science themes and topics in multi-messenger and compact-object astrophysics. 
We conclude by discussing the observing requirements needed to effectively explore these themes and topics.

\subsection{The ground-based GW landscape: LIGO, VIRGO, KAGRA, Einstein Telescope, Cosmic Explorer}\label{subsec:GWdetectors}

Operation of the second-generation of ground-based GW detectors began in 2015, and the first two observing runs of LIGO and Virgo produced the detections of the first ten binary BHs and the first binary NS~\parencite{LIGOScientific:2018mvr}.
From the first half of the third observing run, an additional 39 GW detections were made, 36 of which are binary BHs, one was a binary NS, one was a binary with either the lightest BH or the heaviest NS measured, and one may have been a BH-NS binary~\parencite{Abbott:2020niy}.
The LIGO, Virgo, and KAGRA detectors will undergo additional upgrades during this decade, which will increase their rate of detections of compact-object mergers.
There are also plans to build new facilities (ET and CE), which will dramatically increase the number of detections of these mergers.
We next describe the planned upgrades to the detectors and their capabilities for detecting compact binaries in greater detail.

\textbf{Detectors in the 2020s}:
LIGO and Virgo's third observing run lasted for nearly a year, and it concluded in March 2020.
The KAGRA detector was operational for the final month of this run.
During the next few years, LIGO, Virgo, and KAGRA will undergo upgrades to reach their design sensitivities, after which there are concrete plans to make further upgrades to the LIGO and Virgo detectors (called LIGO A+ and Advanced Virgo+).
There is also a plan underway to build a LIGO detector in India, which will achieve the same sensitivity as the two LIGO A+ detectors in the US.
During the latter half of the 2020s, a five-detector network is expected. It is composed of three LIGO detectors (two in the US and one in India), an advanced Virgo detector in Italy, and KAGRA in Japan~\parencite{Aasi:2013wya}.
We therefore take this five-detector network, which we denote by HLVKI, as a conservative configuration operating at the time of the \textsl{Athena} mission.

\textbf{Detectors in the 2030s}:
There are a number of ambitious plans to dramatically increase the sensitivity of ground-based GW detectors in the 2030s.
These range from substantial upgrades to the existing 
LIGO facilities, known as LIGO Voyager~\parencite{Adhikari:2020gft}, to the design of the next generation of GW detectors.
In Europe, the ET will be an underground detector with three interferometers (two of which are independent) in a triangular shape  with 10km long arms, and it would begin operating in the early 2030s~\parencite{Punturo:2010zz}.
In the US, the third-generation GW detector is called CE, and it is planned to begin operating in 2035 (for the first phase of the detector) with further upgrades in the 2040s (the second phase of CE).
CE is intended to be two above ground, L-shaped interferometers, with 40km arms~\parencite{Reitze:2019iox}.
We consider three possible three-detector networks of different levels of sensitivity: (i) three Voyager detectors, (ii) one ET and two Voyager detectors, and (iii) one ET and two CE detectors.

\begin{figure}[htb]
\centering
\includegraphics[width=0.55\textwidth]{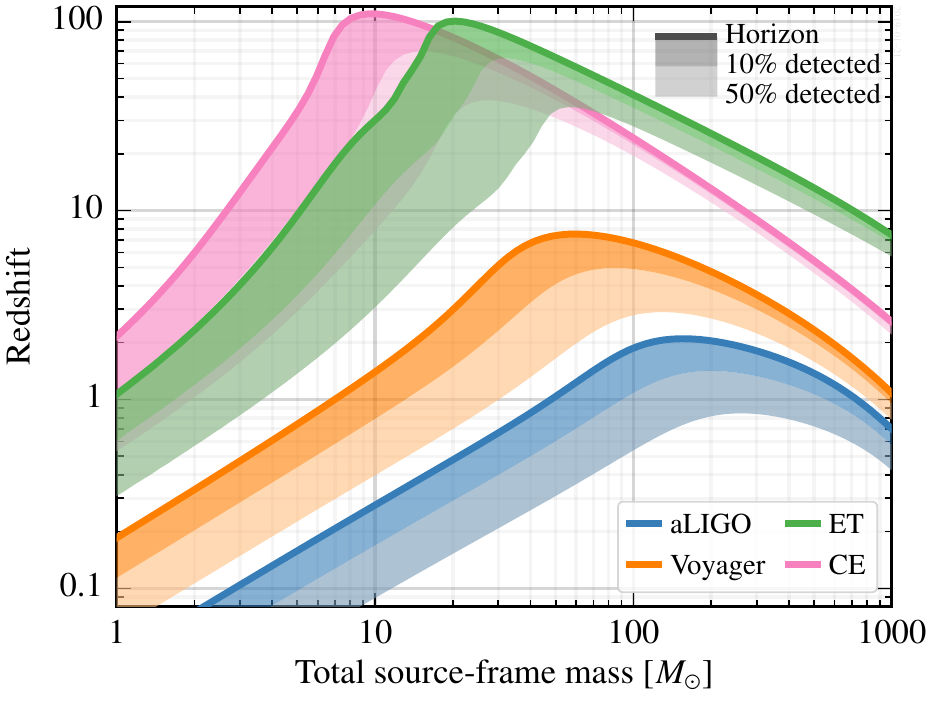} 
\includegraphics[width=0.44\textwidth]{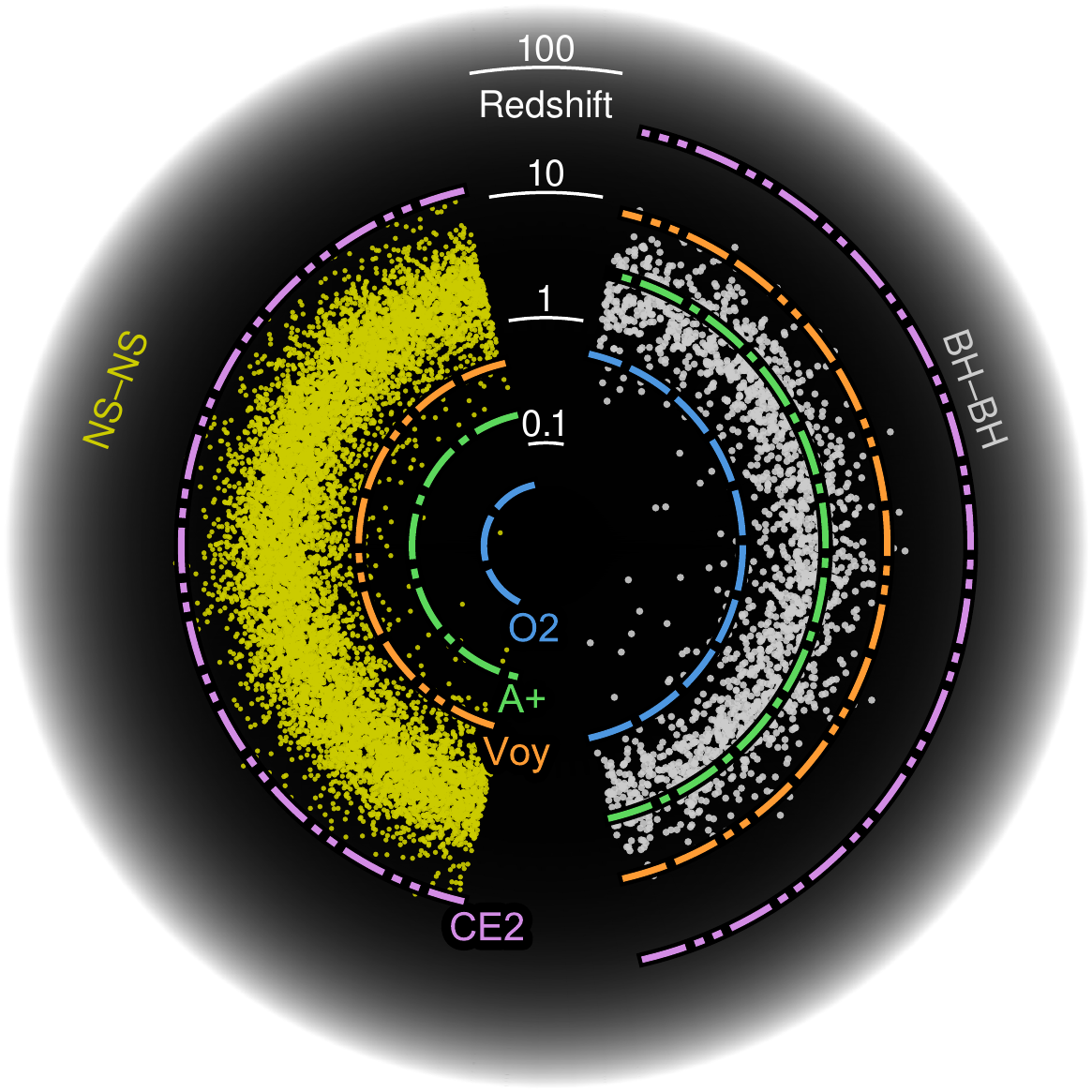}
\caption{\textit{Left}: Detection horizons and 10\% and 50\% response distances for four GW detectors, LIGO A+ (aLIGO in the figure), LIGO Voyager, ET, and CE, as a function of the binary's mass, for equal-mass binaries. Figure available from~\cite{DetectorHorizon} and appears in~\cite{2019CQGra..36v5002H}.
\textit{Right}: Populations of binary NSs (left half) and binary BHs (right half) for binaries that follow the Madau-Dickinson star-formation rate with a characteristic delay time of 100Myr. Also shown are the detection horizons of different GW detectors. The figure is available from~\cite{PopulationHorizon}.}
\label{fig:gw_horizons}
\end{figure}

\textbf{Detection horizons and localization}:
Because all the sources detected by LIGO and Virgo to date have been compact binary mergers, we focus on the detection capacities for this class of sources.
Future capabilities to observe and localize compact binary mergers are highlighted in Fig.~\ref{fig:gw_horizons} and Table~\ref{table:BNSdetect}. 
The left panel of Fig.~\ref{fig:gw_horizons} shows the detection horizons (solid lines) for four different detectors as a function of the total mass of the binary, for equal-mass non-spinning binaries. 
The detection horizon is the maximum distance that an optimally oriented binary can be detected.
The distance here is represented as a redshift which was computed using a $\Lambda$-CDM model of cosmology~\parencite{Ade:2015xua}.
Also shown as shaded region is the 10\% and 50\% response distance, respectively. The response distance is a measure of detector sensitivity defined as the luminosity distance at which 10\% (respectively 50\%) of the sources would be detected, for sources placed isotropically on the sky with random orientations, and with all sources placed at exactly this distance.

The right panel of Fig.~\ref{fig:gw_horizons} illustrates the number of two types of GW sources as a function of redshift:  binary NSs with equal mass of 1.4 $M_\odot$ in yellow, 
and binary BHs with equal mass of 30 $M_\odot$ in white.
The dotted-dashed lines are the horizons of the different ground-based detectors.
The binaries are distributed to follow the Madau-Dickinson star-formation rate with a characteristic time delay of 100~Myr (see~\cite{Vitale:2018yhm} for more details).
Figure~\ref{fig:gw_horizons} shows that the next generation of GW detectors will have a significantly larger reach and thereby detect a much larger number of compact object mergers.

In Table~\ref{table:BNSdetect}, we give numbers of expected NS-NS detections per year, localization areas, and number of detections with less than 1, 10, or 100 deg$^2$ localization areas for four possible detector networks ~\parencite{Sathyaprakash:2019rom,3Gdraft}.
The binaries were distributed in the same way as for the binary NSs shown in Fig.~\ref{fig:gw_horizons} with a local merger rate of 320 Gpc$^{-3}$ yr$^{-1}$, which is consistent with the median rate inferred from LIGO's first three observing runs~\parencite{Abbott:2020gyp}.

The number of detections per year varies by many orders of magnitude depending upon the specific detector configuration.
In the most conservative scenario, there will be tens of NS-NS events localized to better than 10 deg$^2$ per year; in the most optimistic scenario, this becomes of order $10^4$ events with a localization of less than one deg$^2$ per year.
Given the number of well-localized mergers, one can expect that EM counterparts will be associated with a subset of them. 

\textbf{Neutron-star--black-hole binaries}:
NS-BH binaries are also a promising multi-messenger source for ground-based GW detectors and \textsl{Athena}.
In June of 2021, the LIGO, Virgo and Kagra Collaborations announced the discovery of two NS-BH mergers, GW200105 and GW200115~\parencite{LIGOScientific:2021qlt}. The rate of NS-BH mergers was inferred from these observations, and it was found to have a median value of 130 Gpc$^{-3}$yr$^{-1}$. The total masses of the two binaries had median values of 11$M_\odot$ and 7.2$M_\odot$, respectively, and both merged at distances of roughly 300Mpc.  Although we do not provide quantitative forecasts of the number of detections and the localization areas for NS-BH systems, we make a few qualitative comments on their detection prospects. First, given that a NS-BH binary’s total mass falls between the masses of typical NS-NS and BH-BH binaries, from the left panel of Fig.~\ref{fig:gw_horizons}, the detection horizons and response distances are expected to be between that of binary NSs and binary BHs. Second, given that their rate of merger also falls between those of NS-NS and BH-BH binaries, the number of detections and their localizations will also be of a similar order of magnitude as those of NS-NS binaries in Table~\ref{table:BNSdetect}.
Third, EM counterparts from NS-BH mergers are expected to occur and to resemble those from binary NS mergers, which include optical and infrared kilonovae, short GRBs (SGRB) and their afterglows.
The precise similarities and differences between NS-NS and NS-BH merger counterparts is an active area of research.

\textbf{Black-hole--black-hole binaries}: For BH-BH mergers it is not obvious that an EM counterpart exists. However the possibility to estimate the exact time and position of the merger by LISA will allow \textsl{Athena} to plan in advance and  seek for any EM emission at the merging, simultaneously with observations by ground-based GW interferometers (see Sect.  \ref{sec:LISA_other}).

We turn now to discuss the synergistic science themes and topics that can be explored with these multi-messenger detections.

\begin{table}[htb]
\centering
\begin{tabular}{lrrrrr}
\hline \hline
Network & N(detected) & Median loc. &   N(\ensuremath{<}1~deg$^2$)    & N(\ensuremath{<}10~deg$^2$) & N(\ensuremath{<}100~deg$^2$) \\
& [yr$^{-1}$] &   [deg$^2$] & [yr$^{-1}$] & [yr$^{-1}$] &  [yr$^{-1}$] \\
\hline
 HLVKI        &               15 &                      7 &                  0 &                  15 &                   15 \\
 3Voyager         &             800 &                     20 &                 5 &                 170 &                 770 \\
 1ET+2Voyager     &            6,100 &                     21 &                 20 &                960 &                6,100 \\
 1ET+2CE      &           320,000 &                     12 &              4,500 &              130,000 &               310,000 \\
\hline \hline
\end{tabular}
\caption{NS-NS Detections per year, localization, and localization rate estimates for different detector configurations. Numbers were computing assuming a Madau-Dickinson star formation rate, with a characteristic delay time of 100~Myr as in \cite{Vitale:2018yhm}. A local co-moving NS-NS merger rate of 320~Gpc$^{-3}$ yr$^{-1}$ was assumed.}
\label{table:BNSdetect}
\end{table}

\subsection{Synergy Science Themes} \label{subsec:GWsynergies}

The \textsl{Athena} mission's broad aim of understanding the hot and energetic universe has a number of themes that overlap with the ground-based GW detectors broad aims of understanding the gravitational universe.
For example, both 
types of facilities
have as goals to investigate the physics and astrophysics of compact objects in a general sense.
This includes more specific areas such as (i) improving our understanding of the EoS in neutron stars, (ii) measuring the spins of BHs, (iii) looking for BHs at high redshifts, (iv) determining the production sites of heavy elements, (v) improving our understanding of the engines powering SGRBs and the physics of jets, and (vi) learning about the environments around compact object mergers.
Some of these themes can be done independently by \textsl{Athena} and GW detectors, and their results can provide useful complementary information.
However, all of the themes are enhanced by having multi-messenger observations of the relevant events.
In the following, we explore three more detailed science topics that highlight the benefits of combining GW and X-ray observations.

It is important to realize that \textsl{Athena} will only be able to follow-up a fraction of these events. 
However, in a few nearby cases and especially when LISA is operational as well (see Section~\ref{subsec:athenalisa_sciencecase}), pre-merger early warning could allow \textsl{Athena} to slew and observe the sky location where the merger will happen (\citealt{2018arXiv181207307A}).


\section{Synergy Science Topics}
\label{GWground:science}
We focus in this part on  three topics that range over three different timescales: the possibility of early-time follow-up (including precursors) of GW sources, the monitoring of off-axis afterglows, and the search for late-time remnants of kilonovae.
Thanks to the combination of superior sensitivity and rapid response mode, these observations would explore 
pristine phases in the evolution of GW counterparts, 
providing  new insights on  synergy science themes such as the production sites of heavy elements, the engines powering short GRBs and the NS EoS, the physics of relativistic jets, and the environments around compact objects.

\subsection{Early-time rapid follow-up and precursors} \label{subsec:early_follow_up}

Merger events involving stellar objects that are not both black holes (e.g. two neutron stars) will involve a range of physical processes producing EM counterpart emission, a number of which have already been demonstrated by the rich observational data set produced by GW170817. In X-rays, these include a resonance flare from NS crust shattering \citep{Tsang:2012} as the merger approaches. When the merger takes place, further observables include a neutron star break-up signal, early magnetar emission, a merger flash \citep{Xie:2018}, and the X-ray part of the prompt emission spectrum of the GRB (the latter potentially visible for hours on account of the additional travel time for high-latitude emission, see e.g. \citealt{Ascenzi:2020}). The accreting engine formed during the merger then produces a kilonova and the first signals from any collimated ejecta, and ultimately a post-merger afterglow.

The resonant shattering flare (RSF) is a predicted precursor event which may occur up to 20~s before a binary neutron star coalescence \citep{Tsang:2012}, releasing $\sim 10^{47}$ erg of energy in $\sim 10^{-1} $~s. The delay time between RSF and coalescence depends on the resonant frequencies of the neutron star and provide a constraint on the nuclear EoS. If a NS merger occurs within 100~Mpc in \textsl{Athena}'s line of sight, either serendipitously or via gravitational wave pre-warning, the RSF will be observable for X-ray radiative efficiencies $\gtrsim 10^{-10}$. If the RSF accelerates a relativistic outflow of neutron star material, its synchrotron afterglow may be observable minutes to hours after the merger.

Before transferring its kinetic energy to a decelerating forward shock passing through the circummerger environment, a successfully launched relativistic ejecta will leave a direct imprint on the observable afterglow emission. During the first hours following the merger, this ejecta is expected to delay the onset of forward shock deceleration by pushing against the shocked circummerger medium, and to give rise to emission produced by a reverse shock (RS) running back into the ejecta. Rapid follow-up on the timescale of hours will potentially be able to measure this imprint, which is expected to lead to a temporary (1-3 hrs) flux enhancement by a factor of 5-10 \citep{Lamb:2019}. Characterization of the RS can provide independent constraints on the Lorentz factor and magnetization of the GRB jet (see the discussion on \emph{jet geometry and orientation} below for further context). 







\subsection{Off-axis afterglow monitoring} \label{subsec:off_axis}


\begin{figure}
    \centering
    \includegraphics[width=0.49\columnwidth]{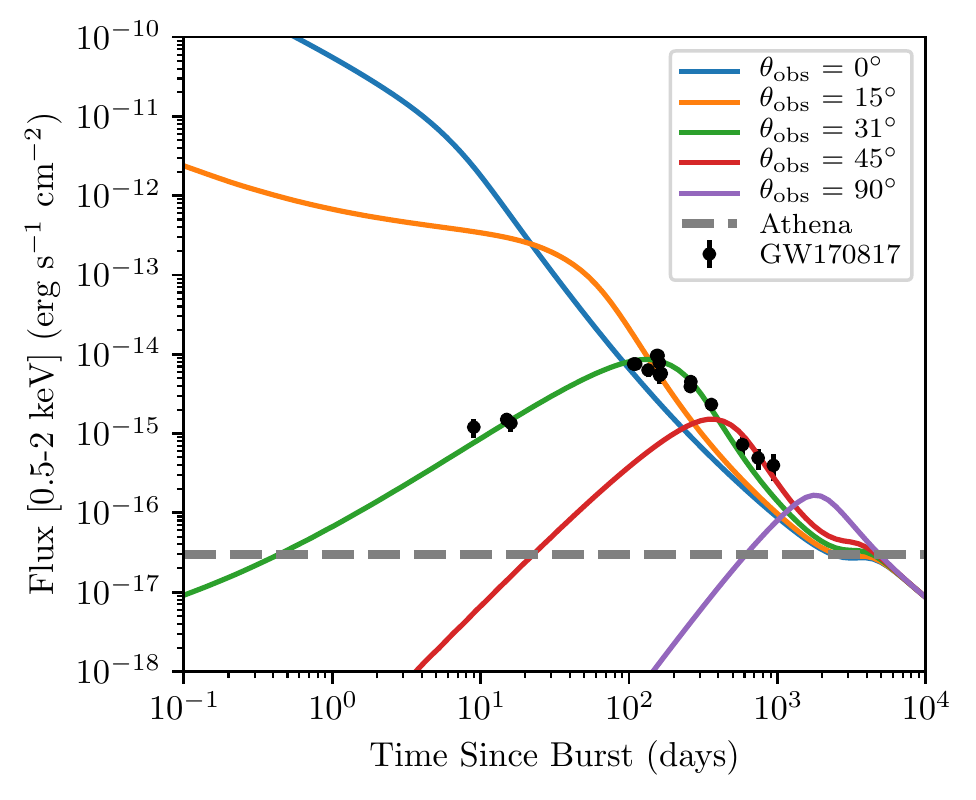}
    \includegraphics[width=0.49\columnwidth]{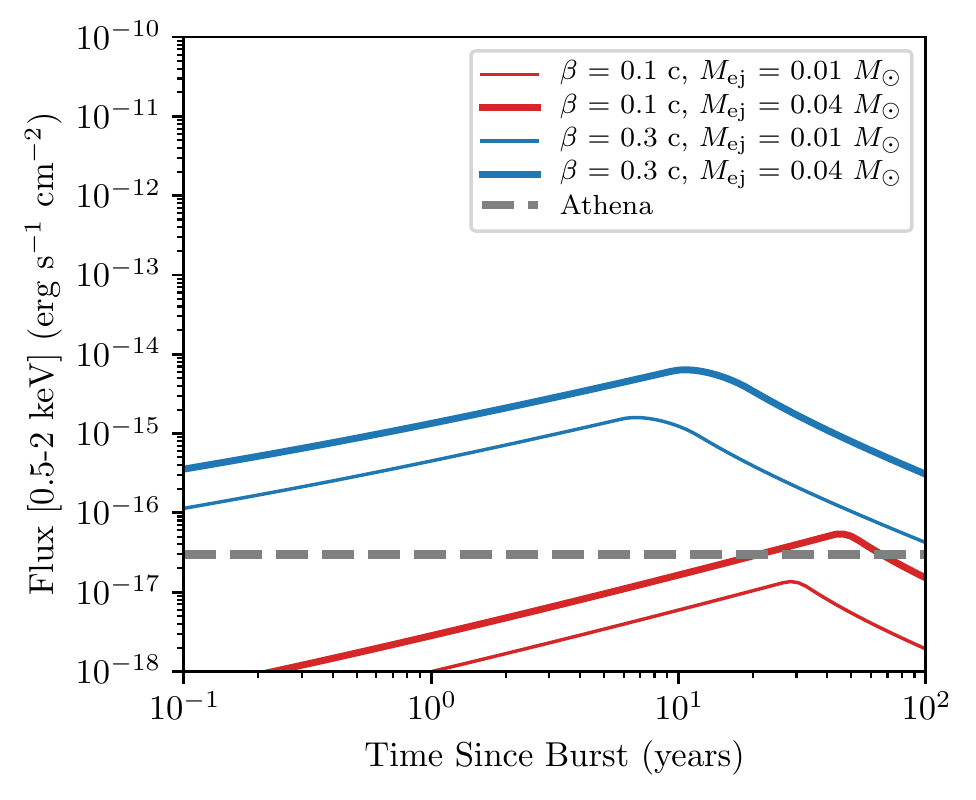}
    \caption{Left: light curves of a Gaussian 170817A-like jet at a distance of 41 Mpc at various viewing angles, parameters and X-ray data taken from \citet{Troja:2020GW170817}.  Right: light curves of a kilonova afterglow at the same distance at different characteristic ejecta velocities and ejecta masses. The kilonova material has velocity stratification $k=5$, the ambient medium is taken to have density $n_0=10^{-2}$ cm$^{-3}$, fiducial synchrotron parameters $p=2.2$, $\epsilon_e=10^{-1}$, and $\epsilon_B=10^{-3}$. Both: fiducial \textsl{Athena} sensitivity of $3\times 10^{-17}$ erg s$^{-1}$ cm$^{-2}$ in 0.5-2~keV band.}
    \label{fig:jet_and_kn}
\end{figure}

\begin{figure}
    \centering
    \includegraphics[width=0.7\columnwidth]{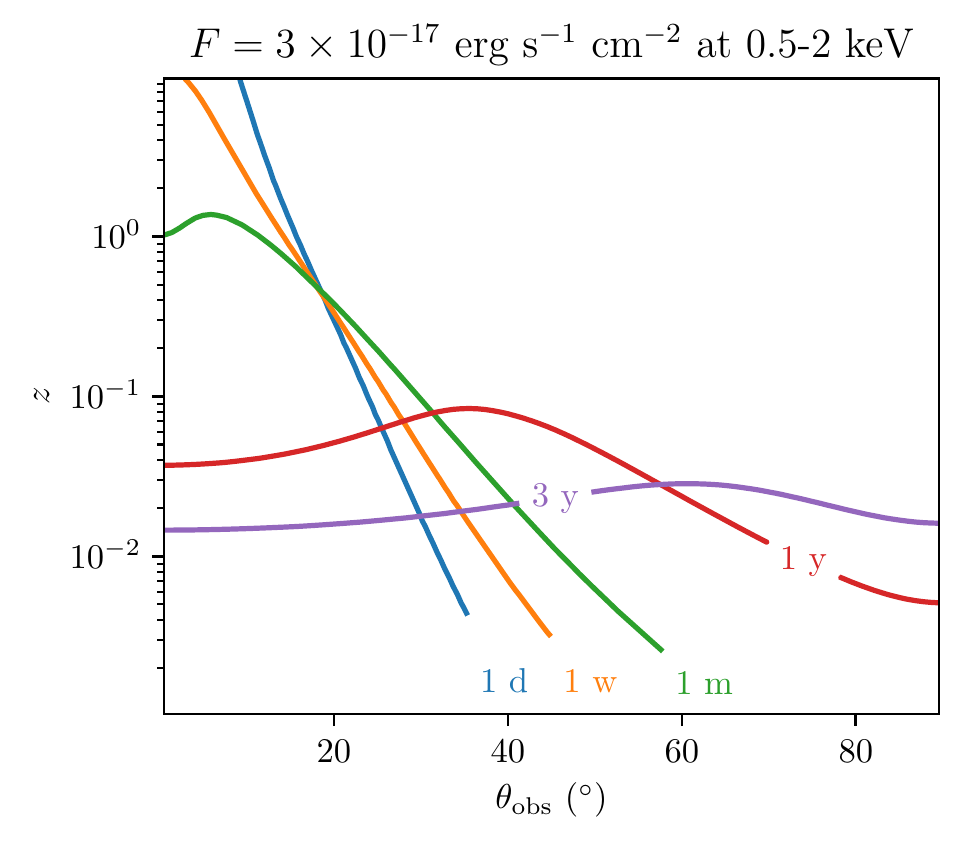}
    \caption{Detectability by \textsl{Athena} of a Gaussian 170817A-like jet at various observing times, as function of viewing angle and redshift. }
    \label{fig:jet_range}
\end{figure}
Thanks to its superior sensitivity, \textsl{Athena} will play a key role in GRB and GW afterglow studies.
With a detection threshold of $3 \times 10^{-17}$ erg cm$^{-2}$ s$^{-1}$ in the 0.5-2.0~keV band (100~ks integration, see Fig.~\ref{fig:Athenasensitivity}), \textsl{Athena} outperforms
current X-ray facilities, such as the \textsl{Swift} X-ray Telescope and the \textit{Chandra} X-ray Observatory, as well as future X-ray observatories 
(see Chapter~\ref{chapter:Theseus}) in the search for orphan afterglows (i.e. those without a GRB counterpart) and
off-axis afterglows (i.e. those seen at an angle from their jet-axis). This is illustrated in Fig.~\ref{fig:jet_and_kn} and \ref{fig:jet_range}, showing the \textsl{Athena} sensitivity compared with an event similar to GRB~170817 at a range of orientations and distances.

It is expected that \textsl{Athena} will observe a multi-messenger event location already pinpointed by a combination of previous GW, GRB prompt emission or kilonova observations. In this case, \textsl{Athena} will provide vital input to address a number of outstanding science questions, by being able to detect more events that are \emph{further distant} and/or \emph{further off-axis} and monitor them at \emph{later times}. 
For mergers within their host galaxy, bright nearby sources may contaminate these measurements. 
For example, both GW170817 and GRB~150101B (the other known off-axis afterglow, \cite{Troja:2018}) 
were found within a massive elliptical galaxy, harboring
a low-luminosity active galactic nucleus (AGN). 
However, based on past observations of short GRBs, 
their heterogeneous environment and broad offset distribution, the issue of nearby polluting sources should affect a limited fraction of events ($\lesssim$35\%).

Fig.~\ref{fig:jet_and_kn} (left) shows characteristic X-ray light curves for a GW170817-like event, demonstrating the diversity of observable off-axis afterglow light curves. We briefly discuss a few specific goals for off-axis afterglow monitoring with \textsl{Athena} below:
\begin{itemize}
    \item \textbf{Determine the rate of choked jets}. If the intrinsic distribution of Lorentz factors of jets produced by NS mergers is a power , similar to  those of other high-energy phenomena such as AGN and blazars, one would expect a substantial population of low Lorentz factor jets to exist (\cite{Lamb:2016}). Furthermore, if the path of the jet is blocked by a sufficiently large amount of material, consisting e.g. of debris from the merger \citep{Nagakura:2014} or NS wind \citep{Murguia-Berthier:2014}, there will be jets that fail to emerge. These end up as \textit{choked} jets that fully transfer their kinetic energy to a cocoon of surrounding material that eventually gives rise to a quasi-spherical shock that is marginally relativistic at most (e.g. \citealt{Nakar:2017cocoons, Lazzati:2017}). The amount of debris in the polar region traversed by the jet is currently not known and different studies predict different angular distributions (e.g. \citealt{Rosswog:1999, Hotokezaka:2013}).
     
    The afterglow light curve can be used to distinguish between successful and choked jets, as shown by \citet{Troja:2018GW170817}, and represents an important means by which \textsl{Athena} can help determine the outflow geometry (isotropic versus collimated) and eventually unveil a new population of X-ray transients produced by choked  jets. Following the X-ray peak time, a slope steeper than 2, similar to the slope following the jet-break of a standard GRB, is characteristic of a collimated flow.
    
    GRB~170817 has indeed been confirmed to involve a successful jet by measurement of the late-time temporal slope of the afterglow light curve \citep{MooleyFrail:2018, Troja:2019GW170817, LambLyman:2019}. Key observations using very-large baseline interferometry (VLBI, \citealt{Ghirlanda:2019Sci, Mooley:2018GW170817}), further established the jet nature. 
    
    While not necessarily conclusive \citep{Zrake:2018}, VLBI does offer an independent means of model verification. However, if GW170817 had happened at a distance above 80~Mpc, VLBI measurements would have been unresolved \citep{Dobie:2020}, whereas \textsl{Athena} observations would still have been able to constrain the jet geometry and its structure by measuring the afterglow temporal evolution. 
    
    
    \item \textbf{Constrain jet geometry and orientation}.  Normally, GRB afterglows are observed close to on-axis, with the bright prompt gamma-ray emission and early observations of a monotonically decaying afterglow 
    providing a bottleneck for detection (see e.g. \citealt{Nousek:2006, ZhangBing:2006}). 
   GW observations are instead less biased towards on-axis events, and herald bursts whose collimated outflows are likely oriented at an angle relative to the observer. Depending on the lateral energy distribution of the jet, the observer angle and the jet opening angle, the afterglow light curve can be expected to show a rising stage first (as was the case for GRB~170817A). The flux during this stage is dominated 
   by emission from progressively smaller angles from the jet core, and the light curve rising slope therefore directly probes the geometry of the jet \citep{Ryan:2020, Takahashi:2020}. By observing the early rising light curve at the $\sim$day time scale, sensitive detectors such as those on-board \textsl{Athena} will be able to probe (1) the outer rim of the jet, which will constrain the jet launching mechanism and the interaction with the torus and/or the merger debris, (2) the early stage of the jet, potentially including a RS \citep{Lamb:2019} or pre-deceleration signatures that constrains the physics of the ejecta, in particular its magnetization and Lorentz factor. The RS emission can lead to a temporary flattening and an enhancement of the light curve by a factor 5-10, strongly dependent on the relative magnetizations between forward and reverse shocks (the factor 10 corresponding to an amplification of about 500 of the RS magnetization, \citealt{Lamb:2019}). Within the context of the off-axis light curves shown in Fig.~\ref{fig:jet_and_kn}, the implication is that for such an event seen at angles up to $\sim 20^\circ$, \textsl{Athena} will be uniquely capable of probing the initial magnetization of the ejecta, which in turn helps to constrain models for jet launching.
    \item \textbf{Probe the fundamental physics of particle shock-acceleration}. GRB~170817A was exceptional among GRBs for a number of reasons, 
    including 
    the remarkable stretch of a single power law of non-thermal emission observed from radio to X-rays \citep{Margutti:2018, Troja:2019GW170817}. Whereas GRBs commonly show at least one spectral break within this range (either the synchrotron injection break due to the lower limit on energy of the  shock-accelerated electron population often seen between radio and optical, or the synchrotron cooling break often observed between optical and X-rays), GRB~170817A allowed for an unprecedented accuracy in determining the electron energy power-law distribution slope,  $p = 2.17$ (e.g. \citealt{Troja:2019GW170817}). It is noteworthy that except for its orientation relative to the observer, the afterglow modeling of GRB~170817A has not required extreme values for the other physical parameters that enter these models (explosion energy, circumburst medium structure, synchrotron efficiency parameters), which suggests that observations of the same spectral regime across a wide range of frequencies are potentially the rule rather than the exception for counterpart observations, and additional tightly constrained measurements of $p$ are to be expected.
    
    This bodes well for the capability of MM counterparts observations to address a number of fundamental open questions in relativistic plasma physics of shock-acceleration. In particular, it remains unknown to date  whether the $p$ values measured thus far are consistent with a universal value for relativistic shocks. While observations suggest
    the opposite
    \citep{Curran:2010, Troja:2019GW170817}, various theoretical models suggest this should be the case \citep{Bell:1978, Kirk:2000, Achterberg:2001}. It is also still unknown whether $p$ will evolve over time as the shock Lorentz factor decreases. Although a subtle evolution towards $p \approx 2$ is theoretically expected, and would be a direct evidence of a fundamental difference in shock-acceleration between relativistic and non-relativistic shocks, this is normally obscured by the uncertainty on estimates of $p$. \textsl{Athena} will be able to observe afterglows deeper into the trans-relativistic transition, increasing the odds of detection shifts over time in $p$ seen in the broadband. 
    \item \textbf{Improve broadband afterglow calorimetry, and circumburst density measurements}. GRB afterglow jet models contain a range of physical parameters \citep{Wijers:1997, Sari:1998}. Of these, energy and circumburst density set the time frame and flux level of the afterglow light curve. Jet geometry and orientation set the slopes of the light curve at different stages. Efficiency parameters for magnetic field generation and particle acceleration at the shock front set the flux level. All affect at which frequencies the transition points between the different power laws of the synchrotron spectrum can be found. With increasing sophistication in light curve modeling \citep{vanEerten:2015}, observations no longer need to be simultaneous in order for them to be combined into constraints on these physical parameters, and  late-time \textsl{Athena} observations will complement earlier broadband observations, while probing a unique dynamical regime of a blast wave seguing into trans-relativistic flow.
    Following the transition into a non-relativistic regime, the emission pattern  of the jet attains isotropy, enabling a direct measurement of the jet energy (and potentially including observable counter-jet emission, although this is likely to be detectable in the X-rays only for strongly off-axis events at distances closer than 100~Mpc; see the late-time upturn of the light curves in Fig.~\ref{fig:jet_and_kn}). \textsl{Athena} thereby further broadens the potential of true MM modelling efforts (including e.g jet orientation constraints directly from GW observations in a comprehensive cross-messenger fit to data) to shed light on the physics of GRBs and on the nature of the engine.
\end{itemize}

\subsection{X-ray emission from kilonova} \label{subsec:late-timeKN}

The merger of a binary NS system can eject a large amount of material $M_{\mathrm{ej}} \gtrsim 0.01\,M_\odot$ at considerable velocity $\beta_{\mathrm{ej}} \gtrsim 0.1 c$ \citep{Bauswein:2013, Hotokezaka:2013}.  This material undergoes radioactive heating to produce a kilonova hours to days after the burst and then continues to expand homologously into the circumburst medium.  In much the same way as the GRB afterglow, this material can sweep up ambient material and drive a synchrotron-producing forward shock wave: a kilonova afterglow \citep{Nakar:2011}.

Figure \ref{fig:jet_and_kn} (right panel) shows X-ray light curves of fiducial kilonova afterglows at a distance of 41 Mpc for various values of $M_{\mathrm{ej}}$ and $\beta_{\mathrm{ej}}$.  It is assumed the kilonova material is stratified in velocity as a power law: $M(\beta > \beta_0) \propto \beta_0^{-k}$.  The initial rise is due to slow material catching up and refreshing the decelerating shock. The slope of this phase depends strongly on the ejecta's velocity profile.  The light curve peaks years after the original burst when the slowest material, containing the bulk of the mass, catches up with the shock.  Afterwards the light curve decays as a standard adiabatic Sedov-Taylor blast wave.  Keeping other parameters constant, faster ejecta peaks earlier while more massive ejecta peaks later.  Increases in either the velocity or mass of the ejecta leads to brighter peak emission.

Direct observations of a kilonova afterglow can help characterize the kilonova material and identify details of its launching mechanism.  The primary observables of an X-ray mission are the rising slope, the peak time, and the peak flux.  The rising slope is a strong function of $k$ and $p$. A clear measurement of the rising slope will constrain $k$, the degree of velocity stratification in the ejecta.  The peak time depends on both $\beta_{\mathrm{ej}}$ and $M_{\mathrm{ej}}$ as well as the circumburst density $n_0$.  If $\beta_{\mathrm{ej}}$ and $M_{\mathrm{ej}}$ are known to good confidence from prompt kilonova observations, then a measurement of the peak time will provide a measurement of $n_0$.  On the other hand, if $n_0$ is known from the GRB afterglow, a measurement of the peak time will provide an independent constraint on $\beta_{\mathrm{ej}}$ and $M_{\mathrm{ej}}$.  The peak flux is sensitive to the same parameters as the peak time, as well as the synchrotron parameters of its forward shock $\epsilon_e$, $\epsilon_B$, and $\xi_N$.  This makes the absolute flux level a difficult observation to draw conclusions from, apart from bounding some combination of these parameters.  The utility of kilonova afterglow observations will be greatest in a combined analysis with the prompt kilonova and the GRB afterglow.  

The peak of the kilonova afterglow occurs $\sim 3-30$ years after the original burst, at a flux level that depends strongly on $\beta_{\mathrm{ej}}$ and $M_{\mathrm{ej}}$ as well as on the circumburst environment density and energy fraction in non-thermal electrons and magnetic field \citep{Nakar:2011, Metzger:2014, Hotokezaka:2015, Ricci:2021}.  The right panel of Figure \ref{fig:jet_and_kn} shows that, for fiducial values of these latter parameters, the kilonova afterglows of flows with $\beta_{\mathrm{ej}} \gtrsim 0.2$ should be detectable with \textsl{Athena}.  These correspond to the afterglow of the ``blue'' kilonova observed in GW170817.  The afterglows of slower (``red'') kilonovae may be observable if they occur in denser environments.


Another science case for the X-ray mission is residual radioactivity of the r-process isotopes in the ejecta from NS mergers, or mergers of a NS with a solar-mass BH~\citep{ripley14, li19, wu19, korobkin19}. 
Detection of this kind could give a unique insight into the nature of the r-process, more direct than a kilonova which re-radiates thermalized radioactive heat. 
The X-ray or gamma-ray detection gives more accurate estimates of the amount and composition of the ejecta, similarly to what has been done for supernovae and their remnants~\citep{diehl14}.

Newly created r-process elements can produce X-rays in a number of ways. 
Some of these processes are discrete and can help characterize the isotopic composition: deexcitation of the daughter nucleus, rearrangement of atomic orbitals after the change of nuclear charge, or radiative deexcitation after one of the innermost electrons is ejected by a $\beta$-particle or via internal conversion. 
All such X-rays bear signatures of individual isotopes and allow to quantify their presence, providing insight into the nuclear properties along the r-process path. 
Discovery of such isotopes may be synergistic with the experimental works such as those currently conducted in e.g. the CARIBU (CAlifornium Rare Isotope Breeder Upgrade) facility at Argonne National Laboratory~\citep{marley13,vanschelt13}, or the Facility for Rare Isotope Beams (FRIB) at Michigan State University~\citep{FRIB18}.

Other processes of X-ray production, particularly continuous ones, tend to complicate identification of individual isotopes. 
Rapid expansion of the merger ejecta broadens spectral lines by ${\sim10-20\%}$ of their energy and blends the neighboring lines together. 
Photoelectric absorption is very strong in the regime {$<100$\,keV}~\citep{hotokezaka16,li19};
additionally, there is a continuum contribution from down-scattering of gamma-rays and other high-energy particles. 
Although simulations indicate that some of the signature r-process X-ray lines may remain identifiable, they are predominantly in hard X-rays~\citep{korobkin19}.
Nevertheless, the overall energetics of the X-ray counterpart appears to be extremely sensitive to the nuclear composition, neutron richness and fissioning in the ejecta~\citep{rosswog18}.

\begin{figure}[h]
\centering
\begin{tabular}{cc}
  \includegraphics[width=0.49\textwidth]{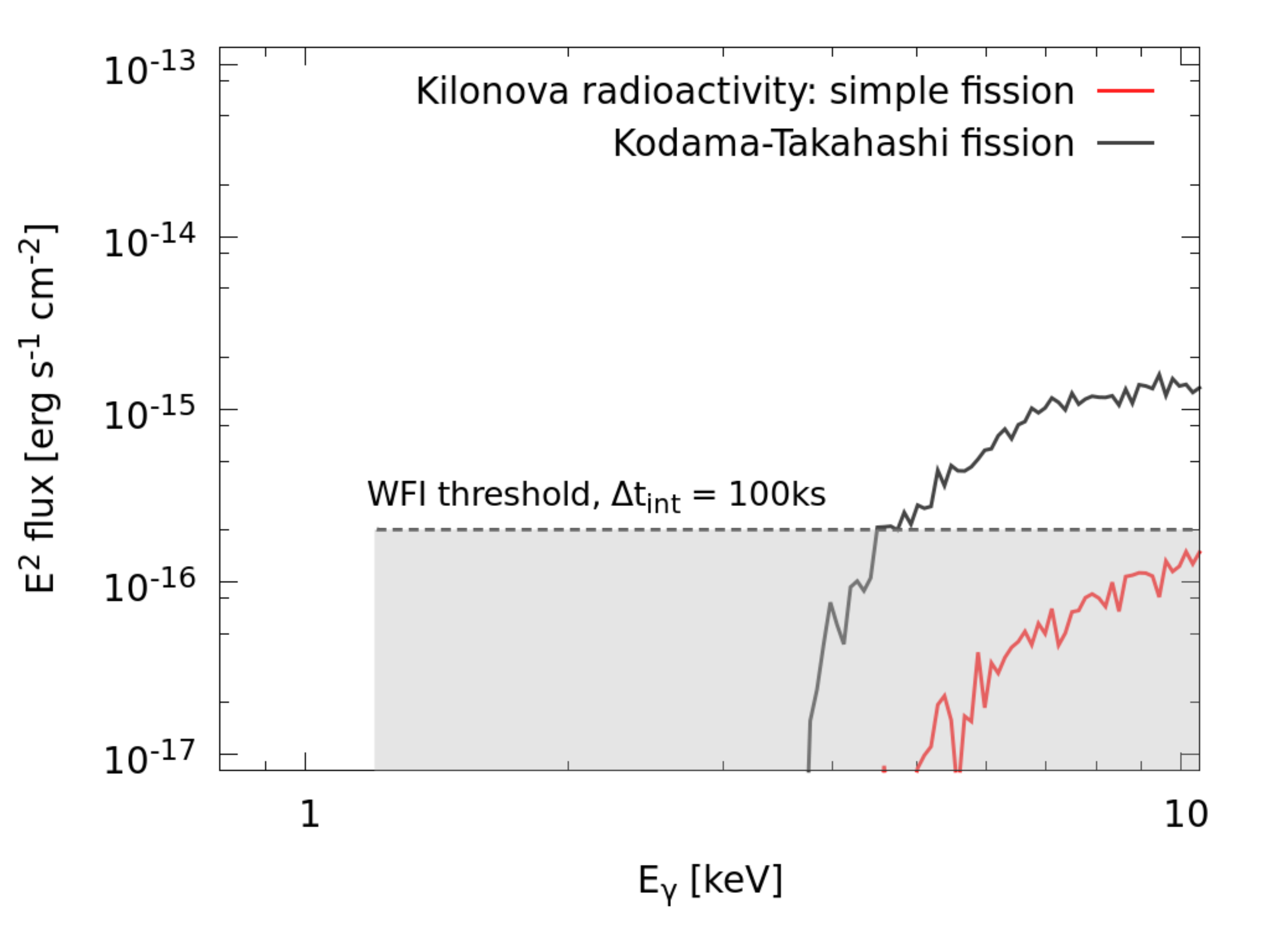} &
  \includegraphics[width=0.49\textwidth]{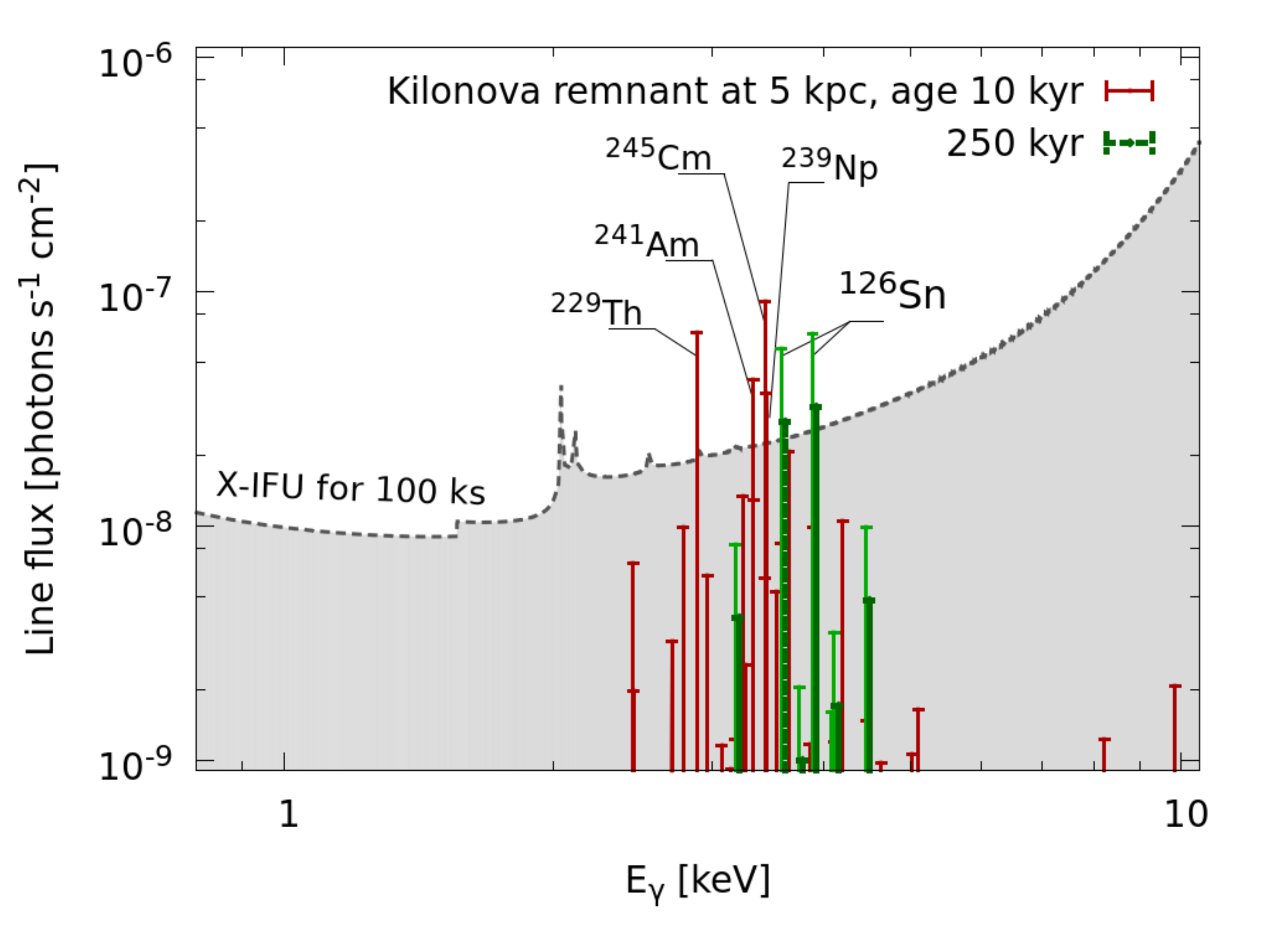}
\end{tabular}
\caption{Left: tentative X-ray spectra of kilonova radioactivity at a distance of 10~Mpc, integrated over 100\,ks, for two representative models of nuclear fission~\citep{korobkin19}.
Right: X-ray line emission from a galactic kilonova remnant at a distance of 5~kpc, for two epochs: 10~kyr (solid lines) and 250~kyr (dashed lines).
 X-ray spectral lines at an age <10~kyr resulting from  actinide isotopes from neutron rich r-process are shown in red.
The green lines highlight X-ray emission from $^{126}$Sn with the half-life of 230,000 years.
The remnant is assumed to have the main r-process composition and a mass of {0.03~$M_\odot$}.
} 
\label{fig:kilonova_flux}
\end{figure}

Figure~\ref{fig:kilonova_flux} (left panel) presents tentative X-ray spectra of the kilonova ejecta with two types of fission for an event 10\,Mpc away, overplotted with the projected \textsl{Athena}/WFI detection threshold (${2\times10^{-16}\,{\rm erg}\,{\rm s}^{-1}\,{\rm cm}^{-2}}$ in 2-10~keV for 100\,ks exposure).
As can be seen, the signal is strongly above the threshold only for one of the fission models, and could remain above threshold for distances up to 20~Mpc. 
Although this strongly limits the number of possible detections, the fact that it could allow to discriminate between different fission models makes it an appealing Target of Opportunity (ToO) case, specifically if higher ejecta masses were to be produced.
In the figure, both models assume total ejecta mass of about $0.04\;M_\odot$, with $0.01\;M_\odot$ corresponding to the extreme neutron-rich component that produces the main r-process~\citep{korobkin19}.
These masses are consistent with the kilonova of GW170817, but higher masses (up to {$\sim0.1\;M_\odot$}) are also possible, especially for mergers of NS-BH binaries.
Only the main r-process component is responsible for the X-ray contribution, while the less neutron-rich ejecta, originating from a hypermassive neutron star or an accretion disc, produces much X-ray emission.
Overall, Fig.~\ref{fig:kilonova_flux} shows very strong sensitivity to nuclear physics inputs.
The lower and upper curves only represent two different fission models and do not bracket all the uncertainty corresponding to the nuclear fission or nuclear masses.
This suggests promising connections with nuclear physics.
Despite the slim chances of such an event happening at distances closer than GW170817, this remains an important potential ToO case for future X-ray missions.



Similarly to supernova remnants, as kilonova ejecta sweeps interstellar material, its expansion slows down from homologous flow to a Sedov solution, and finally to the snow-plow regime.
In this last phase, which is expected to happen 100-1000 years after the merger depending on the kinetic energy of the ejecta and density of interstellar medium, the expansion velocities are only a few hundred km/s~\citep{korobkin19}.
As a result, individual spectral lines are broadened by about 1~eV, which is comparable to the spectral resolution of the X-IFU.

The right panel of Figure~\ref{fig:kilonova_flux} shows the X-ray line spectrum of a nebulous galactic kilonova remnant 10~kyr and 250~kyr after the merger.
As can be seen, such remnant at a distance of 5~kpc can produce line fluxes reaching ${10^{-8} - 10^{-7}{\rm  photons} {\rm cm}^{-2}{\rm s}^{-1}}$, which for some lines is sufficient to pass a significance threshold. 
The gray area indicates 10-photon threshold when collecting the flux over 100~ks with the projected effective area of the X-IFU.
A young kilonova remnant (<10 kyr) with an abundance of heavy r-process contains several long-lived actinide isotopes.
Their X-ray spectral lines are shown in red.
Older remnants with lighter r-process are expected to be rich in $^{126}$Sn, which is located near the second r-process peak and has half-life of 230~kyr~\citep{ripley14}.

No kilonova remnant has been discovered so far, but the situation could change in the next decade. 
For instance, the pair of X-ray lines of $^{126}$Sn can serve as a ``smoking gun'' in searches for such remnants. 
Conducting a systematic survey in search for the signatures of this isotope in the catalogue of supernova remnants \citep{ferrand12,green19} with high galactic latitudes could be one possibility. 
Another possibility is presented by the mysterious Odd Radio Circles \citep[ORCs, see][]{norris20} with their shape and extremely high galactic latitudes~\citep{wu19}.



\section{Observational strategy}

\subsection{The number of targets}

In line with the reasoning behind the \textsl{Athena} science requirements document,
we propose to obtain \textsl{Athena} observations of a minimum of three sources
in each bin in parameter space relevant for the case at hand. The
gravitational wave sources of interest include NS-NS mergers and BH-NS mergers where the
black hole mass/spin combination is such that the black hole will not
swallow the neutron star whole, but instead disrupt it outside the
innermost bound circular orbit, which is a necessary condition
for the existence of an EM counterpart.
The parameter space associated with these objects include; i) the
inclination angle to the line of sight, ii) the spin of the newly
formed NS or BH, iii) the initial black hole spin (for BH-NS mergers).

The first of these parameters is relevant for the detection of X-ray
emission associated with the jets and for the general dependence of
X-ray emission on inclination angle (the initial super-Eddington
fall back accretion disc is expected to block the line of sight
towards the inner disc and any X-ray emission coming from that region
is therefore blocked in higher inclination sources). We expect the
inclination angle dependence between $0-90^\circ$ to be sampled in
intervals of $30^\circ$, hence we propose to fill at least 3 bins for
this parameter. The second parameter is relevant for the Blandford --
Znajek power of the jets and/or spin down energy available for a
magnetar. The jet power is a strong function of the dimensionless
BH spin parameter $\mathrm{a}$, and therefore, assuming all spin values are
possible and occur in nature we propose to use 5 bins over $\mathrm{a}$.  The third parameter will, together
with the BH mass and the NS EoS,
determine if EM emission is expected in a BH --
NS merger. For nearby high signal-to-noise ratio
gravitational wave sources, the BH spin and mass before merger
will be determined by the gravitational wave detectors. Therefore, we
suggest to only sample several sources where those values are such
that standard theory predicts the existence of an EM
signal to test such theories. We envisage sampling this
(dimensionless) BH spin parameter that ranges from $-1<a<1$ in
intervals of 0.4. Also, depending on the existence of very fast
spinning black holes, these bins might not be distributed in a linear
grid, instead a logarithmic grid might be better. Smaller bin sizes
are unlikely to be feasible due to the limited accuracy of the
(instantaneous) spin determination of the gravitational wave
detectors. This would imply 5 bins to fill with \textsl{Athena} observations in
this parameter.

The \textsl{Athena} observing strategy will be based on GW observables that can be measured robustly from the gravitational-wave signal. While the spin of the final object (ii) might be the relevant parameter for understanding jet physics, it is not a quantity that can be determined directly from the gravitational waves with high accuracy. Fortunately, combining the masses of the binary components prior to merger and numerical simulations the spin can be determined. The masses of the two objects prior to merger can be measured directly from the gravitational-wave signal more robustly.
For NS-NS mergers, the masses for the tens to hundreds of nearest events (and thus brightest events in X-rays) are expected to be measurable with percent precision with third-generation detectors~\citep{Grimm:2020ivq}.
As with BH-BH mergers, the largest errors are expected to arise from the uncertainties on the distance (which propagate to uncertainties on the source-frame masses; see, e.g.,~\cite{Vitale:2016icu}).
Measurement of the initial spin of the BH in a BH-NS system by third-generation detectors has not been studied in as great detail.
The results, however, will likely be similar to those for BH-BH systems, where ten-percent accuracy on the spin for individual events is likely~\citep{Vitale:2018nif}.

In total this would imply observing $3 \times 5=15$ double NS mergers. For BH -- NS mergers given the extra
parameter to cover (iii) the number of events to follow seems
prohibitively large: $3 \times 5 \times 5 = 75$. However, the second
and third parameter are not completely independent. If necessary, one can limit the binning in the BH spin to the 3 bins covering the positive and negative maximal rotations and the intermediate values. Hence, we
estimate that $\approx$ 25 black hole -- neutron star merger events should be
followed with \textsl{Athena} to cover the total parameter space effectively. 

For the \textsl{Athena} observations we distinguish trigger observations which
can be used to search for an X-ray EM counterpart and
follow-up observations which will be obtained once the
EM counterpart and a arcsecond localisation are
known. For the latter observations, the nature of the EM
counterpart, i.e., optical, near-infrared, radio and/or X-ray is not
relevant in first instance.

\subsection{An observational strategy for science topics}

Depending on the GW detector network deployed in 2030s, a few to thousands of NS mergers per year are expected to be detected and localized within 1~deg$^2$  (Table~\ref{table:BNSdetect}), possibly with  an additional and comparable number of NS-BH mergers (Sect~\ref{subsec:GWdetectors}). Therefore, there should be enough targets to allow a selection aimed at optimizing the \textsl{Athena} coverage (e.g., based on the distance and the expected brightness of the EM counterpart). 
We expect that the location accuracy of a 
non-negligible fraction of these events 
will 
be good enough to fall in the WFI FoV (0.4~deg$^2$), while for the others a WFI tiling coverage of the error box can be implemented. Once the EM counterpart is identified (either by the \textsl{Athena}/WFI or by other EM observatories), subsequent observations with \textsl{Athena} will be carried out with the X-IFU.


\textbf{Early X-ray precursors}. As described in Sect.~\ref{subsec:early_follow_up}, early-time observations (few hours) can provide key signatures of the nature of the remnant and the launch of an relativistic jet, with predictions ranging over orders of magnitude. Furthermore, early time observations can disclose the radioactive imprints from the newly formed kilonova especially for low-inclination sources (Sect.~\ref{subsec:late-timeKN}). Fast \textsl{Athena} ToO observations with a typical duration of 50~ks on a sub-sample of events selected to be close ($<$200 Mpc) and with a GW error box small enough that the 90\% confidence region can be covered with one WFI pointing will allow to explore this regime to a flux limit low enough to cover a significant parameter space.
A sub-sample of about 10 such events would require about 0.5~Ms.

\textbf{Jet evolution.} \textsl{Athena}  will allow to extend the study of jet emission  up to $z\gtrsim 1$ (the range of the third generation GW detectors:  ET,CE) for NS mergers, if the jet viewing angle is within $\approx 15^{\circ}$ and assuming a GW170817-like jet. Likewise, jets pointing in the orthogonal plane, can be observed up to $\lesssim$ 100~Mpc (see Figures \ref{fig:gw_horizons} and \ref{fig:jet_range}).  We aim at covering the period from  few days to few years after the merger with about 5-10 observations each. Exposure times and number of observations will be optimized according to the distance, luminosity and off-set angle of the event. A typical duration of 50~ks would e.g., allow to derive spectral information for a GW170817-like events (at $30\deg$ off axis) at distances $\lesssim$ 200~Mpc and flux measurements up to $z\approx0.2$.
Assuming an average of about 250~ks per event, \ several Ms will be needed to cover the selected population of 40 objects. 
This preliminary assessment needs to be optimized taking into account the expected distribution of distances, luminosities, angles and model parameters.

\textbf{Late-time kilonova emission.}
Late time emission by shock interaction of kilonova ejecta with the environment is expected to be detectable with \textsl{Athena} from a few years 
up to 10 years of more after the merger, and up to $z\approx0.2$ (Fig.~\ref{fig:kilonova_flux}, left).  The sample will be then mostly based on the second generation interferometers, and possibly on the first years of operations of the third generation interferometers and would require an already identified EM counterpart. One can tentatively assume that about 10 such events could be available, and plan for one 100~ks observation, repeated after a few years in case of positive detection (total of about 1~Ms). Nearby kilonova remnant candidates, e.g., selected by radio observations, should also be targeted with a similar or longer duration (Fig.~\ref{fig:kilonova_flux}, right).

\chapterimage{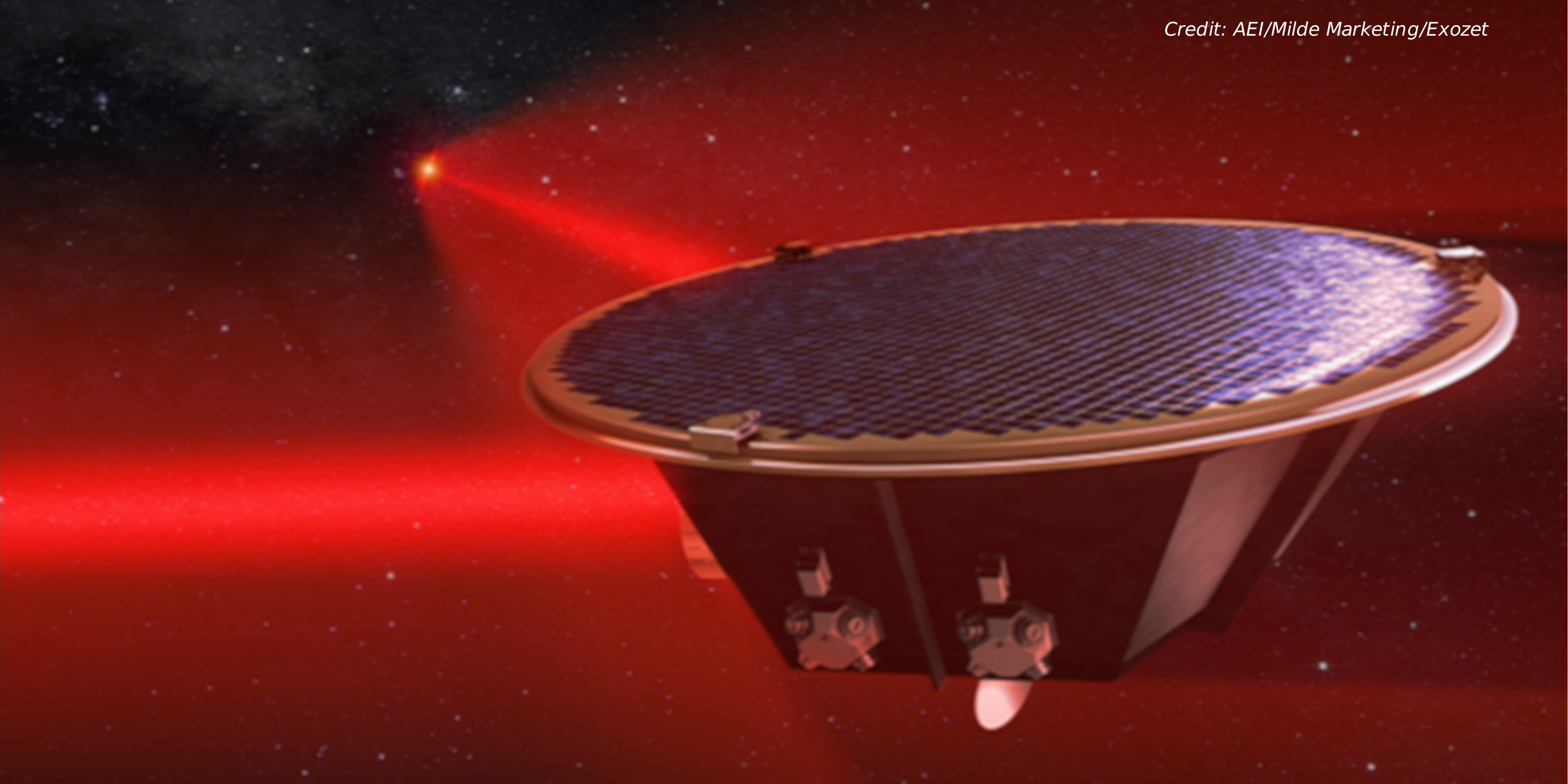} 
\chapter{ \textsl{Athena} \& LISA}\label{chapter:LISA}
\textit{{\color{purple}{By M. Colpi, M. Guainazzi, P. McNamara, L. Piro, J. Aird, A.C. Fabian,   V. Korol, A. Mangiagli, , E.M. Rossi, A. Sesana. N. Tanvir.}}}

\section{Science Themes}\label{subsec:athenalisa_sciencecase}
\subsection{LISA}

\href{https://sci.esa.int/web/lisa}{LISA} \citep{LISA17} is the third large class mission of the ESA Cosmic Vision Program. It will explore the gravitational wave (GW) Universe in the 0.1 to 100 mHz frequency interval, where it is expected to detect the signal  from various classes of sources including: merging massive black holes  of  $10^4-10^7\,\msun$, the inspiral of stellar black holes around intermediate-mass and massive black holes, the early phase of inspiral of stellar black hole binaries  and the nearly monochromatic signal emitted by ultra compact binaries, mostly double white dwarfs, in the Milky Way and its satellite galaxies. 
\textsl{Athena} will reconstruct the accretion history of  black holes  including the intrinsically dim active galactic nuclei (AGN) at low redshift, while LISA will reveal the yet unknown merger history of massive black holes in binaries, i.e. a new population predicted to form during the cosmological assembly of galaxies. The science cases of \textsl{Athena} and LISA are complementary and  outstanding {\it per se}. This Chapter describes the {\em additional science} which can be achieved by the synergy of \textsl{Athena} and LISA observations.

\subsection{Synergy Science Themes}
Concurrent observations in GWs and X-rays by LISA and \textsl{Athena}  can  address a number of open questions in the domains of astrophysics, fundamental physics and cosmology:

\begin{itemize}
\item
 Accretion flows in violently changing spacetimes, formation of an X-ray corona and jet launching around newly formed horizons;

\item
Testing General Relativity as theory of gravity and measuring the speed of GWs and dispersion properties;

\item
 Enhancement of the cosmic distance scale using GW sources as standard sirens.
\end{itemize}

The achievement of these synergistic science themes between LISA and \textsl{Athena} relies on a number of prospected GW sources. These are massive black hole coalescences in gas-rich environments; extreme and intermediate mass ratio inspirals (EMRIs/IMRIs) where a stellar black hole is skinning the horizon of a large black hole surrounded by an AGN disc; interacting double white dwarf systems present in large numbers in the Milky Way Galaxy.
The possibility of performing these observations depends  on LISA’s capability to localize the sources, and on \textsl{Athena} to identify possible X-ray counterparts. In the following we discuss the opportunities opened by such observations, as well as the challenges associated to synergistic observations at the best of our current knowledge of the performance of the two missions.

\section{Synergy Science Topics} \label{subsec:athenalisa_science}

\subsection{Massive Black Hole Coalescences}

\noindent
 Theories of galaxy evolution predict that galaxies started to form within dark matter halos at redshifts  $z\sim 15-20,$ and grew  through repeated 
 mergers and accretion of matter from filaments of the cosmic web. 
 In these pre-galactic, low-metallicity structures, {\it black hole seeds} of $10^2\,\rm M_\odot$ up to $10^{5}\,\rm M_\odot$ are expected to form  (\cite{Volonteri10,Inayoshi2019}) and to grow in mass through accretion and mergers to explain the presence of the luminous quasars at redshifts as early as $z\sim 7$ (\cite{Banados18,Yang2020}), when the universe was only  $\sim 800$ Myr old. They represent  the tip of 
 an underlying  population of fainter AGN 
 that is the least known in terms of basic demographics, birth and growth, which \textsl{Athena} aims at discovering (\cite{Matsuoka18}).
 
 Electromagnetic (EM) observations have revealed the occurrence of tight empirical relations between the black hole mass and quite a few host galaxy properties in the today universe (\cite{KormendyHo2013}). It is now widely accepted that during quasar/AGN activity, the launch of powerful winds by the black hole engine affected their accretion cycles and star formation jointly, self-regulating their growth in the host galaxy. One of the best explanations for these correlations invokes galaxy mergers conducive to massive black hole coalescences in gas-rich environments. These processes are central for establishing a key synergy between \textsl{Athena} and LISA. 
 
 LISA is expected to detect the GW signal from the coalescence of massive binary black holes in the largely unexplored interval between $10^{4}$ and $10^7 \,M_\odot,$ forming in the aftermath of galaxy mergers, with a rate of a few to several tens per year (\cite{2011PhRvD..83d4036S,Bonetti18,Barausse20}).
 LISA detections are likely to be dominated in number by lower mass systems at redshift $z>5$, with low signal-to-noise ratio ($S/N$). However, up to several detections of black holes with masses $\ge 3\times 10^5$~M$_{\odot}$ at  $z<2$ are expected per year. These events deliver the highest $S/N$ in GWs, with a median error box small enough to be observable by \textsl{Athena} (\cite{Mangiagli20}). These are the most promising candidates of multi-messenger emission.   

\begin{figure}[!t]
\centering
\includegraphics[width=1.0\textwidth]{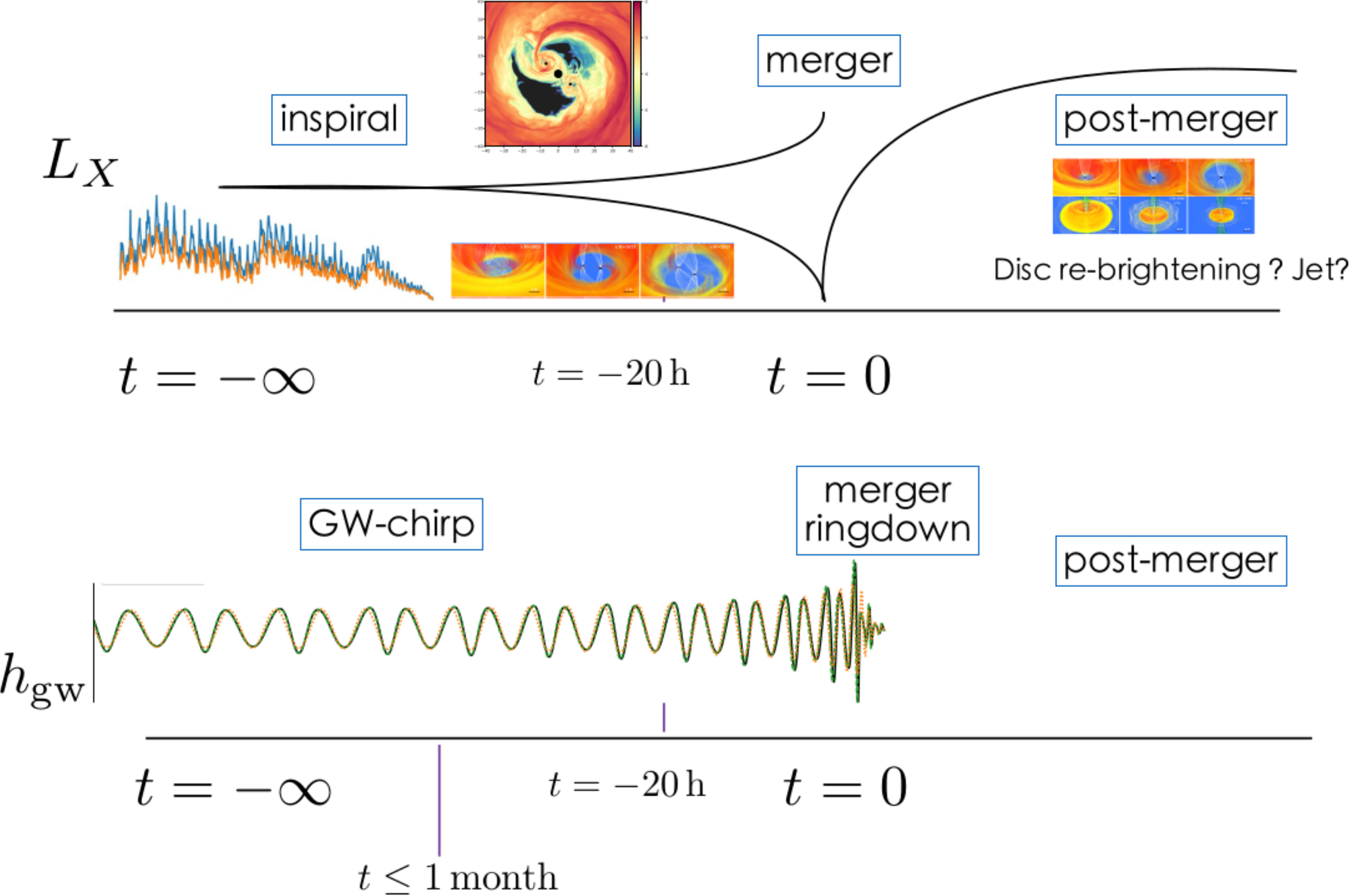}
\caption{A cartoon depicting a possible trend of the X-ray luminosity during the inspiral (pre-merger), merger and post merger phase of two accreting massive black holes in a binary system. The insets in the upper panel are: from \citet{Bowen18} (top) illustrating the relativistic flow pattern around two non-spinning black holes close to merging; from \citet{Tang18} showing the model X-ray light curve; from \citet{Khan18}
showing the circumbinary disc and the incipient jets that naturally form both prior to and after the coalescence. The bottom panel shows the GW amplitude as a function of time. The GW emission dies out when the ringdown phase has ended. }
\label{fig:LISA__1} 
\end{figure}

 
The detection of X-rays emitted by gas orbiting around coalescing massive black holes  {\it contemporary} to the detection of the  GW signal will let us correlate for the first time, the black hole masses and spins encoded in the GW waveform with the X-ray light and spectrum emitted by the surrounding gas. This will shed light on the behaviour of matter and light in the violently changing spacetime of a merger.  Thus, the {\it additional science} resulting from joint observations will have a large impact on our knowledge of massive black holes 
 as sources of both EM and GW radiation. The scientific return  from concurrent observations can be summarized as follows: \\
{\bf Environment around the massive black hole binary.}
The {\it pre-merger} phase, associated to
 the binary  inspiral, might lead to an EM {\it precursor}. 
As the massive black holes spiral-in, X-ray emission is expected to be modulated in time, with characteristic variability correlating with the binary orbital motion or/and with relativistic fluid patterns rising in the non-axisymmetric circumbinary disc surrounding the two black holes  \citep{Tang18,Khan18,Bowen18} (cf. Fig.~\ref{fig:LISA__1}). This should be the distinguishing feature of a binary in the verge of merging. The X-ray window is particularly  favorable as
X-rays are known to come empirically from very close to the black hole horizon, i.e. a few Schwarzschild radii (\cite{Ricci20}). During the last phases of the inspiral gas can be tightly bound to each black hole in the form of two mini-discs. The system can thus be viewed  as a superposition of two rapidly moving quasars  almost all the 
way to the merger.  X-ray Doppler modulations and relativistic beaming would characterize the emission. \\
{\bf Testing General Relativity: speed of gravity.}
In General Relativity, GWs travel with a speed equal to the speed of light (the graviton is massless) and interact very weakly with matter. 
If X-ray variability is expected to evolve in tandem with the GW chirp, i.e. the rise of the amplitude and frequency of the GW, this would enable  measurements of the speed of gravity relative to the speed of light to a precision of one part in $10^{17}$ (\cite{Abbott17gw-gamma,Haiman17}), allowing for a novel test of  theories with massive gravity or extra spatial dimensions \citep[e.g.][]{Kocsis08}.

\noindent
{\bf Cosmic distance scale.}
\noindent
Coalescing binaries are standard sirens as the GW signal 
provides the direct 
measure of the luminosity distance to the source (\cite{Schutz86}). By contrast the signal does not carry any information on the redshift, that can be recovered from the EM observation of the host galaxy. The optical follow-up of the X-ray source will make possible to identify the host galaxy and infer the redshift. The resulting distance-redshift relationship then provides a measure of the Hubble parameter  (\cite{Tamanini16}). This will help  to arbitrate tensions between the late- and early- universe probes of the cosmic expansion from the Planck data and Type Ia supernovae, respectively.\\
{\bf AGN physics.}
The post merger observation of the AGN opens the door to the study of the mechanisms triggering accretion around a newly formed black hole. 
LISA observations provide the mass and spin of the black hole while X-ray observations measure the luminosity and spectra coming from disc re-brightening, corona emission and jet launching.

\subsubsection{Predictions of the EM emission from massive black hole binaries}
\label{sect:SMBH_predictions}

All the science themes discussed above rely on the existence of EM emission from coalescing massive black hole
binaries. So far, no transient AGN like emission that could be attributed to the coalescence of a massive black hole binary detectable by LISA has ever been observed in the variable sky. Thus, we have to resort to theoretical models to infer characteristics of their light curves and spectra, during the inspiral and merging phase.

Joint and contemporary production of the GW and EM signals in a merging system requires on one side the presence of a rich reservoir of gas, possibly in the form of a {\it circumbinary disc} surrounding the binary. On the other side, a key prerequisite for a successful identification of precursor EM emission is a sky localization uncertainty by LISA of 0.4~deg$^2$ (the size of the \textsl{Athena}/WFI Field of View -FoV-),
a few hours prior to coalescence.  

The flow pattern around a black hole binary surrounded by a circumbinary disc shows distinctive features. The binary excavates a cavity of size about twice the binary separation,  which is filled by a hot, tenuous gas and by streams that constantly feed  two mini discs that are seen to form around each black hole \citep{Farris14,Tang17,Bowen17}. High angular momentum gas leaking through the cavity impacts on the inner rim of the circumbinary disc creating an asymmetry in the density pattern, which might give rise to  distinctive periodicities and radiation features. They could be detected in those nearby binaries - called \textit{Platinum Binaries} -
for which LISA sky localization within the \textsl{Athena} FoV is possible during the inspiral phase, hundred to tens of orbital cycles prior to coalescence. 

The {\it precursor emission}, hours prior to coalescence, is expected to come from the 
circumbinary disc, the mini discs around each black hole and the cavity wall, each contributing at different wavelengths to a different extent.  The accretion rate is expected to be modulated and it is highly non stationary being driven by pressure gradients at least as much as by internal stresses \citep{Bowen18}.
When the accretion rate makes the flow optically
thick, the soft X-ray band (around 2~keV), becomes dominated by thermal radiation from the inner edge of the circumbinary disc and the mini-discs, and a modulation may set in with a periodicity   similar to the  inspiral period (\cite{Tang18}).  Harder X-rays  (around 10~keV) 
come from coronal emission \citep{dAscoli18}. 
In the imminent vicinity of the merger, the mini-discs thin as their tidal radius shrinks to a size comparable to the hole's effective innermost stable circular orbit, leading to a dimming of the  X-ray luminosity \citep{Paschalidis21}.
Additional X-ray variability may arise from 
refilling/depletion episodes caused by periodic passage of the black holes
near the overdensity feature at the edge of the circumbinary disc.
Also Doppler beaming (\cite{Haiman17}) and gravitational lensing (\cite{2018MNRAS.474.2975D}) can modulate
the observed light flux seen by near-plane observers. The emission is in general highly anisotropic, especially
when the binary is seen edge-on, and thus with the lowest GW amplitude.

Concerning the {\it prompt, post-merger emission}, which is the most relevant for \textsl{Athena}, the launch of a jet from the spinning black hole  might 
spark X- and gamma-ray emission \citep{Shapiro17}.
The time at which the jet emerges, its duration and the time of emergence of the afterglow emission are now being studied
\citep{2021ApJ...911L..15Y}. 
As at coalescence the GW signal reaches a peak luminosity of the order of a few $\nu^2 10^{57} \,\rm erg\,s^{-1}$ (with $\nu$  the symmetric mass ratio, equal to the reduced total mass ratio of the binary) one could speculate that even if a  minuscule fraction of this luminosity emerges as EM radiation, this could give rise to an observable transient.

At merger the new black hole acquires a gravitational recoil due to linear momentum conservation in unequal mass binaries and in binaries where the black holes carry large, misoriented spins. The interaction of the newly formed, recoiling black hole with the surrounding circumbinary disc can give rise to disc rebrightening in the form of a late afterglow.  Fluid particles  which remain bound to the  black hole modify their orbits not only in response to the change in the underlying gravitational potential, due to the mass loss induced by GW emission, but also in response to the completely new arrangement that is imposed by the kick, particular when the recoil velocity has a large component in the disc plane, as the black hole excites shocks in the fluid. 
This gives rise to an EM transient rising months  to a few years after the merger proper, depending on the extent of the recoil, disc mass, and gas cooling (\cite{rossi10}).
 
\subsubsection{Caveats}
\label{subsec:LISA_caveats}
 There are many uncertainties involved in forecasting the number of binaries that can be localized within the \textsl{Athena}/WFI FoV at the end of the coalescence, and in particular the number that will generate a detectable  X-ray flux.  We call these binaries  {\it Gold Binaries}.  ``Known uncertainties" are discussed here, based on current observations of AGN. 
This being an uncharted territory, any {\it prediction} on the rate of GW mergers and on the EM emission, in particular in the X-rays, has to rely on theory only, with a rather uncertain and widespread range of predictions (\cite{Roedig14,Tang18,dAscoli18}).

For X-ray emission to be generated during, or by a merger of a pair of massive black holes, gas must be present during the merger process and indeed may be instrumental in bringing the black holes to a radius where gravitational radiation drives their inspiral. Thus, accretion is likely to occur before the merger takes place and therefore those binaries that will merge in the 2030s may currently be AGN.

It should first be recognised that the X-ray emission detected from AGNs by \textsl{Athena} is dominated by the X-ray corona, which is generally considered to be magnetically powered by an accretion disc orbiting about the black hole. The corona is relatively compact and contains energetic electrons with temperatures of tens to hundreds of keV that Compton upscatter blackbody photons from the accretion disc into a power-law X-ray continuum. The observed fraction $f_{bol}$ of the bolometric accretion power emerging in the 2-10~keV X-ray band  ranges from about 10 to 2 percent or less, as the bolometric power increases to the Eddington limit (\cite{vasudevan09,lusso12}). There is as yet no predictive theory of the corona nor of $f_{bol}$.  Additional 2-10~keV X-ray emission is seen if the object has a jet  (\cite{blandford18}).  There is no observationally based predictive theory for jet occurrence in AGN; a rough guide is that approximately ten percent of quasars are radio-loud due to jets. 

A complication to observing AGN is obscuration. The flat shape of the X-ray background spectrum in the 2-10~keV band, which is largely the summed emission from all AGN,  demonstrates that most accretion is obscured. Obscuration can occur in all types of AGN, but  simulations suggest that both obscuration and luminous black hole accretion peak in the final merger stages when the two black holes are separated  by less than 3 kpc (\cite{hopkins05}). This is borne out by recent observations by \citet{koss18} who find that there is significant excess (6/34) of nuclear mergers (i.e. a counterpart within 3 kpc) hosting obscured luminous black holes compared to a matched sample of inactive galaxies (2/176). The obscuration most affects the soft X-rays below 2-5~keV.  Prolonged AGN emission at close to the Eddington limit can blow away most of the obscuring gas (\cite{fabian09b,ricci17}). 

Violent accretion events such as tidal disruption events (TDEs) could be an alternative template for accretion in the late stages of a super massive black hole (SMBH) merger. If so, then coronal emission may be weak or absent, with most of the accretion power emerging from a quasi-thermal blackbody disc, sometimes with jetted emission. Unless jets are formed, X-radiation from such objects is mostly confined to the soft X-ray band \citep{saxton20}.
If we assume that accretion takes place in the late inspiral  phase of a pair of massive black holes, so that they appear as AGN, we can use the number densities of observed galaxies and AGN to predict the number of final mergers to be expected within a given interval of time. Concentrating on binaries with masses of $10^6$ to $10^7$\,M$_\odot$ within redshift $z=2$, we first consider  the number densities of their host galaxies, which will have stellar masses of  $\sim  10^9$-10$^{10}$\,M$_\odot$. \citet{Ilbert13} gives number densities of $10^{-2}$--$10^{-2.5}$~Mpc$^{-3}$ at $z=1$ and $10^{-1.5}$--$10^{-2.5}$ at $z=2$. 
The probability $p$ that a galaxy is an AGN within 1 percent of the Eddington limit ($\lambda>0.01$) as a function of black hole mass and redshift has been estimated from observations by \citet{aird18}, giving $p=0.003$ for $10^6$\,M$_\odot$ and $p=0.01$ for $10^7$\,M$_\odot$ black holes.  The intrinsic galaxy merger rate is about $4\times10^{-10}$ yr$^{-1}$  (\cite{Lotz11}) which means that over a 10 yr period of LISA observations the total number of galaxy merger events is 10$^{-9}$ and $0.6\times10^{-9}$ at z~=~1 and 2 respectively. The number densities are per comoving Mpc and the comoving volume out to $z=1$ is 157 Gpc$^3$ and out to $z=2$ is 614 Gpc$^3$. Gathering all these factors together, we predict that, per dex in mass and for an observation period of 5~yr, the number of black holes of mass 10$^6$  merging is 5$\times$10$^{-3}$ within $z=1$, and within z=2 it is $2\times10^{-2}$. For black holes of mass 10$^7$ the corresponding numbers are 5$\times$10$^{-3}$ and $1.2\times10^{-2}$ detectable mergers per 5~yr interval. 
The above predictions assumes that the probabilities of a galaxy having an AGN and of it having had a merger are independent. If however we assume that all mergers lead to AGN, we can eliminate $p$, which raises the number to those listed in Tab.~\ref{tab:rates}. These are the maximum predicted values, whether or not there is gas in the nucleus. If the black hole merger takes place before gas has been blown away by localised feedback effects, then these more optimistic numbers are appropriate. However, if the merger takes place after this phase and there is no gas present then we do not expect to detect  X-ray emission.
\begin{table}
\centering
\caption{Observational-based predicted number of expected SMBH merging events visible by \textsl{Athena} and LISA over 5~years.}
\begin{tabular}{lcc}
& M=10$^6$~M$_{\odot}$ & M=10$^7$~M$_{\odot}$ \\ \hline
$z~=~1$ & 1.5 & 0.5 \\
$z~=~2$ & 12 & 1.2 \\ \hline
\end{tabular}
\label{tab:rates}
\end{table}

How likely is it that the X-ray glow is sufficiently bright to be detectable by the WFI? The answer is in Tab.~\ref{tab:agn} and Tab.~\ref{tab:agnobs},
where we show the expected fluxes and required
\textsl{Athena} exposure time to detect an AGN at the Eddington limit, assuming an X-ray to bolometric luminosity ratio of 30, and the sensitivity of the WFI averaged over its full FoV (see section 1).
These results indicate that follow-up of Black Holes (BH) of mass 10$^6$~M$_{\odot}$ could require considerable \textsl{Athena} observing time, particularly if the source is obscured. 
These results show that an unobscured AGN associated with a merger of SMBHs of masses $\sim10^{6}-10^{7}\msun$ at $z>1$ can be detected anywhere within the \textsl{Athena} FoV in $\sim$a few ks, increasing to $\sim$70~ks for lower mass super massive black hole mergers (SMBHMs) at $z=2$. 
If the associated AGN is obscured (and thus is most efficiently detected at 2-10~keV energies) then the exposure times increase, 
requiring day-long exposures except for the
most massive, and therefore potentially X-ray brightest, SMBH pairs. 
Lower mass, SMBHMs at $z>2$ that are associated to obscured AGN are likely to remain undetectable, even in extremely deep exposures, due to the impact of source confusion.
These numbers provide the rationale for
searching for the X-ray counterpart of a SMBHM event even prior to the merging occurs, as described in Sect.~\ref{sec:LISA_strategy}.
\begin{table}
\centering
\caption{Fluxes (0.5-2~keV) in erg~cm$^{-2}$~s$^{-1}$  and exposure times (in brackets) to detect a X-ray unobscured AGN at the Eddington limit with the current configuration of the \textsl{Athena} mirror+WFI.}
\begin{tabular}{lccc}
& M=10$^5$~M$_{\odot}$ & M=10$^6$~M$_{\odot}$ & M=10$^7$~M$_{\odot}$ \\ \hline
$z~=~1$& 5.3$\times$10$^{-17}$ (250 ks) & 5.3$\times$10$^{-16}$ (7 ks) & 5.3$\times$10$^{-15}$ ($<$1 ks) \\
$z~=~2$& 1.1$\times$10$^{-17}$ ($\gtrsim$1 Ms) & 1.1$\times$10$^{-16}$ (70 ks) & 1.1$\times$10$^{-15}$ (3 ks) \\ \hline
\end{tabular}
\label{tab:agn}
\end{table}
\begin{table}
\centering
\caption{The same as Tab.~\ref{tab:agn} but giving the 2-10~keV fluxes for an AGN obscured by a column density N$_H$=10$^{23}$~cm$^{-2}$.}
\begin{tabular}{lccc}
& M=10$^6$~M$_{\odot}$ & M=10$^7$~M$_{\odot}$ \\ \hline
$z~=~1$ & 8.6$\times$10$^{-17}$ ($\gtrsim$1 Ms) & 8.6$\times$10$^{-16}$ (270 ks) & 8.6$\times$10$^{-15}$ (8 ks) \\
$z~=~2$& 1.9$\times$10$^{-17}$ ($\gtrsim$1 Ms) & 1.9$\times$10$^{-16}$ ($\gtrsim$1 Ms) & 1.9$\times$10$^{-15}$ (70 ks) \\ \hline
\end{tabular}
\label{tab:agnobs}
\end{table}

\subsection{Other classes of BH mergers}
\label{sec:LISA_other}

Other classes of BH mergers will be prominent emitters of GW in the LISA band and, notwithstanding  the large uncertainties of the theoretical predictions, they could be potential targets for possible synergies between \textsl{Athena} and LISA.

\begin{itemize}

\item \textbf{Extreme  and Intermediate Mass Ratio Inspirals}. EMRIs are low-mass compact objects (neutron stars -NSs- and stellar mass BHs) spiralling into massive BH (MBH) (\cite{2004PhRvD..69h2005B}). 
Events occurring in AGNs discs are potentially an ideal probe for accretion theories.
 In the IMRI case, the inspiralling black hole is massive enough ($>100\msun$) to strongly perturb the surface density of the disc \citep{Derdzinski21} close to merger which, in turn, is going to affect the intensity and shape of the K-$\alpha$ reflection line \citep{McKernan14} in a way that depends on the extent of the corona and on the emissivity profile of the disc. Detecting such signature might therefore provide new insights on the structure of the MBHs corona and the innermost regions of the accretion flow, which are poorly understood. Conversely, in the EMRI case, the inspiralling object is too small to open a gap and it is thus unlikely to leave a direct signature in the AGN spectrum. The drag from the disc, however, might be strong enough to leave a distinctive signature in LISA data, due 
 to the dephasing of the GW signal with respect to vacuum predictions  \citep{Kocsis11,Barausse14} and the error volume might be small enough to host a single AGN that can be spotted by \textsl{Athena} \citep{Mcgee20}. A detection of such event would be a milestone in experimental tests of accretion theory. In fact the EMRI is essentially a probe mapping the midplane of the disc, which is inaccessible by any other means (being accretion discs optically thick). Combining information about the disc midplane extracted by the GW dephasing to the disc surface emission reconstructed from EM observations will allow unprecedented tests to disc structure models.

LISA is expected to detect about a hundred EMRI per year (with an uncertainty of about one order of magnitude in each direction) with a location accuracy ranging from few to sub-deg$^2$ \citep{Babak17}.
A few pointings lasting a few kiloseconds each could be sufficient to probe if there is an EM counterpart to an EMRI at z\,$\le$\,0.3 \citep{Mcgee20}. Furthermore at least 1\% of EMRI should take place in the AGN phase (see section \ref{subsec:LISA_caveats}), so one can expect a few such events to be observable by \textsl{Athena} in a year, possibly more if the probability of forming an EMRI is larger in AGN \citep{Levin07,Dittmann20}.

\item \textbf{Stellar mass black holes binaries coalescence}. 

The discovery of the first BH binary coalescence and  their routine observations during the first three observing runs by LIGO-Virgo \citep{Abbott:2020niy} open the possibility of detecting  some of these objects
 by LISA months to years before they transit in the high-frequency LIGO-Virgo sensitivity window.
LISA will detect tens to hundreds stellar BH binaries  during their long-lived  inspiral phase, with many being localised to better than 1~deg$^{2}$, with merger times predicted with an error of 10 seconds, weeks before coalescence. A few will enter into the advanced LIGO-Virgo bandwidth where the merger is taking place \citep{Gerosa19}, enabling a \textit{coincident}
(concurrent) GW-EM detection, whereby \textsl{Athena} and other EM facilities can simultaneously stare  at the position and predicted time of the merging event.

It is not obvious if an EM counterpart exists, although a tentative EM counterpart has been recently proposed for a LIGO-Virgo stellar black hole merging event \citep{graham20}.
 Some of them may take place in the Milky Way (\cite{2018MNRAS.480.2704L}). For a binary of a combined mass of $\approx 60$M$_\odot$, at 10 kpc, the Eddington luminosity in X-ray corresponds to a flux of about $10^{-6}$erg\,s$^{-1}$cm$^{-2}$. So \textsl{Athena} will be sensitive to X-ray luminosities down to $10^{-10}$L$_{\rm Edd}$.

\end{itemize}

\subsection {Accreting stellar binaries} 
\label{sect:wd}

\begin{figure}[!t]
\centering
\includegraphics[width=0.7\textwidth]{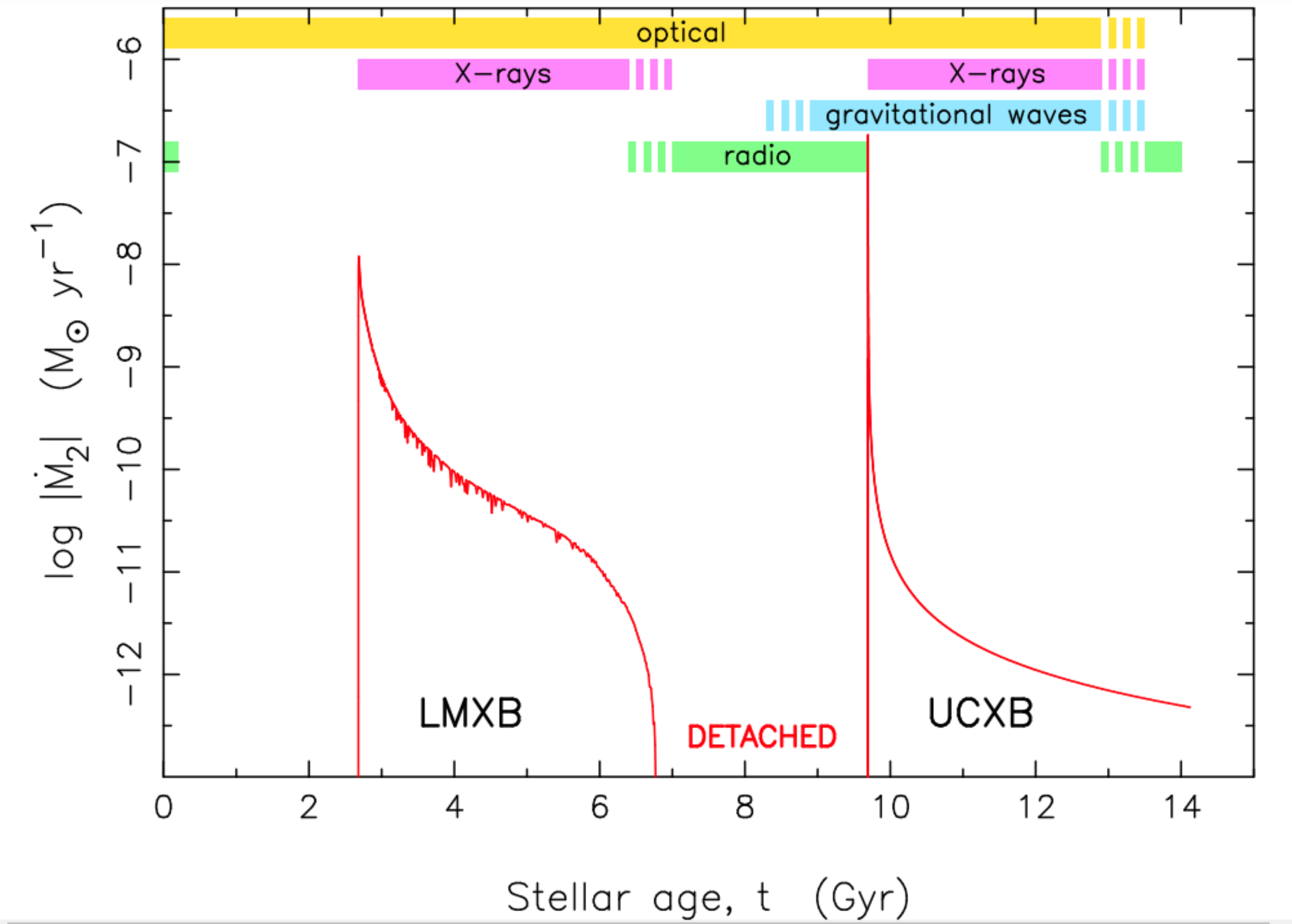}
\caption{ The evolution of the initial main sequence (MS) star + NS binary with respectively 1.40\,M$_\odot$ and 1.30\,M$_\odot$ mass components. The system evolves through two observable stages of mass transfer: an LMXB for 4\,Gyr, followed by a detached phase lasting about 3\,Gyr where the system is detectable as a radio millisecond pulsar orbiting the helium WD remnant of the donor star, until GW radiation brings the system into contact again, producing an UCXB. The colour bars indicate detectability in different regimes resulting in synergies between electromagnetic bands and LISA. The figure is adopted from \citet{Tauris:2018}.
}
\label{fig:LISA__2} 
\end{figure}

Since X-ray observations became available it has been realised that the Milky Way harbors a large variety of stellar X-ray binaries. These are Roche-lobe filling main sequence (MS) stars transferring matter to a compact companion (NS, BH or white dwarf) and are traditionally classified in the high- and low-mass X-ray binaries (HMXBs and LMXBs) and cataclysmic variables (CVs). When the MS star evolves into a compact object, newly formed double compact object binary can further shrink orbital separation via GW radiation reaching orbital periods of less than a few hours; at which point the binary may be detected with LISA. Currently, two types of stellar binaries emitting both X-rays and GW have been observed and unambiguously identified: ones with a neutron star accretor, called ultra-compact X-ray binaries (UCXBs), and ones with a white dwarf accretor (called AM CVn stars, named after the first detected system of this type). Figure~\ref{fig:LISA__2} shows an example of the evolution of a  main sequence - NS binary system in the accretion rate - stellar age parameter space \citep{Tauris:2018}.

{\bf Accreting white dwarf binaries (AM CVns).} Our understanding of stellar evolution and of the stellar initial mass function imply that over 95\% of all stars in the Milky Way will end their lives as white dwarfs (WDs). Between $\sim 5\%-10\%$ of all WDs are in double WD (DWD) binaries (\cite{maxted99,toonen17,maoz18}), with simulations predicting today over $10^8$ systems in total (\cite{nelemans2004}). In their evolution, DWD binaries do not always remain detached. A small and highly uncertain fraction ($\sim 10^{-3}$) of DWD survive the onset of mass-transfer and currently stable mass transfer occurs (called ``AM CVn"). These short ($<30$ min) period binaries are strong UV/X-ray and mHz GW emitters. Actually, the most compact binaries known today, HM Cnc (with orbital period of $5.4$~min) and V407~Vul (9.5 min), are semi-detached AM CVn systems and currently the loudest guaranteed (i.e. we know they exist) LISA sources. 
We expect around
150  systems that can  be detected both   in GWs and in X-rays, with $F_X\gtrsim $10$^{-13}$~erg~s$^{-1}$~cm$^{-2}$ in the 0.5-2~keV energy band, and about 10 times more at  $F_X\gtrsim $10$^{-15}$~erg~s$^{-1}$~cm$^{-2}$(\cite{nelemans2004}). Due to the high $S/N$ and the short period ($\gtrsim$ 1 mHz), a large fraction of these sources will be localized within the \textsl{Athena}/WFI FoV by LISA. In addition, the brightest in X-ray will be precisely identified from pre-existing  X-ray survey data, enabling observations with the \textsl{Athena}/X-IFU. Combined  EM (from optical to X-ray)-GW observations of accreting DWDs would allow us to uniquely study fundamental physical processes, related to WD accretion physics and mass transfer stability. For example, the largest uncertainties in predictions of the final fate of DWD binaries (merger versus stable mass transfer) and hence of supernova (SN) Ia rates is the treatment of the onset of mass transfer when the larger WD fills its Roche Lobe and starts to accrete onto its companion. So far the space density of observed systems in EM is several orders of magnitude below the theoretical predictions, which challenges our theoretical understanding of their formation and evolution (\cite{carter13,ramsay18}). The large statistical sample of accreting and non-accreting DWDs detected by LISA will provide us with space densities and system properties of each group individually. This in turn gives a direct measure of how many DWDs prevent the merger as well as what are the system properties of the surviving AM\,CVn binaries.
Moreover, the angular momentum transport in accreting DWDs is a combination of GW radiation, which tends to shrink the system, and mass transfer, which typically widens a binary. GW observations alone can hardly resolve the degeneracy. However, the system properties derived from combined EM-GW data can disentangle the contribution from GW radiation and mass transfer from the overall period evolution ($\dot{P}$) and study, for the first time, the transport of angular momentum in accreting DWDs on a statistical significant number of systems. Indeed, 
the amount of transport of angular momentum is closely related to how much mass is being accreted in the system. The X-ray (and UV)   luminosity and spectrum will then allow us to constrain the accretion rate, thus solving the degeneracy and getting a deeper insight into the radiative property of matter. In particular we expect that the brightest X-ray sources associated to LISA AM CVns will be localized post-facto by existing X-ray surveys, but high--quality spectral information for a large sample can only be obtained by \textsl{Athena} with relatively short observations ($\approx 10$~ks). 

{\bf Ultra-compact X-ray binaries (UCXBs)} represent short-orbit (typical orbital period of less than 1 hour) X-ray binaries observed with an accreting NS (or BH; however systems containing BHs have not been confirmed yet). Based on the orbit compactness argument, it is expected that the donor star is either a WD, a semi-degenerate dwarf or a helium star \citep{Rapport:1982}. UCXBs are thought to form from post-LMXB systems when the donor star starts to fill its Roche lobe avoiding a possible catastrophic event \citep{Webbink:1979,Nelson:1986,Podsiadlowski:2003,Nelemans:2010,Heinke:2013}. The formation of an UCXB is also possible via stellar exchanges in dense stellar environments \citep{Fabian:1975}. Indeed, a large fraction of the known UCXB population is found in globular clusters. Based on the mass-transfer rate UCXBs are classified into two categories: persistent and transient \citep[e.g.][]{Heinke:2013}. So far only about 14 UCXBs have been confirmed - of which 9 are persistent and  5 are transient - and a comparable number of candidates are known. Thus, UCXBs are either hard to detect or represent a rare population. Both \textsl{Athena} and LISA will be pivotal for discovering and characterising this binary population and for elucidating on the debated formation and long-term stability of these systems. In particular, the X-ray observations are essential for multi-messenger studies because the majority of UCXBs are located in the Galactic plane or in globular clusters, where ultraviolet and optical spectroscopy are challenging or impossible. 
The success of the \textsl{Athena}-LISA synergy is due to LISA’s ability to provide binary orbital parameters such as the chirp mass, orbital period and eccentricity, while X-ray observations yield the chemical composition of the donor star. Currently available observations show that donor stars in UCXBs are either carbon-oxygen WDs or helium WDs. 
More detailed X-ray spectroscopy potentially will be able to make detailed abundance estimates for the donor stars, that may yield the time of the donor formation (through estimates of the non-CNO elements abundances) and its  initial metallicity \citep{Nelemans:2010}. 

{\bf Stripped stars in binaries with compact objects.}
Stars that have been stripped of their hydrogen envelopes are compact enough (although still could be 10 times bigger than white dwarfs) to fit into a compact orbit emitting gravitational radiation in the LISA's frequency band (e.g. CD--30$^\circ$11223 and J2130$+$4420, \citealt{gei13,kup20}). The tightest stripped star systems are also the ones that are most likely to lead to gravitational wave mergers. Thus, finding these binaries is valuable for understanding latest stages of binary evolution immediately prior to the double compact object formation. Although stripped stars are expected to be bright at optical wavelengths (reaching Gaia $G$ magnitude of 10 for the brightest ones, \citealt{Gotberg:2020}), they are located deep in the Galactic disc and thus are heavily affected by extinction that makes them hard to find. In particular, no intermediate-mass stripped star in a binary with a compact object has been found to date. X-ray observations could reveal these systems when they are in the accretion state.
\citet{Gotberg:2020} estimated a hundred stripped star with white dwarf companions and several stripped star with neutron star companions reaching $S/N$ ratio greater than LISA's detection threshold after 10 years of gravitational wave observations. 
Specifically, for stripped stars orbiting neutron stars in the accretion stage, they estimated X-ray luminosity between $10^{33} -- 10^{36}$\,erg\,s$^{-1}$ and could therefore be detectable simultaneously by \textsl{Athena} and LISA \citep[see also][]{Shao:2019}.  As for AM CVns, the combination of GW and X-ray observations will help to understand tidal interaction and stellar distortion crucial for improving theoretical models. 

\section{\textsl{Athena} observational strategy for monitoring massive black hole coalescences}
 \label{sec:LISA_strategy}
\subsubsection{X-ray counterpart identification}

Assuming that the counterpart of a merging massive BHB is a photon-emitter, and that it produces  a flux above the instrumental threshold, the challenge is then to {\it identify} the counterpart in a field that will likely count hundreds to thousands of sources.  In this respect the X-ray band offers the best combination, thanks to the sensitivity and the FoV catered for by \textsl{Athena}. 
Assuming that the broad-band EM spectrum has an overall shape similar to that observed in SMBHs at the center of active galaxies ($\alpha_{\rm OX}=1.3$ in optical; \cite{vasudevan09b}) and $\nu L_{\nu}(5 {\rm GHz})/L(2-10\ \rm{keV}) \lesssim 10^{-4.5}$  for low-luminosity AGNs (\cite{Terashima03,panessa07}), one can relate the X-ray flux to the optical magnitude or radio flux and then compare the number of field sources expected in the three bands.  For example, the X-ray sky at a flux of $\approx 10^{-15}$~cgs in the 2-10~keV range is populated with about 3000 sources deg$^{-2}$ (\cite{Georgakakis08}), while at the corresponding magnitude m$_V\approx 24.3$ and radio flux of $\approx3 \mu$Jy there are about  30 (10) times more contaminating objects in the optical (radio) band (\cite{Smail95,Vernstrom16}).
Still, a proper {\it characterization} of the source behaviour (mostly in the time domain) is thus requested both from theory and observations to pin down the candidate. 

\begin{figure}
\begin{center}
\includegraphics[width=\textwidth]{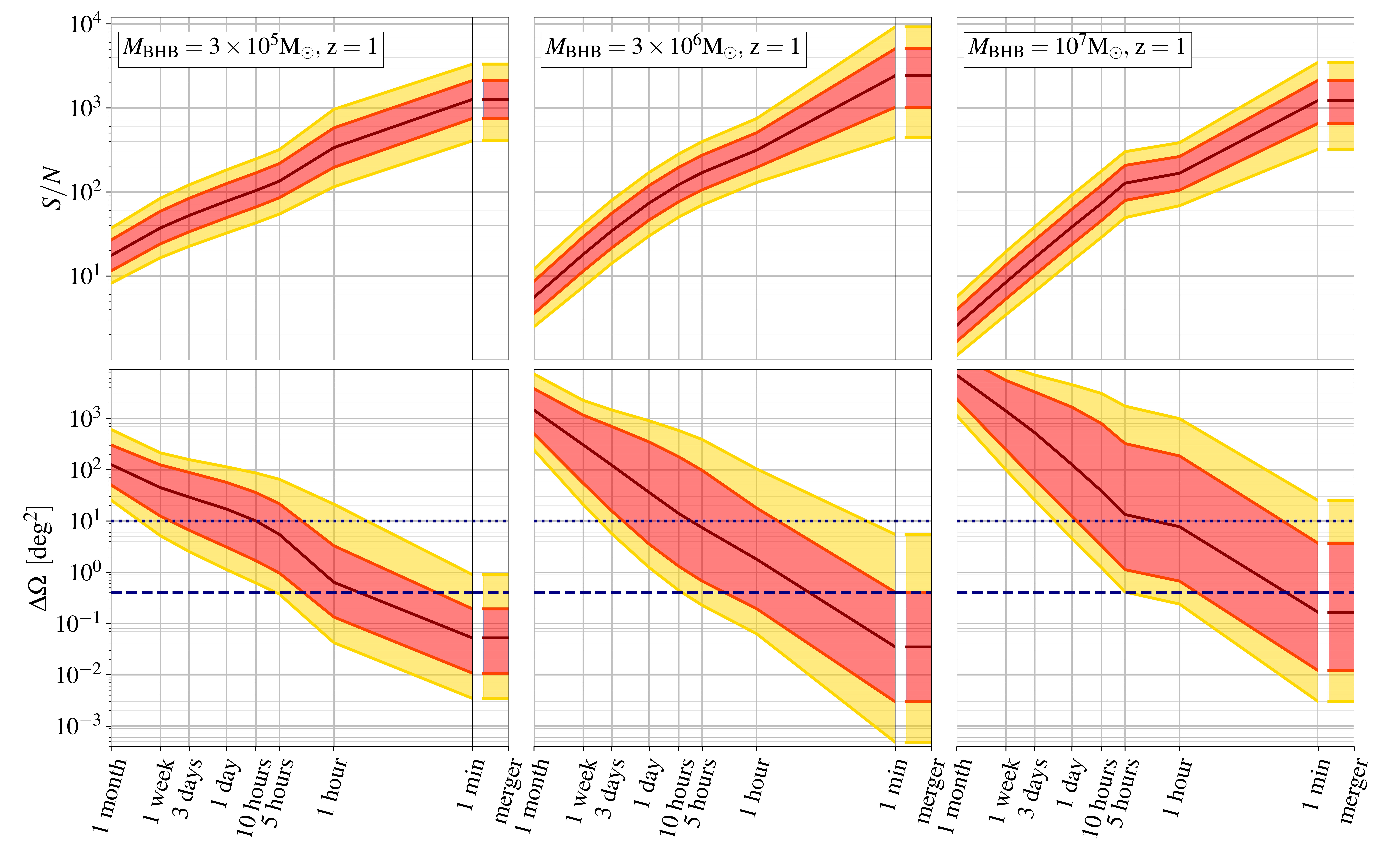}
\caption{Signal-to-noise ratio ($S/N$) (top) and sky localization error $\Delta \Omega$ in deg$^2$ (bottom) versus time to merger for precessing binaries with total  mass equal to $3\times 10^5\,\rm M_\odot$ (left panel), $3\times 10^6\,\rm M_\odot$ (central panel), $10^7\,\rm M_\odot$ (right panel) as computed in \cite{Mangiagli20}.  The sources are located at $z=1$. The binary mass ratio of these sources is extracted randomly between [0.1-1], the spin between [0-1] and polarization, inclination and sky position angles are extracted from a sphere.   Full precessing inspiral with higher harmonics waveforms are used to carry out the analysis up to one hour before coalescing. The signal is extrapolated down to merger using a simplified waveform and scaling the uncertainty on $\Delta \Omega$ with $(S/N)^{-2}$.
Shaded areas are the 68\% and 95\% 
confidence interval computed over $10^4$ systems and the dark solid line 
is the median value.
The horizontal dashed line denotes the Field of View (FoV) ($\sim 0.4$ deg$^2$) of the WFI on-board \textsl{Athena}. The 10 deg$^2$ wider FoV of LSST  is denoted with a dotted line. It is noticeable that at merger a large fraction of the sources is in the \textsl{Athena}/WFI FoV.}
\label{fig:LISA_LOC_vs_time} 
\end{center} 
\end{figure}

\subsubsection{Localization and sample selection}
The uncertainty in sky localization of massive BH mergers by LISA depends on a number of factors: the binary mass, mass ratio, spins, binary inclination, polarization angles, sky location relative to the LISA antenna pattern and redshift. The localization depends on the cumulative signal-to-noise ratio ($S/N$), which improves significantly  with time, from the inspiral phase to the merger. Therefore, the best localization occurs at the end of the merger, with  sources that can be localized with a precision  down to arc-minutes.  In Fig.~\ref{fig:LISA_LOC_vs_time} we show the $S/N$ and sky-position uncertainty $\Delta 
\Omega$ as a function of the time-to-merger for the three systems at $z=1$, as described in the caption. Light binaries live longer in the LISA band and accumulate a median $S/N$ of 20 already 1 month before coalescence, compared to intermediate and heavy systems, which accumulate the same $S/N$ a week and few days before merging, respectively.  Light and intermediate systems appear the best targets for planning \textsl{Athena}-LISA joint observations at the end of the merger. These binaries are extremely loud sources with median $S/N$ at merger in the thousands.

More than 60\% of the binaries with  $M_1=3\times 10^5\,\rm M_\odot$ out to
$z\sim 2$, and 50\% of the mergers with $M_1=10^6\,\rm M_\odot$
up to $z \sim 4$ 
can be observed in the post-merger phase within the \textsl{Athena}/WFI, and even within the X-IFU FoV.  
Also, up to $\approx 30\%$ of more massive systems, such as those with $M_1=10^7\,\rm M_\odot,$ can be well localized in the post-merger phase, out to $z\sim 3$. 
We define a {\it Golden Binary} as being composed of objects such that the error box derived {\it after} the merger is smaller than the WFI FoV   ($0.4$~deg$^2$). This {\it Golden Sample} comprises the majority of binary mergers with mass  within $3\times 10^5\,\rm M_\odot$ and $10^7 \,\rm M_\odot$ up to $z\approx 2$ and allows \textsl{Athena} to search for X-ray emission produced in the post-merger phase. 

\begin{figure}
\centering
\includegraphics[width=0.65\textwidth]{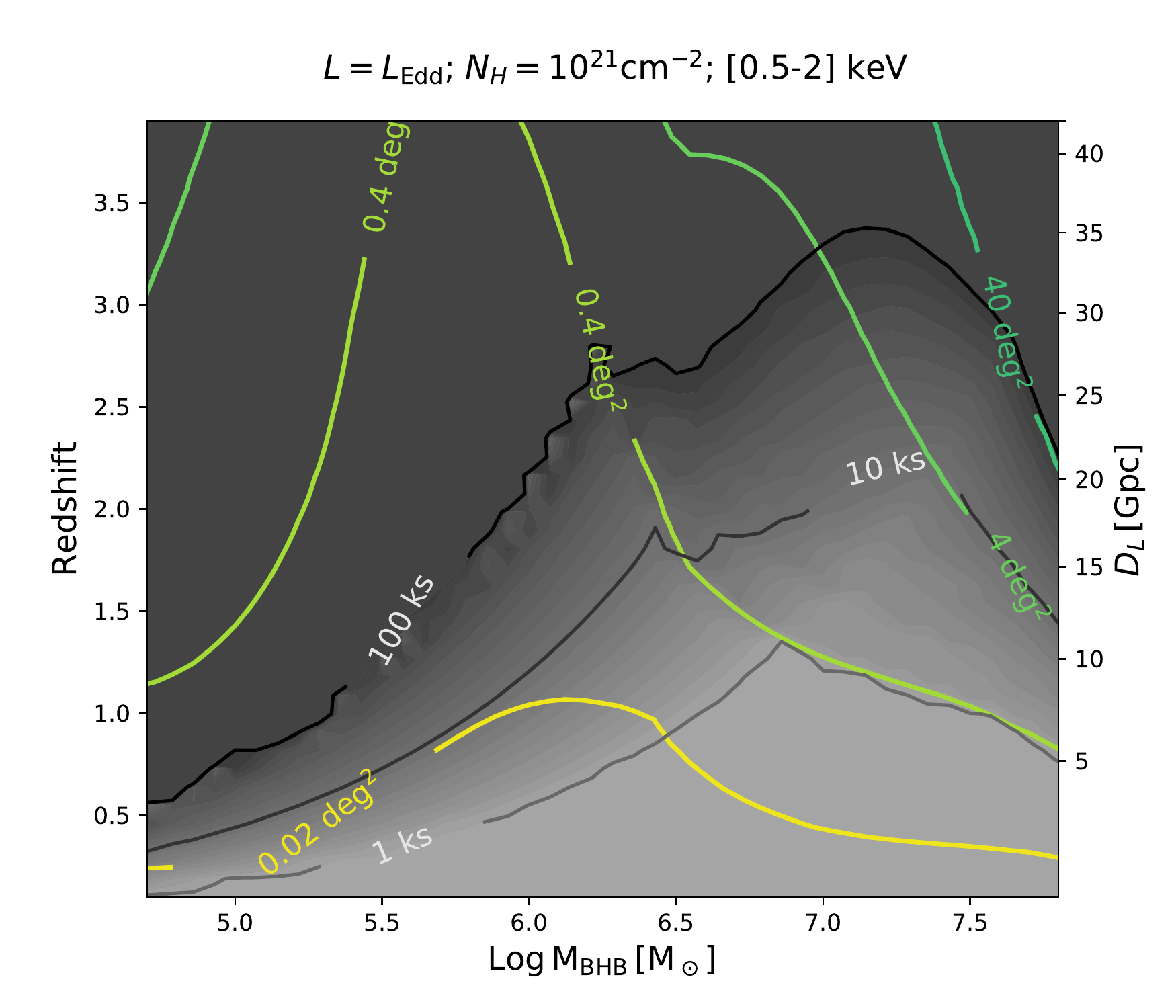}
\caption{
Feasibility of joint GW/X-ray observations of massive BH mergers in the plane redshift versus source-frame mass $M_{\rm BHB}$. 
Green-yellow contours mark the median sky location accuracy of mergers achievable by LISA observations. Grey shaded contours represent the corresponding total exposure time ({\it i.e.} accounting for multiple pointings, when necessary) needed by \textsl{Athena} to cover the LISA error-box with the sensitivity needed to detect the sources, while black lines demarcating the 1, 10, and 100 ks exposure time contours.  In the figure we set the fiducial source bolometric luminosity to the Eddington limit assuming negligible obscuration (i.e., a hydrogen column density of $10^{21}$ cm$^{-2}$). The X-ray portion of the spectrum, between 0.5-2~keV, is modelled as a single power law with spectral index $\alpha$~=~0.7 consistent with the average quasar emission \citep{lusso12}, normalized so that the luminosity in the 2-10~keV energy band is 3\% of the bolometric luminosity.
}
\label{fig:LISA_localizationCurves} 
\end{figure}

For the highest $S/N$ events  the localization  derived in the inspiral phase can allow \textsl{Athena} to repoint {\it before} the merging takes place.
A key element of any strategy to identify EM counterpart of GW events are the localization capabilities of the GW detector. Assuming the current LISA configuration, it can be estimated that 20-40\% of the binaries with primary $M_1=3\times 10^5 \,\rm M_\odot$ and $10^6\,\rm M_\odot$ at redshift $z\le0.5,$ and 10\% at $z=1$  can be localized within the 
\textsl{Athena}/WFI 5~hours before merger. The results are weakly dependent on the value of the mass ratio $q$.
Conversely, only about 10\% of the $M_1=10^7\,\rm M_\odot$ systems at $z=0.5$ can be pre-localized.  
We define a {\it Platinum Binary} as being composed of objects whose localisation error, determined 5 hours before the merger, is smaller than the WFI FoV. The timing is consistent with the \textsl{Athena} capability of carrying out a ToO in 4 hours. The {\it Platinum Binary} thus comprises a fraction of binary mergers with mass  within $3-10\times 10^5\,\rm M_\odot$  below $z\approx 1$, and thus likely to be rare (Tab.\ref{tab:rates}). However, for the {\it Platinum Binaries}  the inspiral and merging phases can both be observed with \textsl{Athena}, opening the intriguing perspective to observe in X-rays the BH merging event in the act.  With a proper observing strategy, \textsl{Athena} can actually start observing a few days before the final binary coalescence. At this time the localisation error of objects in the {\it platinum binary} is $\approx 10$~deg$^2$ (cf. Fig.~\ref{fig:LISA_localizationCurves}), an area that can be effectively covered by tiling WFI observations in about 3 days.
We consider hereby this accuracy as the threshold for the activation
of a counterpart search strategy with the \textsl{Athena}/WFI, and assume 3~days before the merging as the corresponding start time. 

An error box of 10~deg$^2$ can
be covered with the \textsl{Athena}/WFI in 3~days with a raster scan of
at least 23 observations of $\simeq$9~ks each. The ``at least'' caveat is primarily driven by the sensitivity of the \textsl{Athena} telescope decreasing 
gently off-axis due
to the vignetting effect (\cite{willingale13}). 
This implies that a certain overlap between adjacent pointing directions may be required to ensure that a given sensitivity is homogeneously achieved over the whole WFI FoV. 
This may increase the
number of required pointings, decreasing the exposure time available for each of them.
Once the LISA event localization is comparable to, or smaller than the \textsl{Athena}/WFI FoV, \textsl{Athena} could stare to the predicted error box up to the time of the merger. With the improvement of the LISA localization the \textsl{Athena} pointing strategy can be optimized to cover the most likely location of the trigger at any time.

A tiling strategy
of short observations could
allow \textsl{Athena} to detect a significant fraction
of the counterpart of the GW-emitting
SMBHM in {\it one of the WFI observations} prior to the merging
(estimated to be $\ge$20\% for BH masses $\le$10$^6$~M$_{\odot}$
according to the simulations described later), a significantly more challenging
issue is identifying "on-the-fly" which of the hundreds of WFI
sources is the true counterpart of the forthcoming merger. A possible
``smoking gun'' is the variability pattern in the soft and hard
X-ray light curves, mirroring
the GW strain (cf. Sect.~\ref{sect:SMBH_predictions}). The expected variability time-scales could vary from minutes to
hours. This would require that the source should be observed several times
prior to merging in order for the variability pattern to be discernible,
even if the time-dependent quasi-period could be used as a prior in the X-ray light curve analysis
as derived from the measurement
of the GW strain by LISA.
Identifying the correct source during the merging would be made only more complicated by the
commonly observed variability
in the X-ray light curves of
many classes of celestial sources, most notably AGN. 

\begin{figure}
\centering
\includegraphics[width=0.6\columnwidth, angle=90]{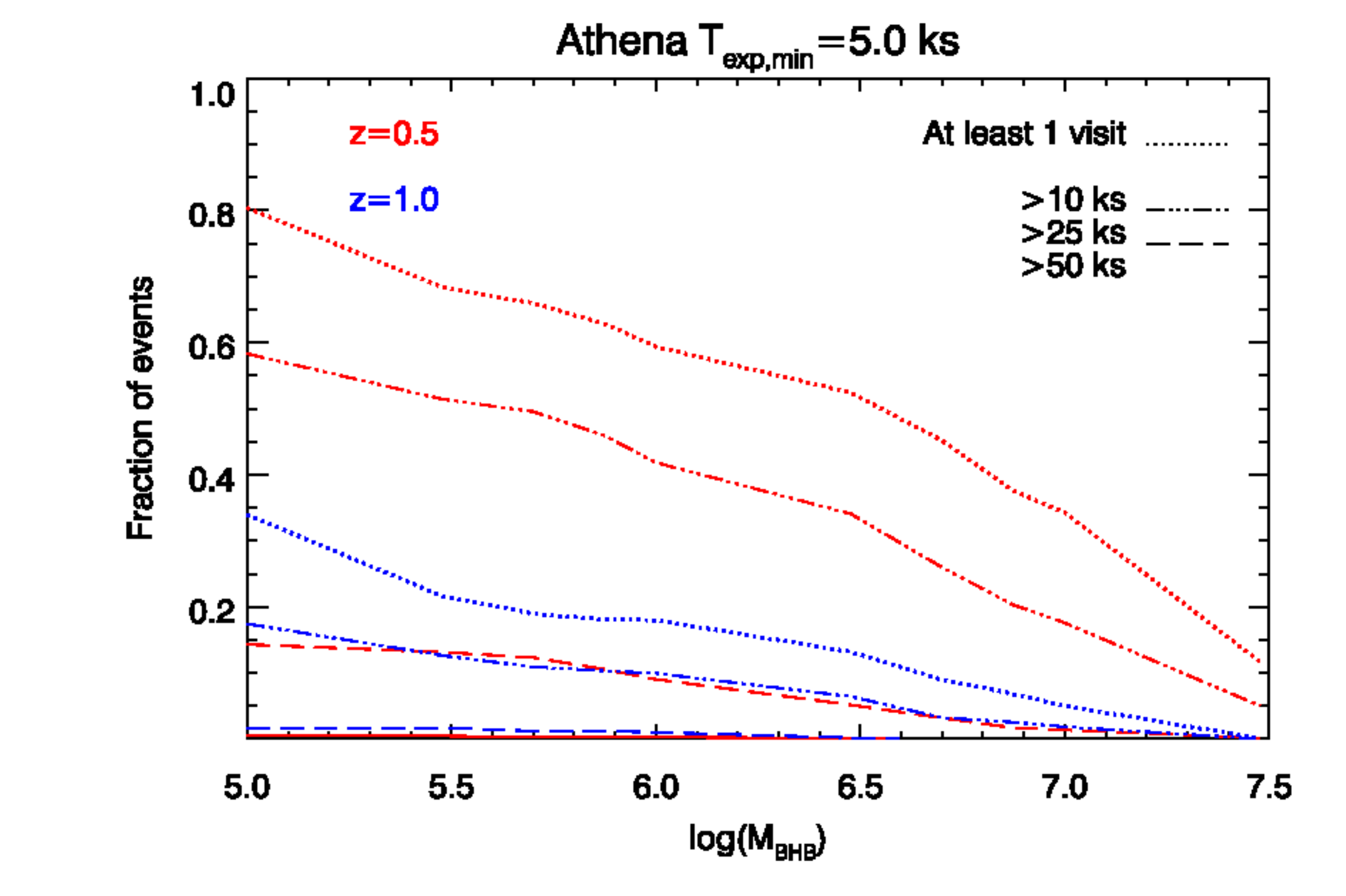}
\caption{Fraction of LISA events observed by the \textsl{Athena}/WFI FoV {\it at least once}
prior to the merging time as a function of
the event total BH mass ({\it dotted lines}).
Different colors represent different SMBH binary redshifts ({\it red}: $z\sim=\sim0.5$;
{\it blue}: $z\sim=\sim1$).
Different {\it dashed line} styles represent the total accumulated exposure time prior to merging, according to the legend in the {\it inset}. The
minimum exposure time of each pointing is assumed to be 5~ks.} 
\label{fig:LISA_athena_summary} 
\end{figure}

\begin{figure}
\begin{center}
\hbox{
\includegraphics[width=0.32\columnwidth, angle=90]{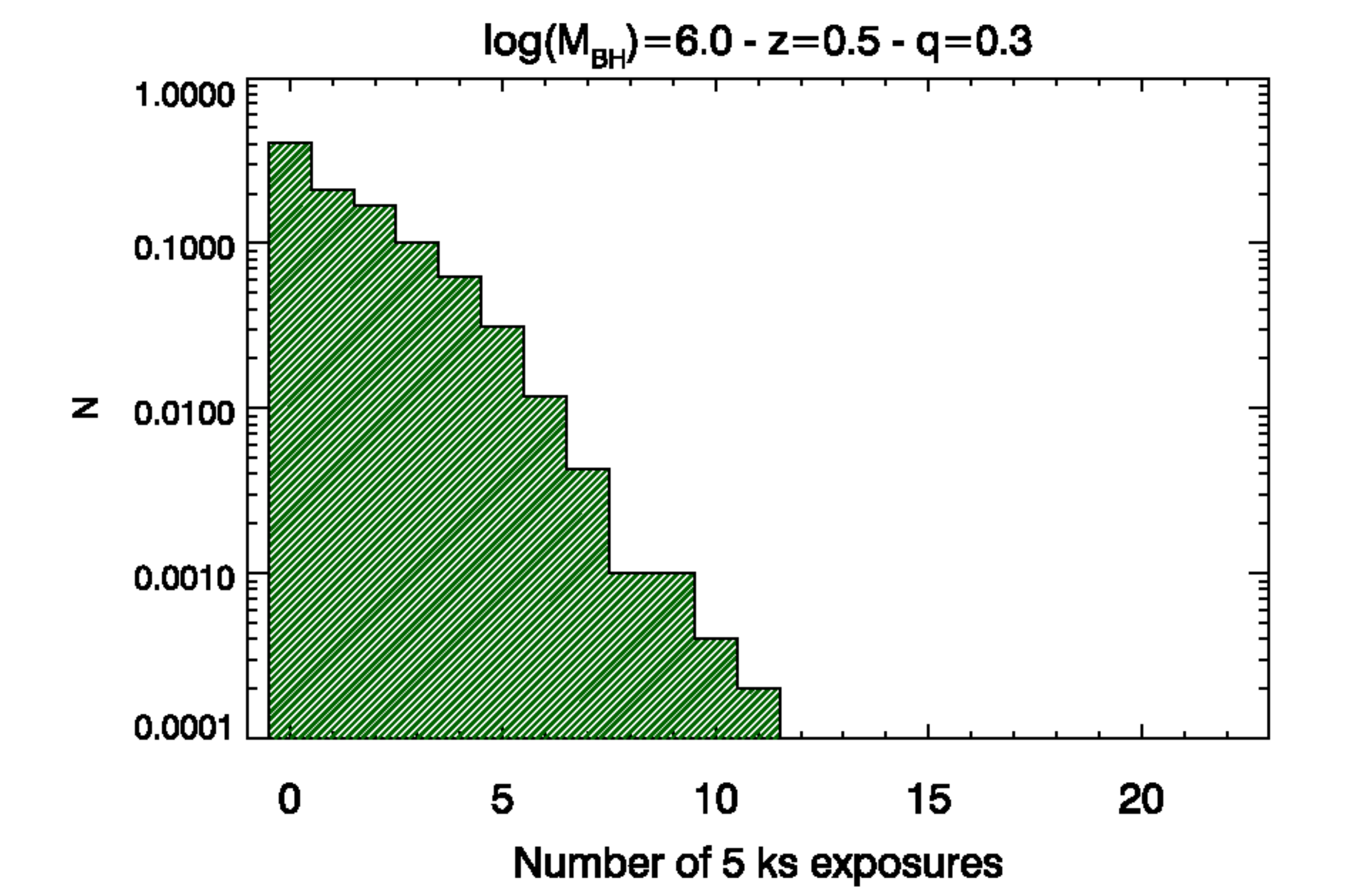}
\includegraphics[width=0.33\columnwidth, angle=90]{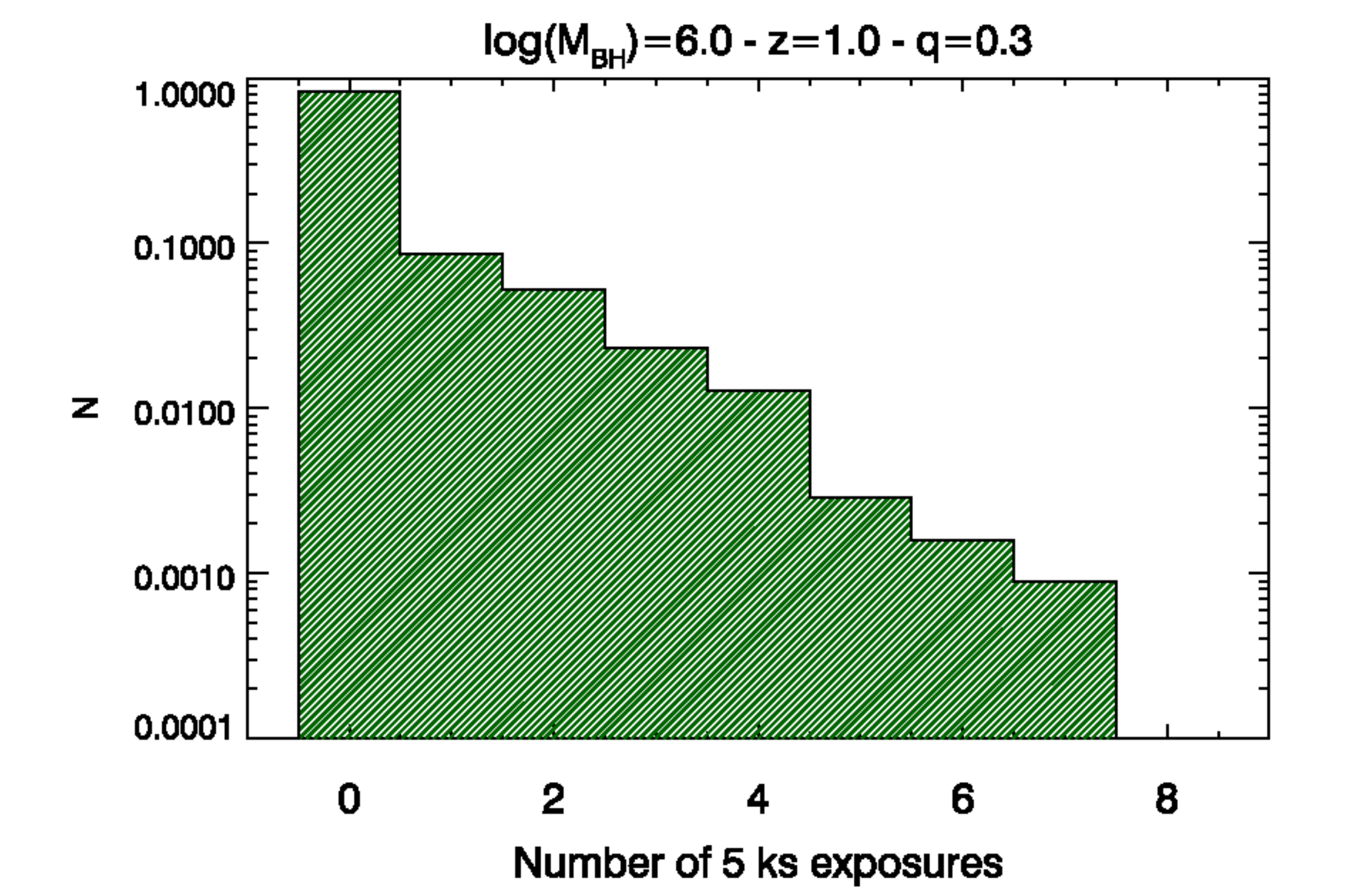}
}
\caption{Distribution of the occurrence fraction of the number of WFI 5~ks snapshots of an X-ray counterpart of a SMBH merger event in the simulated follow-up strategy described in text. The histograms correspond to
a total merger BH mass of 10$^6$~M$_{\odot}$ at z~=~0.5 ({\it left panel}) and z~=~1 ({\it right panel}).} 
\label{fig:athena_visits} 
\end{center} 
\end{figure}

The challenges associated to the identification of a SMBH merging X-ray
counterpart are highlighted by Fig.~\ref{fig:athena_visits}, showing the
normalized distribution of the number of visits that a SMBH merger
event falls into the \textsl{Athena}/WFI FoV prior to the coalescence time. 
This distribution is the result of 10000 Monte-Carlo simulations, assuming that: \textsl{Athena} starts following the LISA-detected event once the localization is better than 10~deg$^2$ with a raster of WFI pointings of equal exposure time until the LISA localization is smaller than the WFI FoV, at which point \textsl{Athena} can stare at this position up to the time of the merger; that the center of the tile is updated whenever a new position  is available from LISA, and that a new pointing tile cycle starts from this position. Compared to a tiling pattern that covers homogeneously the full error box, this strategy maximizes the  total exposure time on the source at the expense of not covering a few sources (Fig.~\ref{fig:LISA_athena_summary}).  We assumed also 1~hour overhead time for the transmission and calculation of the LISA coordinates, a 4~hour response time for \textsl{Athena} to reach the initial position, and an \textsl{Athena} agility of 4~degrees per minute during the raster. 
Assuming a minimum exposure time of 5~ks (sufficient to detect a $\gtrsim$10$^6$~M$_{\odot}$ unobscured AGN at $z\le$1; cf. Tab.~\ref{tab:agn})
only $\le 10$\% of z~=0.5 events can be covered with a number of visits larger than 5 (Fig.~\ref{fig:LISA_athena_summary}. The fraction of these events that can be observed 10~times with a total exposure time $\ge$50~ks is $\le$0.1\% (Fig.~\ref{fig:athena_visits}). 
In summary, our current understanding of the localization capability of LISA, of its operational constraints, of possible mechanisms producing X-rays in circumbinary discs and mini-discs, and of the possible variability pattern of this emission conspire in making a measurement
X-ray variability patterns, as predicted by models of
space-time induced variation of the accretion flow onto the
merging SMBHs, an extremely challenging, albeit exciting, possibility.
While the detailed strategy will need to be defined and refined based on future improvements of the modeling, and once the \textsl{Athena} observational plan has been established, an exploratory strategy can be conceived whereby {\bf \textsl{Athena} follows-up about five sky fields where high-$S/N$ Platinum Binary events are bound to happen}. The field selection shall be based on the best $S/N$ LISA candidates. This strategy would imply a relatively modest ($\le$1~Ms) time investment.

For a large fraction of SMBH merging events,  LISA will be able to localize the
position using the merger and the ring-down well enough to locate it within 
the WFI FoV (the Golden sample).
With predictions of about tens of events over the mission lifetime,
several could be followed after the merging occurs, to
trace the re-brightening of the disc or the shock-heating of the interstellar medium
by a prompt jet, or a late afterglow due
to gravitational recoil.
This indicates the truly exciting opportunity to witness
the birth of an AGN. 
If \textsl{Athena} and LISA would be operated simultaneously,
a strategy is conceivable whereby a certain numbers of Golden and Platinum Binary fields
are monitored periodically after merging to search for X-ray counterparts, coupled with deep 
ToO observations, if/after a counterpart is detected. 
The confusion limit ($\sim$2$\times$10$^{-17}$~erg~cm$^{-2}$~s$^{-1}$)
would be reached in about 4 days.
One can therefore conceive an exploratory strategy {\bf involving the $\sim$5 most promising $S/N$ Golden Sample LISA candidates, limiting the corresponding time investment to $\le$1-2~Ms}. These targets may include some of the Platinum sample followed-up during the inspiraling, although priority should be given to those events whose localization accuracy at merging time is consistent with a follow-up observation with the X-IFU FoV($\le$2').

\chapterimage{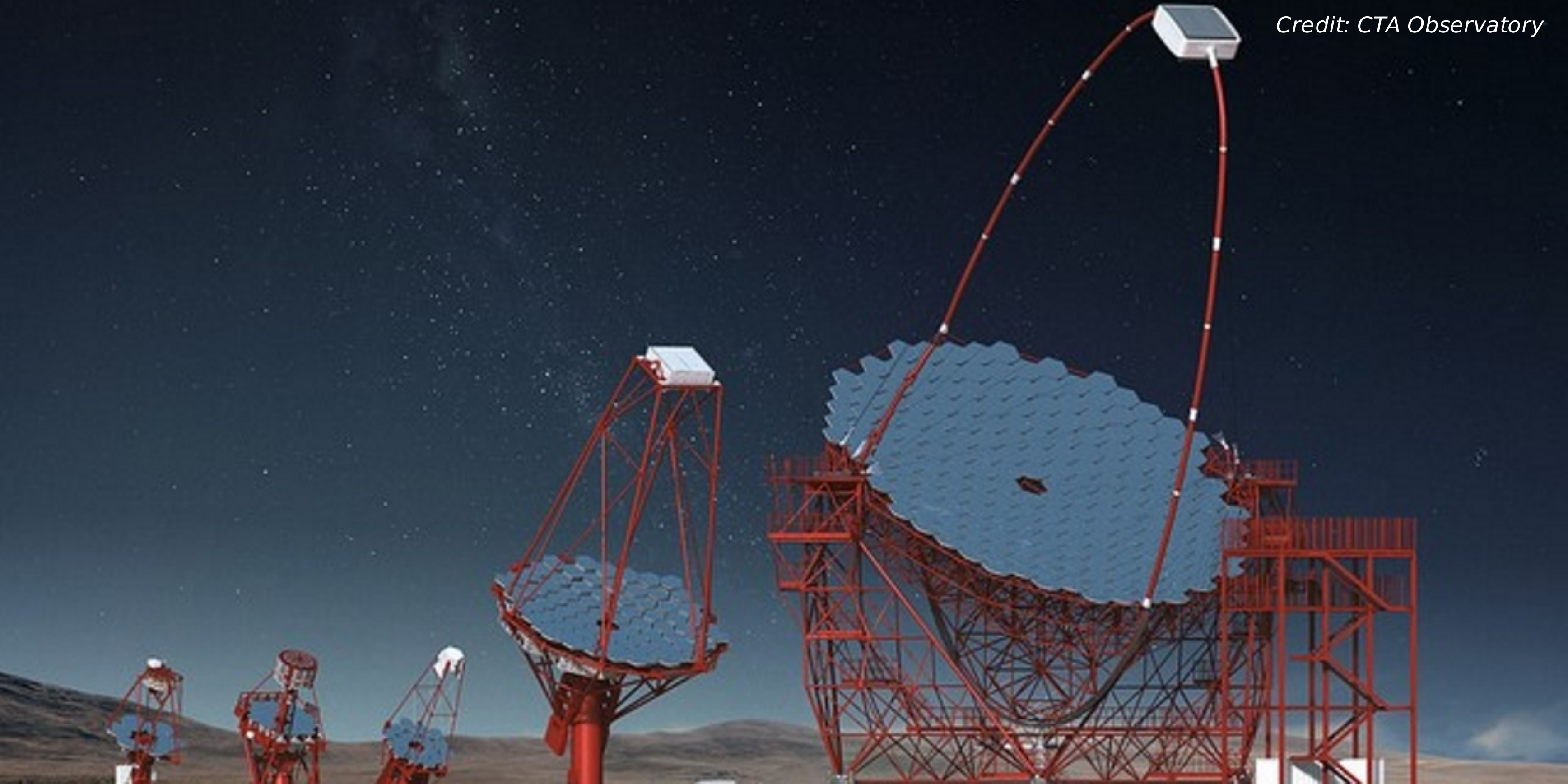} 
\chapter{ \textsl{Athena}, neutrino and VHE gamma-ray observatories: ICECUBE, KM3NET \&  CTA}
\label{chapter:HE}

\textit{{\color{purple}{By M. Ahlers, A. Coleiro, E. de O\~na Wilhelmi, J. Vink, P. Padovani.}}}
\section{Science Themes}\index{}\label{sec:intro_nonthermal}

Many X-ray sources have non-thermal radiation components over a wide range of the electromagnetic spectrum.
In some cases their spectral energy distribution is even dominated by non-thermal radiation.  
Examples are active galactic nuclei (AGN) in all forms, gamma-ray bursts (GRBs), supernova remnants (SNRs), and pulsar wind nebulae (PWNe).  These non-thermal emission components  are invariably the results of particle acceleration by collisionless shocks, or through magnetic reconnection processes.  In the radio and X-ray  bands the non-thermal emission is most likely synchrotron radiation, caused by relativistic electron and/or positron populations. X-ray emission in particular is important for detecting the highest energy electrons/positrons (typically $>10$~TeV for magnetic fields below 1 mG).
 These electron/positron populations, collectively
referred to as leptons, also cause inverse Compton scattering and non-thermal bremsstrahlung in gamma rays.

However, the acceleration processes themselves are often not confined to leptons; in fact in many cases the dominant population
of accelerated particles consists likely of atomic nuclei, i.e. hadrons. The relativistic hadronic populations may be 
energetically more important, but their presence can only be inferred from gamma-ray observations and/or neutrino detection.
The latter is a unique signature of accelerated hadrons, whereas gamma rays can either be hadronic, or (as stated above) leptonic in origin.

The accelerated particles will eventually escape their acceleration sites, thereby filling the Galaxy and even intergalactic space with high energy hadrons. We know of these hadrons as we observe them on Earth as CRs; yet, deflections in turbulent magnetic fields before arrival do not permit to directly trace their origin. Less than 1\% of the CRs are leptonic, so in order to understand the origin of $>$99\% of CRs we need gamma-ray and very-high-energy (VHE) neutrino observations. Revealing the presence of CRs, and thereby pinpointing the astrophysical sources of CRs, are the main science drivers for present and future gamma-ray observatories, and high-energy ($\gtrsim TeV$) neutrino detectors. 

Since the acceleration mechanisms for hadrons and
leptons are the same, X-ray observations are important to probe the active acceleration conditions, since the X-ray synchrotron emitting electrons lose their energy on short time scales. Moreover, X-ray observations
provide superior angular resolution and statistics, and the thermal X-ray component, if present, helps to understand the
conditions in which particles are accelerated, as well as provide a probe of the local plasma and radiation energy densities, 
with which accelerated particles interact.

This in a nutshell shows the synergy between \textsl{Athena} on the one hand, and the CTA gamma-ray observatory, and the
HE neutrino detectors \href{https://icecube.wisc.edu/}{IceCube} and \href{https://www.km3net.org/}{KM3Net} on the other hand: {\em X-ray observations} give a handle on the {\em accelerated leptons} 
and the properties of the local environment 
of CR sources, 
and {\em gamma-rays and HE neutrinos} detections are able to reveal the presence and spectrum of
(ultra)relativistic {\em hadrons}. The X-ray observations are then essential to translate the gamma-ray and neutrino measurements into the CR energy budget of these CR sources.

By the time \textsl{Athena} is launched we expect that CTA-North and South (\href{https://www.cta-observatory.org/}{the CTA observatory}) will be the most sensitive VHE gamma-ray observatories, and that the most sensitive HE neutrino detectors will be IceCube, perhaps in an upgraded form, and KM3NeT. The combination of \textsl{Athena} X-ray observations with neutrino and gamma-ray detections will lead to better disentanglement of non-thermal leptonic and hadronic populations. Moreover, the much better spatial resolution in X-rays and event statistics will help to accurately pinpoint the acceleration sites, as well as characterise the overall environment in sources of CRs.

Here we will briefly describe the present and future gamma-ray observatories and neutrino detectors, and discuss specific synergies that we expect from combining \textsl{Athena} observations with observations and detection of gamma rays and neutrinos.

\subsection{IceCube, KM3NeT, CTA and the VHE gamma-ray messenger landscape}\label{subsec:icecube_km3net_cta}

\vspace*{0.5cm}
{\noindent\bf IceCube:}\\
The IceCube Observatory located at the South Pole is an optical Cherenkov telescope sensitive to neutrino emission in the TeV-PeV energy range. The main detector consists of one cubic kilometer of deep ultra-clear glacial ice instrumented by 5,160 optical modules hosting photomultiplier tubes. The sensors are distributed along 86 read-out and support cables (``strings'') between 1.5~km and 2.5~km below the surface~\citep{Aartsen:2016nxy}. 
High-energy neutrinos are detected in IceCube via the Cherenkov light emission of charged particles produced by neutrino interactions in the vicinity of the detector. The most valuable events for the purpose of neutrino astronomy are those producing high-energy muons traversing the detector, allowing for a good angular resolution of less than 0.4~degrees above 100~TeV. These events can follow from deep inelastic scattering of muon neutrinos with nucleons in the ice, where the secondary muon inherits most of the neutrino's moment. All other neutrino interactions are also visible via charged particles created in electromagnetic and/or hadronic cascades. Whereas these cascade events have a poorer angular resolution in IceCube, at about 10-15 degrees, they allow for a good estimator of the initial neutrino energy with a resolution of better than 15\,\%.

The IceCube Collaboration is presently preparing a detector upgrade, that consist of seven additional strings with  advanced instrumentation. The scientific scope of the IceCube Upgrade includes a calibration program that is designed to improve the knowledge of the natural ice medium, making it possible to enhance reconstruction algorithms and control current leading systematic uncertainties. The ability to re-calibrate the existing and future IceCube data is expected to permit typical angular resolutions for high-energy neutrinos of better than 0.3 degrees and that for cascades to better than 5 degrees. The IceCube Upgrade can be considered a first step towards a full next-generation instrument, the IceCube-Gen2 Observatory, that is planned to become fully operational in the early 2030s~\citep{Ackermann:2019ows,Ackermann:2019cxh,Aartsen:2020fgd}. This future facility envisions the combination of different low- and high-energy sub-detectors, including an extended in-ice array with an instrumented volume of up to ten times the size of the present IceCube. This will allow to improve present event rates of astrophysical neutrinos by a factor of 4-10, depending on channel, and angular resolution by factors of approximately 3.

\smallskip
{\noindent\bf KM3NeT:}\\
KM3NeT is the next generation neutrino telescope in the Mediterranean. Composed of two distinct detectors, it will not only contribute to the observation of the HE sky but also to the understanding of neutrino physics. Each detector consists of three-dimensional arrays of optical modules deployed in large volumes of water, deep in the sea. The ARCA (\textit{Astroparticle Research with Cosmics in the Abyss}) telescope, currently under deployment offshore Sicily, is optimized to detect TeV-PeV astrophysical neutrinos. When \textsl{Athena} flies, it will be composed of 230 detection units over an instrumented volume of about 1 cubic kilometer. The good angular resolution provided by sea water properties (<0.1$^\circ$ for muon-track events at 1 PeV, \citealt{Adrian-Martinez:2016a}), together with the high-depth location of ARCA, will allow for a precise and complementary measurement of the diffuse cosmic flux of high-energy neutrinos detected by IceCube. While muon track events are the best detection channel for point source searches, cascade events in KM3NeT/ARCA will allow for an energy resolution of 5\% at energies above 60 TeV which is particularly useful for diffuse flux analyses. Located in the Northern Hemisphere, KM3NeT will have a good sensitivity toward the Galactic Center which is crucial to identify potential Galactic neutrino sources. The geometry of the second detector, ORCA (\textit{Oscillation Research with Cosmics in the Abyss}) will be optimal for the detection of GeV atmospheric neutrinos, whose oscillation patterns through the Earth will allow for a measurement of the neutrino mass hierarchy. Along with its core science case, ORCA will also contribute to the study of astrophysical GeV neutrino sources. Both detectors will have a good sensitivity to MeV neutrinos emitted by the next Galactic core-collapse supernova \citep{Aiello:2021a} and will benefit from a real-time infrastructure allowing for a fast reconstruction of neutrino events of all flavours and an efficient alert sending system to distribute triggers to the astrophysical community \citep{Assal:2021a}. 

\smallskip
{\noindent\bf CTA:}\\
The Cherenkov Telescope Array (CTA) will be the next generation gamma-ray observatory. 
CTA will consist of two observatories one in Northern Hemisphere (La Palma, Spain) and one in the Southern Hemisphere (Paranal, Chile). Construction of CTA
has started on La Palma, and will start in Chile around 2021/22. The site construction and maintenance of CTA will be done by the CTA Observatory, with
headquarters in Bologna (Italy), whereas the instruments, telescopes and also the science program will be delivered by the international CTA Consortium. 
More information on CTA and its science case can be found in respectively \citet{cta_observatory,cta_science}.

CTA will detect gamma rays
from $\sim 20$~GeV to up to above 100~TeV, and it is a Imaging Atmospheric Cherenkov Telescope (IACT) array. IACTs work through the imaging of the Cherenkov light from air showers which are induced by gamma rays entering the atmosphere. These air showers have narrower emission cones than the ones induced by CRs, which is used for separating CRs from photons. The effective detection area of CTA effectively increases by using a large array of telescopes covering a large fraction of the atmosphere. Also, the simultaneous detection of an air shower by multiple telescopes will provide
a better determination of the original arrival direction of the gamma-ray photon through triangulation. Both the Northern and the Southern observatories will consist of a mix of telescopes, small-sized telescopes (SSTs, $\sim 5$~m), medium-sized telescopes (MSTs, $\sim 15$~m) and large-sized telescopes (LSTs, $23$~m), spread on a large 4~km$^2$  area in the South site, and in a more compact (1~km$^2$) area in the Northern site.
The rationale behind having different-sized telescopes is that the highest-energy gamma rays ($\gtrsim 10$~TeV) are rare, but produce bright Cherenkov signals. Hence, one does not need large diameter telescopes
to register the Cherenkov light, but one needs an extensive array in order to cover a large area of the atmosphere. For the lower-energy end of the VHE gamma ray band ($\lesssim 500$~GeV) the opposite holds: there are more gamma-ray photons, but their air showers produce fainter Cherenkov signals, requiring larger telescopes to detect them.

\begin{figure}
\centerline{
\includegraphics[width=0.5\textwidth]{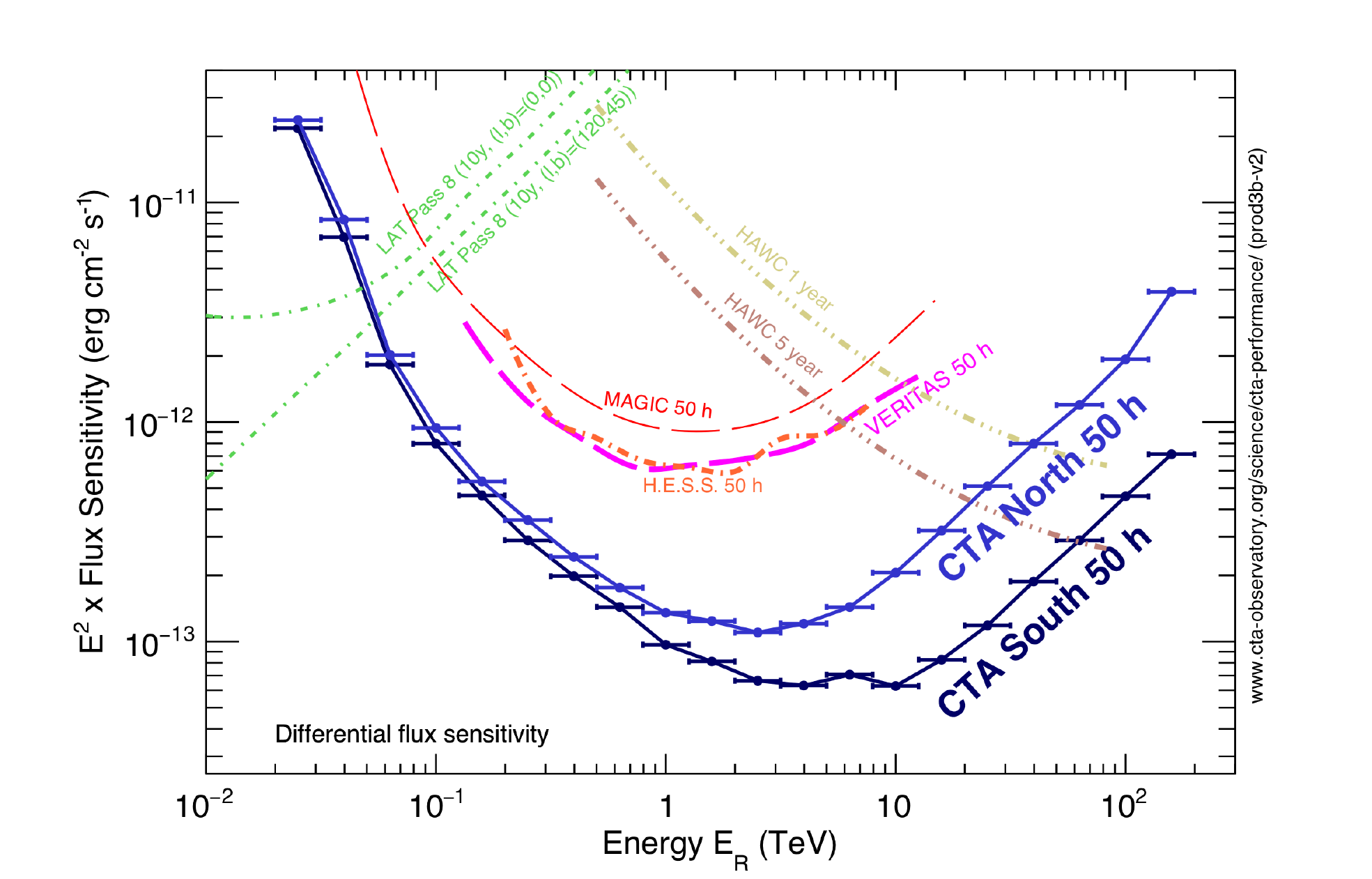}
\includegraphics[width=0.5\textwidth]{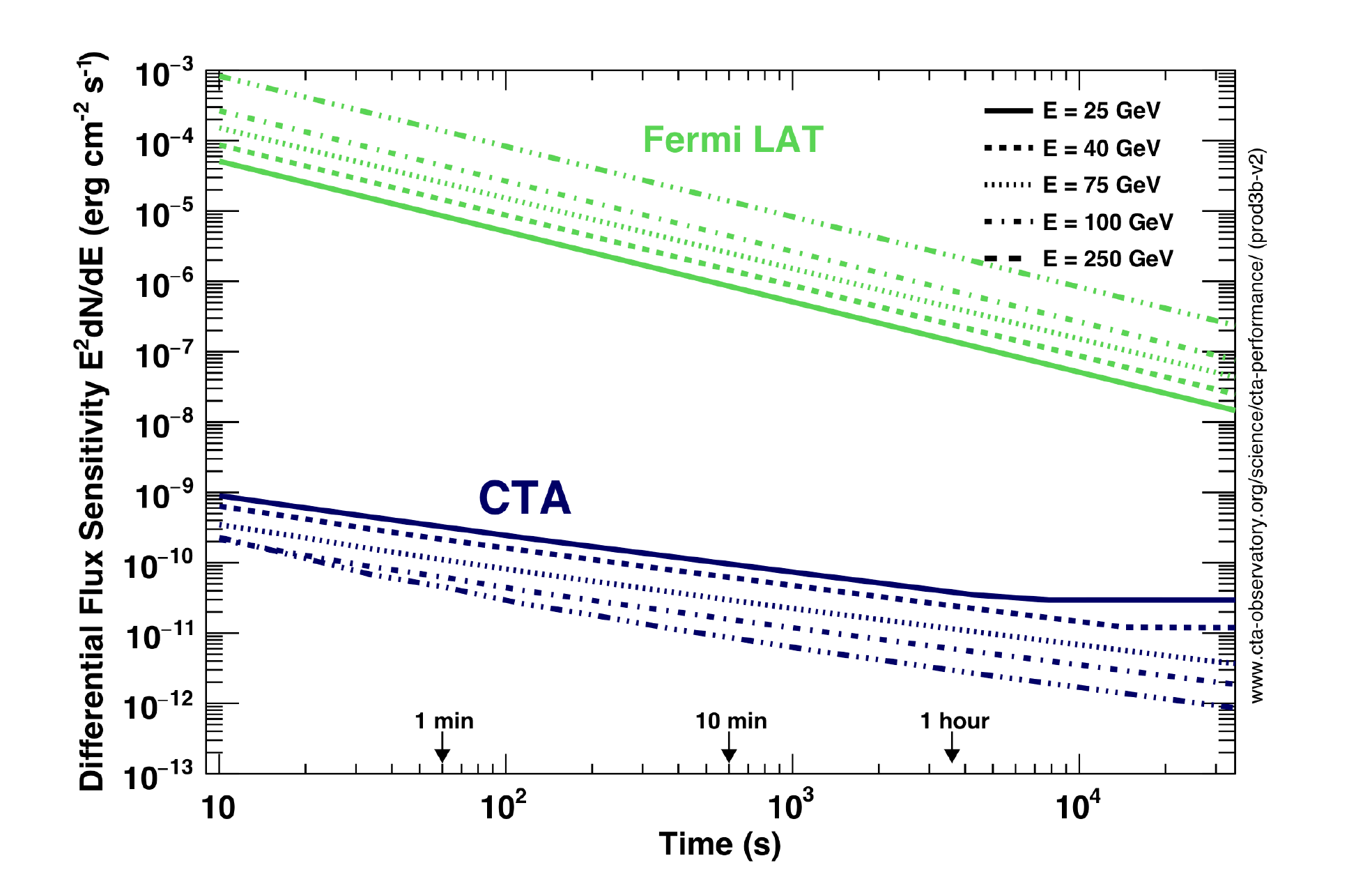}
}
\caption{Sensitivity of CTA compared to other gamma-ray observatories. Left: 
Comparison for persistent sources taking into account that \textsl{Fermi}-LAT and HAWC are wide-area detectors surveying large part of the sky, 
accumulating data over long periods, whereas CTA requires pointed observations. Hence the different integration times. Right:
The sensitivity of CTA compared to \textsl{Fermi}-LAT as a function of integration time, which is relevant for transient sources like GRBs, novae, X-ray binaries,
and neutron star-neutron star mergers.
(Source:~\url{https://www.cta-observatory.org}.)
\label{fig:cta_sensitivity}
}
\end{figure}

The sensitivity of CTA will be an order of magnitude better than that of the current IACTs (H.E.S.S., MAGIC, VERITAS), see Fig~\ref{fig:cta_sensitivity}. In the range of 25-100~GeV CTA compares
favourably with  NASA's \textsl{Fermi}-LAT (angular resolution $< 0.15^\circ$ at $>10$ GeV, ~\citealt{Ackermann:2013wqa}). This holds in particular when one takes into account that \textsl{Fermi}'s sensitivity is the result of the accumulation of data over a long
period of time, being a wide-field survey instrument, whereas CTA is an observatory operating in pointing mode, having a Field of View (FoV) of $\sim 8^\circ$. 
The angular resolution of CTA will be $\approx 7$\arcmin\ at 100~GeV, rapidly improving to $\approx 3$\arcmin\ at 1~TeV, and even being smaller than 2\arcmin\ above 10~TeV (all numbers are 68\% point spread function radii).
The absolute pointing accuracy will be a few arcseconds.
As a result below 100~GeV CTA will provide in 50~hr a  sensitivity that is comparable to the \textsl{Fermi}-LAT 8 year survey, but CTA will be several orders of magnitude
more sensitive than \textsl{Fermi}-LAT for transient sources. 
A similar situation exist for the High Altitude Water Cherenkov (HAWC) telescope, which detects gamma-ray induced showers using water tanks. 
Due to its detection principle, HAWC can operate day and night, and is sensitive to gamma rays arriving from 2$\pi$ steradian directions. But the effective
area is much smaller than CTA, so its sensitivity is similar to CTA only above 10~TeV, and for persistent sources, after accumulating several years of data.

Currently, no new gamma-ray satellite is secured that will cover the gamma-ray energy range targeted by the \textsl{Fermi}-LAT instrument.
This leaves CTA to cover the 25-100~GeV range, and may result in a gap in coverage of the electromagnetic spectrum for gamma rays below 25~GeV.

\subsection{Synergy Science Themes} \label{subsec:}

One of the main scientific drivers for CTA, IceCube, and KM3NeT is the question: "what is the origin of cosmic rays?" Cosmic rays were discovered by Victor Hess in 1911, and since the 1930s it was realised that the mysterious "rays" were in fact highly-energy charged particles ($>10^8$~eV) entering the Earth's atmosphere. The cosmic-ray (CR) spectrum has roughly a power-law distribution with index -2.7 from $10^9$~eV up to $\sim 10^{19}$~eV, but a spectral softening around $3\times 10^{15}$~eV (CR "knee"), and a subsequent hardening around $5\times 10^{18}$~eV (CR "ankle") which is usually explained by the idea that sources in the Milky Way are responsible for accelerating protons up to the knee, whereas particles in excess of the ankle must be of extragalactic origin. The range between the knee and the ankle is a transition between the two components. Apart from the question of what the sources of Galactic and extragalactic CRs are, a related question is how these sources are capable of accelerating particles to very high energies, and what fraction of the energy budget goes to the acceleration of particles. 

For the energy budget it is also important to distinguish the leptonic and hadronic components (see \S~\ref{sec:intro_nonthermal}). Leptons  are radiatively efficient, and produce synchrotron radiation from radio ($\sim$ 10 MHz) to X-ray frequencies ($\sim 10^{19}$~Hz), but the latter only if lepton acceleration is very fast. The same electrons/positrons may also Compton up-scatter background photons, causing non-thermal X-ray to gamma-ray emission. However, the CRs on Earth consist of less than 1\% of leptons. It is, therefore, likely that the non-thermal populations in sources of CRs consist primarily of hadronic CRs. These hadronic CRs do not produce a direct tracer of their existence in traditional astrophysical wavelength bands like radio, optical or X-rays, but instead produce gamma rays and neutrinos.

\begin{figure}
    \centering
    \includegraphics[width=0.9\textwidth]{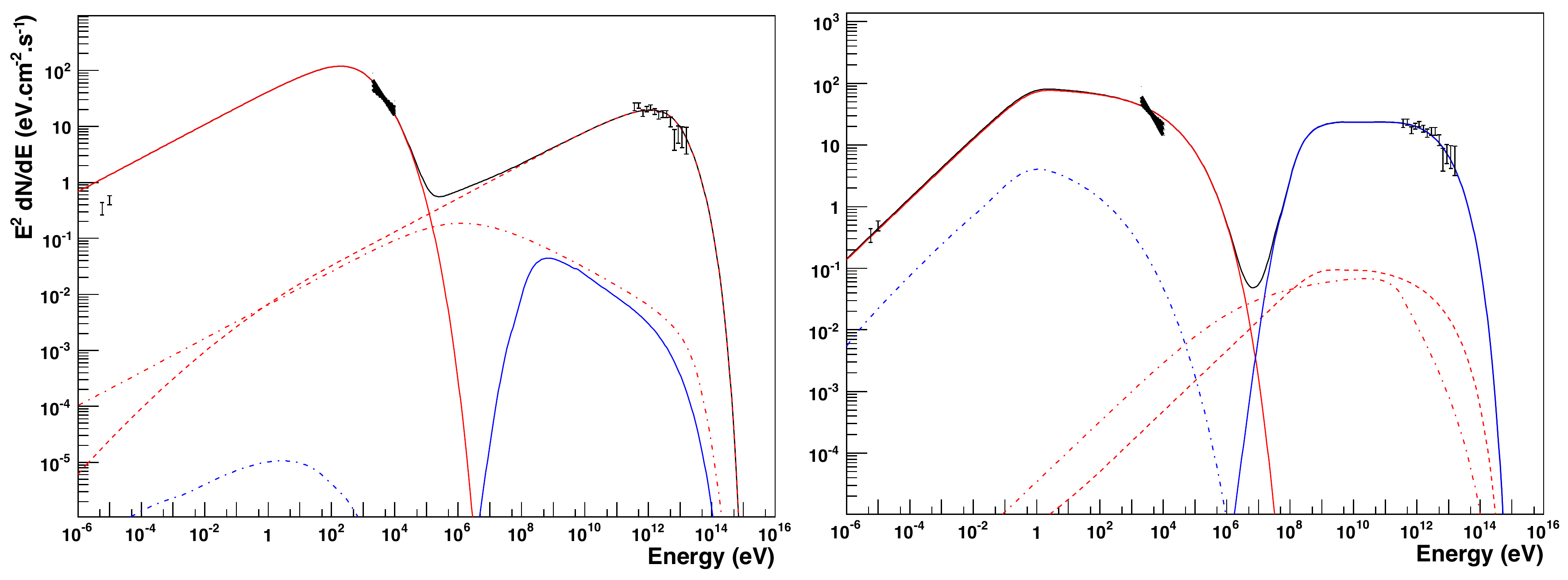}
    \caption{Example of a broad spectral energy distribution displaying the non-thermal radiative out from the SNR RX J0852.0-4622 \citep[taken from][]{HESS2007Velajr}.
    In both panels the emission below 1~MeV is synchrotron radiation, whose brightness depends on the population of electrons/positrons and magnetic-field strength. 
    The  component above 1~MeV (red) is dominated by inverse Compton scattering (red dashed line) from local radiation fields in the left-hand panel and pion decay (blue) in the right-hand panel. 
    Inverse Compton scattering depends on local radiation fields such as cosmic-microwave background, stellar light, or --- in the case of GRBs and AGN --- synchrotron emission from the source itself.
    In the case of blazars and GRBs, relativistic effects should be taken into account. Pion production depends on the local gas density. In the case of SNR shells these can be determined from thermal
    X-ray emission. For blazars it is theorized that there is a pion production from proton-photon collisions. Identifying such a component would prove that ---apart from electrons--- also protons are accelerated.
    }
    \label{fig:broad_sed}
\end{figure}

The process by which they do this is the production of secondary particles caused by the interaction of energetic hadrons with gas 
($pp$) or radiation ($p\gamma$). The most common secondary particles thus produced are neutral or charged pions ($\pi^0$, $\pi^+$, $\pi^-$). These pions subsequently decay via $\pi^+\to\mu^++\nu_\mu$ followed by $\mu^+\to e^++\nu_e+\bar\nu_\mu$ and the charge-conjugated processes, and $\pi^0 \to \gamma + \gamma$. The charged pions therefore result in neutrino emission, whereas the neutral pion results in gamma-ray emission. On average, an individual neutrino or gamma-ray photon from these production channels receives about 5\% and 10\%, respectively, of the energy from the initial CR nucleon. Note that the charged pions also result in an additional population of high-energy electrons/positrons, which may result in an additional X-ray synchrotron component. 
Fig.~\ref{fig:broad_sed} shows the non-thermal spectral-energy distribution of 
a young supernova remnant, with the characteristic two peaked structure, one
in X-rays and one in TeV gamma rays. In this case it is not clear whether
the gamma-ray peak is caused by leptonic or hadronic radiation processes.

The detection of high energy neutrinos can be directly linked to CR hadrons. For gamma-ray emission it is not immediately clear whether the emission is caused by hadronic or leptonic CRs, as leptons also produce gamma-ray emission through inverse Compton scattering, and (less likely)  non-thermal bremsstrahlung. There is one spectral feature around 200~MeV that is a clear signature of hadronic CRs, the so-called pion bump. However, this feature is outside the spectral range of CTA, and has so-far only been identified in a few sources with the \textsl{Fermi}-LAT instrument~\citep{Ackermann:2013wqa}. Further evidence of CR interactions in our Milky Way can be inferred from diffuse gamma-ray emission in the Galactic center region~\citep{Abramowski:2016mir} and extended diffuse emission along the Galactic Plane~\citep{Ackermann:2012pya}.

Identifying X-ray synchrotron emission can help to disentangle hadronic from leptonic gamma-ray emission, because X-ray synchrotron emission is caused by roughly the same energetic leptons ($\gtrsim 10$~TeV) as VHE inverse Compton emission. The X-ray synchrotron brightness depends on the number density of leptons and strength of the local magnetic field, whereas the gamma-ray emission also depends on the number density of leptons, but also on the radiation energy density. The latter can often be estimated, so that there is only one free parameter left to be determined: the strength of the magnetic field. Under the assumption of purely leptonic emission (and a single-emitting zone scenario) one can estimate the magnetic field. If this magnetic field estimate is unphysically low, one can conclude that the gamma-ray emission
is likely not leptonic in nature, and an additional hadronic component is necessary. However, what are plausible magnetic field strengths is open to debate. So the ultimate proof for the presence of contribution by hadronic processes to the gamma-ray spectrum of an astrophysical source is to detect high-energy neutrinos from this source.

A problem for TeV gamma-ray observation is that for sources at cosmological distance the gamma-ray photons interact with extragalactic background light (EBL), in particular with the infrared background originating from the integrated star formation in the Universe. The attenuation of gamma rays is caused by pair formation ($\gamma+\gamma \rightarrow e^+e^-$)
and is more severe for high-energy gamma rays. As a result distant AGN and GRBs 
will appear to have softer gamma-ray spectra than they
intrinsically have. TeV observations of AGN are in fact used to probe the EBL, as the infrared background is difficult
to detect directly, due to infrared emission from the galaxy \citep[e.g.][]{hess07ebl,franceschini08}. A problem with assessing
the effect of the EBL on the gamma-ray spectra is that the intrinsic spectrum is not constrained. However, if the gamma-ray emission is leptonic in origin, the radio to X-ray ratio provides information on the spectral index of the electron population, 
which can be used to infer the intrinsic gamma-ray spectral slope. Ideally this requires simultaneous X-ray and gamma-ray observations as the spectral slopes are often time varying.

In 2013, the IceCube Observatory reported the first evidence of TeV-PeV astrophysical neutrinos in an analysis of two years of data based on the high-energy starting event selection~\citep{Aartsen:2013bka,Aartsen:2013jdh}. First evidence of this astrophysical flux in the up-going muon neutrino sample was found shortly afterwards. With continued observations, both analyses have now reached a significance of more than $5\sigma$~\citep{Aartsen:2014gkd,Aartsen:2016xlq,Kopper:2017zzm,Haack:2017dxi}. The origin of the emission is presently unknown. All analyses are consistent with a power-law extrapolation of the neutrino emission above 1~PeV. Neutrinos at these energies are associated with CR nuclei with an energy of the order of 20~PeV in the case of protons or 1~EeV in the case of iron. These CRs are therefore located in the CR spectrum between the knee and ankle -- the transition region between the dominance of Galactic and extragalactic sources. However, the arrival direction of neutrino events is consistent with an isotropic diffuse emission in the Universe, which indicates that extragalactic sources are the most likely origin; see e.g.~\citet{Ahlers:2018fkn}.

The high relevance of X-ray observations to resolve this puzzle can be seen by the following argument~\citep{Murase:2015xka}. Strong X-ray emission by the region that is also responsible for CR acceleration allows for efficient neutrino production via CR/photon ($p\gamma$) interactions. The energy of secondary neutrinos from pion decay can be estimated by the center-of-mass energy of the $\Delta$-resonance in the shock environment, assumed to move at a bulk Lorentz factor $\Gamma$. The secondary neutrino energy $E_\nu$ and target photon energy $\varepsilon_t$ observed in the observer's frame are then related as $\varepsilon_t \simeq 6~{\rm keV}(\Gamma/10)^2/(E_\nu/100~{\rm TeV})$. In other words: strong X-ray sources are excellent candidate sources for the neutrino emission observed in IceCube. At the same time, the strong X-ray background would allow to absorb gamma rays produced in hadronic emission ($\gamma\gamma  \rightarrow e^+e^-$ interactions), making these sources less prominent in GeV--TeV gamma-ray observatories~\citep{Murase:2015xka}.
The minor contribution  to the diffuse neutrino emission observed by IceCube by  gamma-ray blazars \citep{Aartsen:2016lir} is another motivation for gamma-opaque AGN as neutrino sources.

\section{ Synergy Science Topics}
Many classes of X-ray sources are also known, or suspected to be, gamma-ray and/or neutrino sources.
When it comes to science synergies between \textsl{Athena} on the one hand and CTA, IceCube and KM3NeT on the other hand,
it is worth remarking that one should distinguish between direct synergies, in which simultaneous multiwavelength/multi-messenger
observations are needed to address the science questions, and more indirect synergies. 
Clearly for transient sources such as gamma-ray bursts and highly variable blazars simultaneous observations are needed.
For persistent sources such as SNRs, and PWNe the observations do not need to be simultaneous.
But even on these targets one could distinguish between direct links (for example the need to assess the broad non-thermal
spectral energy distribution to understand the relativistic particle populations), and more indirect
synergies, for example the fact that X-ray observations allow to better understand the source energetics or the local
densities, which are needed to correctly interpret the particle acceleration properties of a source.
In the particular case discussed below, that of large-scale cluster formation shocks, \textsl{Athena} may be able to prove their existence and determine
their properties, but the expected gamma-ray and neutrino emission may be too low to detect from individual sources,
and one has to resort to stacking gamma-ray and neutrino data, or merely use the X-ray observations to establish
these shocks as contributing to the gamma-ray and neutrino background.

Below we will discuss in detail various sources classes and their X-ray/gamma-ray/neutrino synergies.

\subsection{Topic 1: Galactic sources: supernova remnants, pulsar wind nebulae, pulsars and star forming regions} \label{subsec:VHE_galactic}

Most of the CRs observed at Earth are of Galactic origin. Given the typically energy density of CRs in the Milky Way (about 1 eV/cm$^3$), the volume of the Galactic CR halo, and the typically residence time of CRs in this halo ($10^7$~yr, as modeled from CR compositions) show that the total injection rate of energy into CRs in the Milky Way has to be about $10^{41}$ erg s$^{-1}$. This energy rate is about 10\% of the total power that supernovae produce in a typical galaxy like our Milky Way. Supernovae have, therefore, been considered the primary energy source of CRs in the Milky Way, and the evidence for non-thermal emission from radio to gamma rays from SNRs suggest that these objects are indeed the primary sources of galactic CRs. However, there is no conclusive  evidence yet that SNRs regularly convert 5-10\% of the explosion energy into CRs, nor is there evidence that SNRs can accelerate protons up to the CR ``knee'' at $3\times 10^{15}$~erg. In fact, a prime candidate CR accelerator like Cas A seems to fall short of both the energy content in CRs, as well as in the maximum energy particles are accelerated to, and so does Tycho's SNR (SN1572).

In order to estimate the energy content of CRs in a SNR one needs to measure the gamma-ray luminosity to determine its nature (leptonic or hadronic), and model accurately the density and composition in the SNR shell. Hopefully in the future also neutrino measurements will be included, as this will establish clearly the hadronic nature of the accelerated particles.

For the conversion of gamma-ray and neutrino flux to CR energy in the SNR, X-ray measurements, as obtained by \textsl{Athena}, are crucial: accurate modeling of the thermal emission will provide the density and composition of the SNR as a function of location, important for the hadronic interaction model, needed for converting gamma-ray fluxes into total hadronic CR energies. In this context one also has to mention those SNRs or parts of SNRs where the X-ray emission is dominated by X-ray synchrotron radiation, indicating fast acceleration of electrons. But in these cases it has been difficult to constrain the hadronic contribution to the gamma-ray emission, as the thermal X-ray emission is (nearly) absent precluding a good density estimate of the plasma. With the high-spectral resolution measurements possible with the \textsl{Athena}/X-IFU it will be easier to pick out the line emission above the strong non-thermal continuum. Prime examples of SNRs with dominating X-ray synchrotron emission are RX~J1713.7-3946 \citep{slane99,katsuda15b} and Vela Jr \citep{slane01a}, but also for some regions in RCW~86, SN1006 and Tycho's SNR the X-ray synchrotron emission appear to dominate over thermal emission near the shock front in some locations. For Tycho and SN1006  a complete mapping of the objects is expected to be part of the \textsl{Athena} observing program, and it may well be that the programmed observations will reveal the
faint thermal component from synchrotron dominated regions.
Note that RX~J1713.7-3946 is the brightest TeV gamma-ray source, and the nature of gamma-ray emission (leptonic or hadronic) is still a topic of debate.

The synergy discussed so far could be called a strong synergy, as X-ray observations are needed to interpret the gamma-ray and neutrino signal, although the observations do not need to be simultaneous given the time scale of the evolution of SNRs. However, thermal X-ray emission can also inform us by other means about the CR acceleration properties of SNRs, which will provide information that complements gamma-ray and neutrino signals. In this case one has to use \textsl{Athena}/X-IFU's ability to measure line broadening. In SNRs line broadening can be either due to line of sight bulk motion, or due to thermal Doppler broadening. The latter will dominate at the edges of SNRs. Thermal Doppler broadening provides a way to measure the ion temperature. Traditional CCD spectra can only measure electron temperatures, but it is well known that in collisionless shocks the electron and ion temperatures may be very different (\cite{vink03b,ghavamian07,vink15}). If that is the case, the ion temperature dominates the energy budget of the thermal plasma.

Only with high-resolution spectroscopy and imaging can the ion temperature be measured and only \textsl{Athena} can do that sufficiently close to the edge of the remnant. The \textsl{Athena}/X-IFU provides the spatial resolution to isolate regions close to the SNR edge, as well as a spectral resolution sufficient to measure line broadening of a few hundreds of km/s and separate thermal line broadening from bulk motions. The link with CR acceleration is that the ion temperature is expected to be lower if CR acceleration is more efficient, which is a consequence of energy and momentum conservation across the shock boundary.  It can be shown that for the overall post-shock temperature we have (e.g. \cite{vink10}) $\langle kT\rangle = (1- w){\chi}^{-1}(1 - {\chi}^{-1}) \langle \mu\rangle m_\mathrm{p}V_\mathrm{sh}^2,$ 
with $\chi\gtrsim 4$ the shock compression ratio (which may be larger than 4 for very efficiently accelerating shocks) and $w=P_\mathrm{cr}/P_\mathrm{tot}$, the ratio of postshock CR pressure to the total pressure, a measure for CR acceleration energy. 
$<\mu>$ is the average particle mass in units of the proton mass $m_\mathrm{p}$.
This means that by accurately measuring ion temperatures we can measure or set stringent limits on the CR acceleration efficiency of shocks, in case the shock velocity has been measured through proper motions (as has already been done for many young SNRs, e.g.\cite{vink08b,patnaude09,carlton11,williams16,yamaguchi16}). 


As discussed there is overwhelming evidence that SNRs accelerate particles (at least electrons) up to several TeV from both gamma-ray emission and X-ray synchrotron radiation, but no evidence that protons can be accelerated beyond $10^{15}$~eV. For that reason several other sources have been proposed. 

The first source class is superbubbles, hot regions created by the multiples winds and supernovae in a star forming region. Indeed, the superbubble 30 Doradus C in the Large Magellanic Cloud is a TeV gamma-ray source, although its nature (hadronic or leptonic) is not certain. CTA will obtain a much better spectrum and localisation of the TeV emission, and \textsl{Athena} will be an important instrument in characterising the local plasma properties, important for modeling the role of hadronic gamma-ray emission. 

The other model to accelerate particles beyond the knee is to have a subclass of SNR accelerate particles beyond the knee in a very early stage of the evolution, in the first year to perhaps a few decades after the explosion \citep{marcowith18}. Since Galactic supernovae are rare, this model is more likely to be tested by joint observations of \textsl{Athena} and CTA of nearby supernovae ($<10$~Mpc) in the first year after the explosion.  Key to this scenario is that the supernovae are interacting with the dense ambient medium of a dense stellar wind created by the progenitor star. Typically these supernovae emit radio synchrotron radiation and X-ray emission early on
\citep[e.g.][]{chandra18}, and are often associated with wind-stripped  progenitors, giving rise to Type IIb supernovae. A prime example was SN1993J.  Again in this case the \textsl{Athena}/X-IFU will be crucial to measure the dynamics of the supernova, and measure the densities and (electron) temperatures, which are needed to convert gamma-ray fluxes into a CR energy budget. 
Supernovae are Target of Opportunity (ToO) sources both for CTA and \textsl{Athena}. The gamma-ray flux is expected to decline with time,
but near maximum light the gamma-ray flux is attenuated by gamma-gamma interactions of supernova photosphere emission
with the gamma-ray photons. So typically gamma-ray observations will start a few days after maximum light, to typically
a year after explosion. Regular X-ray and radio observations are needed to probe the interaction of the supernova blast
wave with the wind. In this case coordinated CTA/\textsl{Athena} observations are preferred. Note that current generation IACTs did already observe supernovae a few days to a year after outburst, 
but so far results have only yielded flux upper-limits, which 
could be translated in upper-limits on the progenitor wind-loss parameters \citep{hess19sne}.
 For the evolution of particle acceleration and thermal X-ray emission from a supernova into the SNR stage we must also mention SN~1987A, whose complex thermal X-ray emission has been closely monitored over the last 33 years, and whose gamma-ray detection over the next five to ten years is eagerly anticipated.

Apart from supernovae and their remnants, several other Galactic sources are known to accelerate particles. Pulsars and PWNe are TeV emitters \citep[e.g.][]{Ansoldi16,HESS18} and are thought to predominantly accelerate electrons/positrons, and may be responsible for locally measured electron/positron CRs~\citep{Aguilar:2014mma}.
 Indeed the \textsl{Athena} science program expects to include a small sample of PWNe to provide insight about transport and particle acceleration and the magnetization of ultra-relativistic plasmas.


\subsection{Topic 2: Jets and outflows from active galactic nuclei} \label{subsec:VHE_jets}

Recently, the IceCube observatory found first evidence of neutrino emission from a gamma-ray blazar~\citep{IceCube:2018dnn,IceCube:2018cha}. This observation was triggered by a high-energy neutrino event observed on September 22, 2017 that was found to be in spatial coincidence with the known gamma-ray blazar TXS~0506+056 at $z=0.3365$ undergoing a period of enhanced gamma-ray emission, first observed by the 
\textsl{Fermi} satellite and later also by the MAGIC telescopes. The chance correlation of the neutrino alert with the gamma-ray source is at the level of $3\sigma$~\citet{IceCube:2018dnn}.

The blazar's spectral energy distribution at the time of the outburst can be well modeled by lepto-hadronic and proton-synchrotron models using one-zone models~\citep{Gao:2018mnu,Keivani:2018rnh,Cerruti:2018tmc}. It was pointed out that these models predict a low neutrino flux ($\ll 1$~event) for the 2017 outburst that is limited by the observed level of the X-ray emission~\citep{Gao:2018mnu,Keivani:2018rnh}, unless one considers enhanced pion production efficiencies in alternative models~\citep{Ahnen:2018mvi,Righi:2018xjr,Murase:2018iyl,Liu:2018utd}. In 
this respect, \textsl{Athena} observations nearly simultaneous with neutrino events can be very 
useful as the proton, and therefore neutrino, luminosity cannot be arbitrarily large because 
synchrotron emission from Bethe-Heitler pairs \citep[e.g.][]{Petropoulou:2015} would otherwise 
overshoot the X-ray data \citep{Keivani:2018rnh}. 

The low model predictions might not be surprising. First, the emission of TXS~0506+056 has to be discussed in the larger context of the blazar population as argued by \cite{Strotjohann:2018ufz}. The collection of blazar flares might individually only contribute a low neutrino flux, but their sum can still provide observable event numbers; which flare is eventually responsible for the observed neutrino 
becomes then simply a matter of chance. Second, the issue could be simply solved if the photon and the 
neutrino emissions originate from different regions of the jet, as in multi-zone models. 

Motivated by the correlation seen by \textsl{Fermi}, IceCube investigated archival data for evidence of past neutrino emission of TXS~0506+056. This analysis revealed that during a period from September 2014 to March 2015 the source showed a prolonged outburst with an estimated $13\pm5$ neutrino candidates. A chance correlation of this type of neutrino outburst can be excluded at a confidence level of $3.5\sigma$~\citep{IceCube:2018cha}. Together with the earlier observation, this provides compelling evidence that this blazar is a source of high-energy neutrinos.

The 2014/15 outburst of neutrinos is however not accompanied by a flare in the archival \textsl{Fermi} data. The corresponding neutrino luminosity is about four times larger than the gamma-ray luminosity in the quiescent state. \citet{Padovani:2018acg} have shown that the \textsl{Fermi} spectrum during the time of the neutrino outburst was at its hardest, which could indicate an unobserved outburst of TeV gamma-rays (see also \citet{Aartsen:2019gxs}). In any case, the different photon and neutrino emissions in the 2014/15 and 2017 observations are presently the greatest challenge in finding a unified model of the neutrino emission mechanism.
Are blazars then responsible for the whole diffuse TeV-PeV neutrino emission? This question was already addressed in an earlier study by~\citet{Aartsen:2016lir} looking for the combined time-integrated emission of blazars observed in gamma-rays. This study could not find evidence for neutrino emission and placed an upper limit on the relative contribution of blazars to the diffuse neutrino flux. For an $E^{-2.5}$ spectrum this upper limit is at the level of 27\%. In light of the recent results, it is therefore plausible that blazars make a contribution, but it is unlikely that they dominate the observed diffuse neutrino flux \citep[see also][for a similar but independently derived result]{Padovani:2015}.

Many AGN show evidence for large-scale outflows of matter driven by the central black hole 
\citep[e.g.][]{Harrison:2017}. These can reach semi-relativistic speeds of up to $\sim$ 50,000 km 
s$^{-1}$ that can drive a shock that accelerates and sweeps up matter \citep[e.g.][]{King:2015}.
The protons accelerated by these shocks can generate low-level gamma-ray  
and neutrino emission via collisions with protons in the interstellar medium 
\citep[e.g.][]{Lamastra:2017,Liu:2018}.  The former has been
detected very recently in galaxies with ultra-fast outflows (UFOs) by \citet{2021arXiv210511469T}.
The latter is expected to be much softer than the neutrino
emission associated with blazars and would therefore be complementary to it by being more relevant at 
lower energies ($\lesssim 1$ PeV). \citet{Padovani:2018_2} have recently tested the outflow 
hypothesis against IceCube data, showing results consistent with a scenario where AGN outflows 
might be neutrino emitters, but at present do not provide a significant contribution, implying 
that only part of the IceCube signal might be due to these sources. 

The synergy with \textsl{Athena} here is straightforward, as one will be able to study the 
neutrino/gamma-ray emission of the AGN outflows that will be well characterised in the 
X-ray band. \citet{Padovani:2017} have discussed the synergy between \textsl{Athena} and ESO/ALMA
facilities for the study of molecular outflows in terms of nuclear winds, shocks, extended 
hot/warm haloes and the circumgalactic medium, and the effects of massive outflows on the heating of 
the first groups and clusters. UFOs, in particular \citep{Cappi:2013}, which are 
observed in the Fe K band through blue-shifted Fe XXV and Fe XXVI absorption lines, seem to be
quite common, being detected in about half of local radio-quiet Seyfert galaxies and quasars, and in 
powerful radio galaxies as well. The unprecedented sensitivity combined with the spatial/spectral 
resolution of \textsl{Athena} will allow us to study in detail UFOs in both local and high-z AGN
and conclusively constrain the location, geometry, and energetics of such outflows with only a few 
percent uncertainties. One will then be able to predict the neutrino/gamma-ray emission associated 
with the UFO and compare it with observations. 

\subsection{Topic 3: gamma-ray bursts}\label{subsec:GRBs}
GRBs are among the most powerful explosions in the Universe \citep[][for a review]{gehrels09}. 
They come in two flavours:
1) long-duration GRBs (LGBRs), which last more than 2 seconds, and 2) short duration GRBs (SGRBs). LGRBs are associated with powerful supernovae (hypernovae) of Type Ib/c, which have
progenitors with hydrogen-stripped envelopes. Their brightness is caused by emission from jets that originate from
the central engine of the explosion, which can be either a rapidly spinning stellar-mass black hole (collapsar model)
or a rapidly spinning, highly magnetic neutron star (magnetar model). The SGRBs were long hypothesised
to be the result of the merger of two neutron stars, which was brilliantly  proven to be the case with the association of the
LIGO/VIRGO gravitational wave event GW170817 and the multiwavelength detection of the aftermath of the explosion
(see Chapter 2).

LGRBs are among the candidate sources for extragalactic CRs. Their prompt gamma-ray emission is
thought to be caused by shocks internal to the jet, with X-ray emission caused by synchrotron emission from
relativistic electrons/positrons. The X-ray afterglow emission, which can be visible for several days is likely caused by the
interaction of the jets with the circumstellar/interstellar medium (CSM/ISM). 
LGRBs were long suspected to also emit VHE gamma-ray emission, but only
recently was this discovered to be the case: MAGIC detected for the first time a LGRB on January 14, 2019,
GRB~190114C at $z=0.42$, 57 seconds to 15912 seconds after the detection by \textsl{Swift}-BAT, and for an energy
range of 0.2-1 TeV \citep{magic2019grb}.
The gamma-ray light curve suggests that the emission is coming from the afterglow, i.e. interaction with the 
ISM/CSM rather than from internal shocks. The high energy of the photons is incompatible with Lorentz-boosted
synchrotron radiation, and is likely caused by synchrotron-self-Compton scattering. This discovery was soon afterward
succeeded by the detection of TeV gamma rays from GRB~180720B ($z=0.65$) ten hours after the outburst \citep{hess2019grb}.
These detections were made possible by improved instrument performances and scheduling conditions. With these discoveries
it is clear that GRB science will be an important component of CTA observations. Moreover, the gamma-ray detection will be important
input for KM3NeT and IceCube to correlate neutrino detections. In particular, the timing of neutrinos will profit from simultaneous
TeV gamma-ray monitoring. This is needed to establish LGRBs as neutrino sources, which in turn would imply that LGRBs are
sources of hadronic CRs.
CTA observations are important to probe the particle acceleration in the jets, in particular by measuring cut-offs in the gamma-ray spectra, which are directly
connected to the maximum energy of the accelerated particles.
The combination of studying non-thermal X-ray emission with \textsl{Athena},
together with gamma-ray emission of CTA can be used to constrain the close environment of the GRB and the key parameters of the relativistic shock, including the magnetic fields in the jet.
 The
X-ray spectra of the GRBs can be combined with the gamma-ray spectra in order to  constrain the emission mechanism(s),
including hadronic emission (ideally in combination with neutrinos) and
to correct for the effects of the extragalactic background light on the gamma-ray spectrum. Note that detection of hadronic emission, or limits on it, are important not only for establishing LGRBs as CR sources, but also to constrain the energy budgets of LGRBs,
which is important to better understand their central engines.
In this regard neutrino observation are particularly important. Present limits on neutrino emission from GRBs already set tight constraints on the hadronic emission, questioning the role of GRBs as main origin of extragalactic CRs \citep{Halzen16} and providing insight on the nature of the constituents (hadronic vs leptonic) of relativistic jets in GRBs \citep{Guetta15}.   
For the SGRBs, multiwavelength observations are important to better understand the aftermath of neutron-star merger events,  their
role in the chemical evolution of galaxies through r-process element nucleosynthesis, and the geometry and energetics of the jets.

\subsection{Topic 4: Other potential extragalactic gamma-ray and neutrinos sources (starburst galaxies and cluster formation shocks)}
\label{subsec:VHE_cluster}

Although AGN and GRBs are most often discussed as sources of ultra-high energy CRs, among the viable alternative sources 
one finds two source classes that are of direct interest to the core science cases of \textsl{Athena}:
starburst galaxies and large scale cluster shocks.

In starburst galaxies the enhanced star formation rate results in increased level of CRs acceleration, due to the high supernova rate and input from stellar winds.
The CRs, but also the enhanced input of kinetic energy from supernovae and stellar winds results in a hotter interstellar medium (producing more thermal
X-ray emission) and the  increased level of interstellar turbulence and CR streaming will amplify the magnetic fields.  The increased magnetic field in turn
will confine the cosmic particles for much longer times. This turbulence may result in further acceleration of the CRs to higher energies, whereas
the longer CR confinement time will enhance the radiative output of CRs: instead of escaping into intergalactic space, CRs may lose their
energy while still in the dense parts of the interstellar medium. It has, therefore, been suggested that starburst galaxies are CR calorimeters, i.e.
the total gamma-ray and neutrino output from starburst galaxies may approach the total energy output in CRs \citep{wang18}.

On the other hand, CRs may also drive galactic winds from starburst galaxies \citep{breitschwerdt91}.  
The reason is that the adiabatic index of the relativistic CRs is $\gamma=4/3$, whereas
for the hot gas  it is $\gamma=5/3$. At the base the wind may be driven by the hot plasma, but as the wind is expanding, adiabatic (and in addition non-adiabatic, i.e radiative) cooling affects the 
hot thermal gas much more than the CRs. The CRs interacting with the gas in the galaxy will result in gamma rays \citep[as already verified for a few galaxies by
current gamma-ray telescopes][]{veritas09_starburst,hess09_starburst} and VHE neutrino emission.
The resulting outflows and the thermal X-ray emission from the hot plasma in the galaxy and outflows will be studied in detail with the \textsl{Athena}/X-IFU.

Clusters of galaxies are the largest gravitationally bound structures in the Universe. Their gravity is dominated by their dark matter content ($\sim 85$\%), whereas
most of the baryonic matter is in the hot, X-ray emitting plasma ($\sim 13$\%), with only a few percent of the baryonic matter being contained in stars \citep[e.g.][]{voit05}.
The hot plasma is roughly in hydrostatic equilibrium, i.e. the hot gas provides enough pressure support to prevent further collapse of the plasma. However,
the hot plasma must have obtained its entropy from a non-adiabatic heating mechanism. Large scale N-body simulations indicate that this entropy was created by the shocks
produced as gas accretes onto the cluster. These shocks have Mach numbers of 10 to 1000 \citep{miniati00}. 
Although these shocks are easily identified in simulations, the locations of these shocks is outside the virial radius of the clusters,
whereas the density is low, and the X-ray emission (scaling with $n^2$) is too faint to yet have been detected. But \textsl{Athena} may have the sensitivity to detect these
accretion shocks. According to simulations these shocks have a high Mach number and may be able to accelerate CRs to energies above $10^{19}$~eV,
provided the magnetic fields are of the order of 1~$\mu$G \citep{kang96}.
The interaction of these CRs with plasma in and around the cluster could contribute to the VHE neutrino background.

The synergy between IceCube/KM3NeT concerning accretion may not so much be direct, but involves a few steps: accretion shocks first need
to be detected in X-rays and the properties of the gas needs to be determined from X-ray imaging as well as spectroscopy. This will establish the existence for accretion shocks
and determine whether their Mach numbers are high enough for particle acceleration. This may lead to establish accretion shocks as viable sources of ultra-high energy CRs.
The neutrinos detected from these shocks are likely too few to link them to individual shocks, but the contribution of these shocks to the VHE neutrino background can be better
constrained, and perhaps the neutrino directions can be statistically compared to the distribution of clusters in the sky.
Note that the accretion shock physics is tied to one of the core science cases of \textsl{Athena}, namely the characterisation of the warm-hot intergalactic medium (WHIM). The reason is that the WHIM consists
of filamentary structures, and clusters of galaxies are located where filaments are connected. The accretion shocks are likely not spherically symmetric but located
where the filaments intersect the clusters of galaxies.

The situation will perhaps be better for directly linking clusters of galaxies to gamma-ray emission. First of all,  the gamma-ray emission from clusters of galaxies
is thought to originate from CRs accelerated during cluster formation plus CRs injected by the jets of giant radio galaxies.  
We know that, at least in some clusters, there are relativistic electrons, which are producing the detected radio synchrotron emission from merger shocks and the radio haloes.
The lack of a gamma-ray counterparts to these structures puts a lower limit on the intracluster magnetic fields of 0.15~$\mu$G \citep{ackermann10}.
The actual magnetic fields may well be larger than this limit. Faraday rotation measurements of polarisation in the radio suggest magnetic fields of 1--5~$\mu$G \citep[e.g.][and references therein]{bonafede20}.
Since electrons with energies in excess of 1~TeV lose their energy fast through synchrotron radiation and inverse Compton scattering, TeV gamma rays are likely
to be only produced by hadronic CRs. For individual clusters we therefore do not expect to detect TeV gamma rays from the intracluster medium nor from
around the accretion shocks, where the density is even lower. However, clusters of galaxies will be observed as part of the CTA extragalactic surveys, and may be established
as VHE gamma-ray sources by stacking events from the directions of clusters of galaxies.
The X-ray detection of accretion shocks around clusters of galaxies may help this analysis, as one can make a pre-selection of targets based on the X-ray measured shock properties.

\subsection{Topic 5: Fundamental physics: Violation of Lorentz Invariance}

Despite the success of the standard model of particle physics in organizing the known fundamental particles and their interactions, it fails to explain the origin of neutrino mass which is evident by the effect of neutrino flavor oscillations. This indicates that the neutrino and its interactions are a gateway to physics beyond the standard model (BSM). Atmospheric and astrophysical neutrinos observed by ANTARES, IceCube, or KM3NeT allow to study neutrino properties under conditions that are presently not accessible in laboratory experiments. High-energy astrophysical neutrinos take on a unique role in BSM searches. Their energy surpasses that achievable by neutrino beams in nuclear reactor or accelerator experiments. The large distance that neutrinos are propagating from Galactic or extragalactic sources, allow to probe even tiny BSM effects that could accumulate up to observable levels~\citep{Ackermann:2019cxh}.

One particular effect that can be tested in combination with X-ray observations is the possibility that Lorentz invariance is only an approximated symmetry. Violation of Lorentz invariance can manifest itself by different energy-momentum relations of neutrinos, photons or gravitational waves~\citep{AmelinoCamelia:2003ex, Christian:2004xb, Diaz:2014yva}. This could be observable as time delays between neutrinos~\citep{Diaz:2016xpw, Murase:2019xqi}, photons\ \citep{Longo:1987ub, Wang:2016lne, Wei:2016ygk}, and gravitational waves~\citep{Baret:2011tk} emitted at the same time from transient sources arriving at Earth at different times.

The potential of multi-messenger observations to probe these effects has been illustrated by the analysis of recent coincident observations of photons and gravitational waves from GW170817 and GRB~170817~\citep{Monitor:2017mdv,Sotiriou:2017obf,Mewes:2019dhj} as well as coincident gamma rays and neutrinos from TXS~0506+056~\citep{IceCube:2018dnn,IceCube:2018cha,Ellis:2018ogq,Wei:2018ajw,Boran:2018ypz}. Neutrino telescopes like IceCube and KM3NeT can timestamp events to within a few nanoseconds~\citep{Aartsen:2016nxy}. IceCube's present realtime multi-messenger program has a median alert latency of less than one minute~\citep{Blaufuss:2019fgv}. However, even a latency of the order of 4 hours, implied by \textsl{Athena}'s ToO reaction time, can produce competitive results on beyond the standard model (BSM) physics. For instance, the coincidence of the 200~TeV neutrino alert IC-170817A {\bf IC-170922A} with the gamma-ray blazar TXS~0506+056~\citep{IceCube:2018dnn} allowed to set stringent limits on Lorentz invariance violation in terms of an energy-dependent velocity difference $\Delta v = -E/M_1$~\citep{Ellis:2018ogq}. Even accounting for the prolonged gamma-ray flaring period within about 10 days, the implied limit on the BSM scale, $M_1\geq 3 \times 10^{16}$~GeV, improved previous constraints by many orders of magnitude.

\section{Observational strategy}

\subsection{Multi-messenger surveys and observation of specific point sources}

The \textsl{Athena} science program expects to include several classes of objects that are promising  sites of particle acceleration (SNR, PWNe, colliding winds  and shock interaction in bubbles and supernovae -SNe-, outflows in AGN, clusters of galaxies, cosmological filaments and starburst galaxies). 
A substantial fraction of these objects should thus constitute the basis for a coordinated observing program with CTA and neutrino observatories. 
Furthermore, a synergy program should also include as \textsl{Athena} targets a  selection of new sources discovered by VHE and neutrino telescopes. 

During the first decade of operation, CTA will function in two modes. A substantial fraction of the observing time in the first years will be dedicated to key science projects led by the consortium, which will ensure that the most important science cases are addressed with a well-defined strategy, and will produce high level datasets (catalogs, skymaps, etc.) which will be made publicly available. The rest of the observing time will be dedicated to guest observers programmes, working through the submission of proposals in response to Announcements of Opportunity or by Director Discretionary Time for what concerns urgent ToOs.
The combination of wide field-of-view and better sensitivity ensures that CTA can deliver surveys 1 or 2 orders of magnitude deeper than existing ones within a few years, once CTA is complete. In particular, the Galactic plane survey will provide a sensitivity between 2 and 4~mCrab that will allow for a deeper investigation of known sources and for the discovery of about 400 new VHE sources including SNRs, PWNe and PeVatrons. In this context, the  \textsl{Athena}/WFI and \textsl{Athena}/X-IFU will be able to provide identification and characterization of the GeV to TeV gamma-ray sources, the majority of which should be bright in X-rays, in order to constrain the nature of diffuse VHE emission and localize particle acceleration as detailed in section \ref{subsec:VHE_galactic}.
On another hand, a dedicated extragalactic survey will cover one fourth of the sky and will be enable to probe new source populations such as extreme blazars (whose inverse Compton peak is located above 100~GeV) as well as investigate further the content of relativistic jets and the galaxy cluster formation shocks. Understanding further these sources will require multi-messenger observing campaigns including both \textsl{Athena} and neutrino telescopes (see sections \ref{subsec:VHE_jets} and \ref{subsec:VHE_cluster}). While these science cases do not require very fast response from \textsl{Athena}, they will certainly require extended observations by \textsl{Athena} to characterize the spectrum and, for  diffuse sources, their spatial and spectral properties, along with multi-instrument observing campaigns that need to be planned and coordinated well in advance.

Despite a low effective area, neutrino telescopes have a large FoV, covering at least 2$\pi$~sr of the sky at any time. Neutrino point sources are searched for by looking for excesses of high-energy (TeV-PeV) neutrino events clustered at one specific position. While no significant persistent neutrino source has been observed so far, evidence of neutrino excesses at almost 3$\sigma$ are currently raising. In particular, the most significant cluster of events observed by IceCube in the Northern sky is spatially compatible with the Seyfert II galaxy NGC~1068. In the forthcoming years, multi-wavelength observations of its spectral energy distribution will be important to explore the possibility that hadronic acceleration takes place either in a relativistic jet or in the corona surrounding the supermassive black hole. By the next decade, with an increase of statistics thanks to both IceCube-Gen2 and KM3NeT, we will certainly enter into an era where significant neutrino point sources will be unveiled and multi-wavelength studies of these sources will become crucial to understand what makes them good candidates for hadronic emissions. \textsl{Athena} will play a major role in this context, in particular with its ability to probe the direct environment of supermassive black holes.
\textsl{Athena} observations will be unique to constrain the spectral shape in the X-ray range that, when combined with the VHE spectrum and neutrino observations , can constrain hadronic processes (Sect.~\ref{subsec:VHE_jets}). The precision required to  determine the spectral index ($\sigma_{\Gamma}/\Gamma\approx 10\%$) can be obtained by \textsl{Athena} with a 10~ks (100~ks) observation for a source as faint as  $F_X\approx $10$^{-14}$ (10$^{-15})$~erg~s$^{-1}$~cm$^{-2}$
(see Sect.~\ref{sec:athena_intro}).
While standard or ToO \textsl{Athena} observations will probe VHE and/or neutrino sources from individual sources, the synergies between \textsl{Athena} and Cherenkov/neutrino telescopes can also work the other way around. In particular, \textsl{Athena} may be able to prove the existence of particle acceleration in specific sources, but the expected gamma-ray and/or neutrino emission may be too faint to be detected from individual sources, and in such a case, a stacking analysis of gamma-ray or neutrino data from a set of sources will be required to detect them. In this context, joint studies of specific sources may benefit from joint CTA/\textsl{Athena} calls for observations would be helpful as e.g. ESO/XMM-\textsl{Newton} calls that are regularly opened.

\subsection{Transient sources and alerts}

By constantly monitoring at least one hemisphere, IceCube and KM3NeT allow for a complementary coverage of the sky with an almost 100\%-duty cycle, and thus are well designed to detect transient neutrino sources. By 2030, both detectors will keep operating extensive programs of nearly real-time multi-wavelength (from radio to gamma rays) follow-up as soon as a high-signal neutrino event is detected. Those multi-messenger programs will exploit the legacy of current real-time frameworks designed for IceCube and ANTARES (Ageron et al., 2012; Aartsen et al., 2017). A similar procedure is being developed for Baikal-GVD \citep{Baikal21}. Most of the triggers  will be sent publicly to the astrophysical community as IceCube already does using GCN notices distributed through AMON (Astrophysical multi-messenger Observatory Network, Smith et al., 2013).

Potential electromagnetic counterparts of high-energy neutrinos will be first detected by large FoV
instruments, but some of them may be too faint to be detected by those instruments. Next generation neutrino observatories will deliver typical angular resolution for high-energy neutrinoS within the WFI FoV, that should be sufficient to trigger
\textsl{Athena} follow-up as ToO observations.  If the candidate is a catalogued source, \textsl{Athena} follow-up observations can use the X-IFU.
Otherwise, a refined position of the source might be obtained first through observations with large FoV X-ray instruments 
such as the Soft X-ray Imager on board THESEUS, discussed here as a case study. 
In the same energy range, if the source is sufficiently bright, its transient behaviour might also be confirmed by large survey archival data such as the  eROSITA survey.
This would be the case for a blazar like TXS~0506+056, with an X-ray flux $\approx$ 10$^{-12}$~erg~s$^{-1}$~cm$^{-2}$. Such a source would be bright enough for \textsl{Athena} to carry out a monitoring campaign with relatively short exposure (a few ks) and with a precise measurement of the the spectral index. The existing IceCube realtime program sends about 10 public ``gold'' alerts per year, that on average have at least a 50\% chance to be of astrophysical origin~\citep{Blaufuss:2019fgv}.


In parallel, the CTA transient program includes VHE gamma-ray follow-up observations of a wide range of multi-messenger alerts (including SNe, gamma-ray bursts, neutrinos, gravitational-wave and multi-wavelength transient sources). Although the chance probability of catching a prompt GRB during a survey is quite small (between 0.08/year and 2/year depending on the pointing strategy), the possibility of repointing the LSTs in less than 50~s gives good opportunities to detect a large sample of new GRBs. In addition to fast reaction, low energy threshold and high sensitivity (see fig. \ref{fig:cta_sensitivity}) CTA will have the capability to analyze data in near real-time, which makes it suitable for transient source searches. In case of a new transient source detection, CTA will thus be able to trigger an alert which might be followed-up by \textsl{Athena}.
The \textsl{Athena} science program is expected to include several fast ToO observations of GRBs, this allowing to optimize the observing time. 
We expect that some of the GRBs with  VHE emission  should be pretty bright and closeby, as those required for the one of the \textsl{Athena} GRB programs (WHIM). However, in order to track the evolution of the blast wave in both X-rays and VHE, additional follow-up observations at later time  will be required.
A similar consideration applies to SNe. Closeby SNe are fast ToO sources for \textsl{Athena}, but the addition of late follow-up observation in the \textsl{Athena} program is required to the study of the interaction of blast wave with the environment.


\chapterimage{Images/THS_WP_bkg} 
\chapter{ \textsl{Athena} and the transient universe with THESEUS}
\label{chapter:Theseus}
\textit{{\color{purple}{By P O' Brien, L. Amati, E. Bozzo, M. Guainazzi, J. Osborne, L. Piro, G. Stratta, N. Tanvir.}}}

\section{Science Themes}\index{}\label{sec:}
\subsection{\textit{Athena} in the landscape of Transient Universe missions}\label{subsection6}

High energy sources are often transient in nature. Sources may appear once, like a gamma-ray burst (GRB), or can in principle repeat, such as tidal disruption events (TDEs). Other sources, such as X-ray binaries or active galactic nuclei (AGN), continuously vary. \textsl{Athena} has a number of science goals, some included in  the core program,  which rely on follow-up observations of high-energy transients identified by external observatories. These include: probing stars in the early Universe; using GRBs as backlights to probe the warm hot intergalactic medium (the WHIM); and probing galactic and extragalactic variable sources, such as TDEs, AGN and stellar binary systems. In this chapter we discuss about the part of the \textsl{Athena} science program specifically targeting  these sources (Sect.~\ref{dawn}, \ref{WHIM} and \ref{transients}). It should be remarked that \textsl{Athena} science goals related to the Transient Universe  that do not rely on external triggers are not discussed here. 

As discussed in previous chapters, \textsl{Athena} can also contribute significantly to the area of multi-messenger astrophysics. 
Although a significant fraction of the localisations from multi-messenger facilities will be good enough to point {\it Athena} directly, their number can be increased by high energy transient facilities, enabling a pre-selection of promising counterparts. This is discussed in Sect.~\ref{multi}.

To maximise the role of \textsl{Athena}, and indeed to enable it to rapidly observe transients, it is essential that \textsl{Athena} is fed information on the location, brightness, redshift and object class for a newly discovered transient as quickly as possible. Like other large astronomical facilities, this cannot be achieved by \textsl{Athena} itself, but rather 
demands the existence 
of separate facilities to find transients. To make use of transients \textsl{Athena} must in some cases be provided with targets (and redshifts), quickly (within a few hours) while others can be provided on timescales of days – weeks. In the following we discuss the required capabilities of a transient discovery mission viewed from the \textsl{Athena} perspective and describe the panorama of facilities that may be operating when \textsl{Athena} flies. 

Extragalactic transients are sufficiently rare that finding large numbers of bright specimen requires intrinsically large Field of View (FoV) instruments. Example instruments in orbit today include the \textsl{Swift} Burst Alert Telescope -- a coded-mask telescope -- and the \textsl{Fermi} Gamma-Ray Burst Monitor (Fermi GBM) -- a scintillator crystal device. These missions can provide source location and brightness information rapidly, but cannot provide  redshifts directly. Furthermore, they do not carry wide-field soft X-ray survey instrumentation (i.e., in the \textsl{Athena} observing bandpass). \textsl{Fermi} GBM also provides location accuracies too poor for a single pointing by \textsl{Athena}. Both \textsl{Swift} and \textsl{Fermi}  are already mature facilities (launched in 2004 and 2008 respectively), and hence it is difficult to predict if they will still be operational in the \textsl{Athena} era. Future transient discovery missions, due to launch in the next few years, particularly \textsl{SVOM} and \textsl{Einstein Probe}, may still be operational (although this would require extending them significantly beyond their nominal operational life).
In addition to wide field hard X-ray monitors, \textsl{SVOM} has X-ray and optical telescopes on board that, complemented  with  on ground robotic telescopes in the IR specifically developed for the project, will allow SVOM to precisely locate the sources and derive  the redshift for a good fraction of them. 
\textsl{Einstein Probe} has an on-board wide-field soft X-ray instrument based on 
Lobster-eye optics that enables wide field observations with unprecedented high sensitivity, with prompt localizations at the arcminute level (further improved to the arcsecond level by follow-up observations with a narrow field X-ray telescope). Although the mission has no multi-wavelength capability on-board, quick distribution of localizations would allow telescope networks to aid source classification.

Constellations of nanosatellites are a new venue for all-sky high energy transient and GRB monitors \citep[nanosat, e.g.][]{Fiore20,Inceoglu20}. 
Constellations of $\approx50-100$ nanosats equipped with hard-X-ray detectors can provide near–real-time localization of GRBs within an error box of $10~\rm{ arcminute}$-$1^{\circ}$.
There are several nanosat experiments already under study or development for the detection and localization of GRBs, including HERMES, BurstCUBE, EIRSAT-1, CAMELOT, GRBAlpha, but the list is increasing as more teams are planning to follow this venue, taking advantage of the relatively simple technology of the instrumentation, the limited initial investment and the scalability of the program. Most of them are already planning to fly pathfinders (i.e. a constellation of a few nanosats in the next years), with the plan to deploy  full constellations after this first step.
It should then be expected that at least some of these facilities will be operational when \textsl{Athena} flies.
A significant fraction of localizations provided by nanosat constellations for bright GRBs will be already good enough to point directly \textsl{Athena}. Nonetheless, their fast distribution of localizations should enable a preliminary characterization of the counterpart for some of the triggers by other facilities, allowing an optimal pre-selection of targets for {\it Athena} observations. 
In fact, an increasing number of robotic telescopes, including some with diameters as large as 4 meters, is being deployed, boosted by the prospects of the nascent multi-messenger era. This will enable identification of some bright GRB targets for \textsl{Athena}. Identification of high-redshift GRBs is more challenging without a more focussed space facility to find triggers.

Although a combination of different facilities can provide the complementary information needed by \textsl{Athena} to trigger follow-up observations, it would be more efficient to have a single mission alongside \textsl{Athena}, which can find GRBs, determine redshifts on-board, conduct a soft X-ray survey for other transients, and be able to rapidly communicate discoveries to the ground for distribution to other facilities, including \textsl{Athena}.
We have adopted as an  example of such mission \href{https://www.isdc.unige.ch/theseus/}{THESEUS}, a candidate for the ESA M5 slot. \uline{Although THESEUS was eventually not selected, its advanced assessment  allowed us to carry out a detailed study of the synergy with} \textsl{Athena} \uline{, and  provide a reference for future missions based on a similar concept, such as} \textsl{GAMOW} \citep{White21}.
THESEUS expanded on previous missions with similar scientific goals to provide a much more sensitive transient detection capability, carried a 70-cm class IR telescope to enable on-board determination of redshifts in the key 5-12 range, and planned to conduct a wide-field soft X-ray survey capable of finding a wide variety of targets suitable for follow-up by \textsl{Athena}. THESEUS was designed to greatly enhance the detection rate and rapid identification of high-redshift GRBs, could provide locations accurate enough for a single pointing with the \textsl{Athena}/X-IFU and, by observing in the same bandpass as \textsl{Athena}, would have enabled to optimise exposure times for a variety of targets. THESEUS would have also enabled a fast search for high-energy counterparts to multi-messenger sources whose locations could then be provided to \textsl{Athena}. 

\begin{figure}[!t]
\centering
\includegraphics[width=0.8\textwidth]{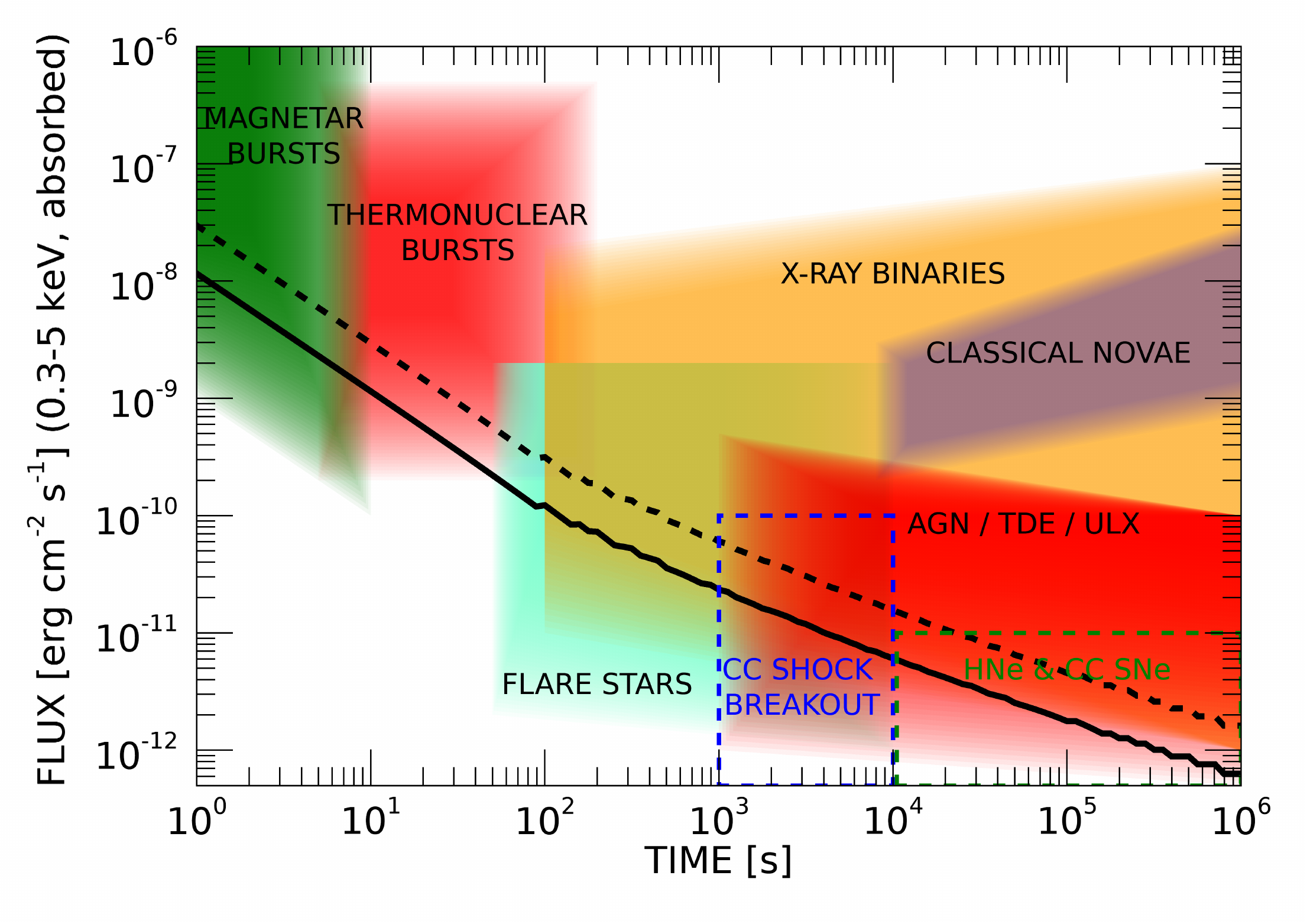}
\caption{An example of THESEUS sensitivity to different classes of celestial objects (figure from \citet{2021arXiv210409533M}). The typical variability time scales of different classes of sources is plotted as a function of their typical soft X-ray flux. These are compared with the SXI sensitivity computed assuming a power law spectrum with photon index 2 and neutral hydrogen column of $N_{\rm H}=5\times10^{20}~cm^{-2}$ (solid black line) and $N_{\rm H}=10^{22}~cm^{-2}$ (dashed black line).}
\label{fig:theseus-sensitivity} 
\end{figure}

THESEUS included two wide-field monitors: (1) the \textsl {X-Gamma-ray Imaging spectrometer (XGIS)}, a coded mask instrument, sensitive from 2~keV - 20~MeV, providing 15 arcminute localisation accuracy over a 2~sr FoV; and (2) the \textsl {Soft X-ray Imager (SXI)}, sensitive from 0.3-5~keV, providing 1--2~arcminute localization accuracy over a 0.5~sr FoV. The \textsl {SXI} was designed to use microchannel plate optics in a so-called 'Lobster-eye' configuration to provide a wide FoV while maintaining X-ray focussing, and hence high sensitivity and localization accuracy. An example of sensitivity to various classes of transients is presented in Fig.~\ref{fig:theseus-sensitivity}. These monitors would feed target location to the spacecraft which then rapidly would slew to place the transient in the FoV of the 0.7~m \textsl {infrared telescope (IRT)} which had both photometric and spectroscopic capability in the 0.7-1.8~$\mu$m band for redshift and arcsecond location determination. The trigger parameters would have been rapidly distributed via a trigger broadcasting unit to the world-wide community of astronomers within a few minutes. While searching for transients, both wide-field monitors would perform a full sky survey.

\subsection{Synergy Science Themes} 

\textsl{Athena} has multiple science objectives that rely on rapid availability of transient targets. Two objectives require the rapid (target sent to \textsl{Athena} within a few hours) identification of bright GRBs: (1) Probe the first generation of stars (Cosmic Dawn), the formation of the first black holes (BHs), the dissemination of the first metals and the primordial initial mass function (IMF). This is achieved by determining the elemental abundances of the medium around high-redshift GRBs. The requirement is to perform X-IFU spectroscopy measurements on 25 $z>7$ GRBs. (2) Measure the local cosmological baryon density in the WHIM to better than 10\%, and constrain structure formation models in the local density regime by measuring the redshift distribution and physical parameters of $\approx$ 100 filaments towards bright GRBs up to $z=1$. 

In addition, the science requirement of \textsl{Athena} dealing with the understanding of BH birth through supernovae (SNe) explosions requires the  identification in the soft X-rays within a few hours of 5 SN shock breakout during the mission lifetime.  
As expected for an observatory class mission, \textsl{Athena} also has multiple science requirements that assume the availability of transients of various types which it can follow-up. Examples include: (a) studying the nature of stellar disruption and subsequent accretion onto super-massive black holes during TDEs, where the requirement is to observe 5 such events. (b)  observing stellar binary systems (BHs, neutron stars and white dwarfs) in both quiescence and outburst to probe the accretion process, where known (or new) examples of such systems are to be observed once it is known they have undergone a state change. Other X-ray transients to be targeted by \textsl{Athena} include Novae, Magnetars, stars and Ultra-Luminous X-ray sources (ULXs). Providing targets to \textsl{Athena} requires an on-going X-ray survey to identify transients as they evolve, either detecting entirely new X-ray sources or strong variability in previously known sources. These goals require continuous monitoring of large areas of sky (such sources are relatively rare) and availability of data on a prompt timescale (days to weeks depending on the detailed object class).

These \textsl{Athena} science requirements are highly synergistic with a mission like THESEUS, which was designed to achieve two overarching primary science objectives: (1) Explore the early Universe and (cosmic dawn and reionization era) by unveiling a complete census of the GRB population in the first billion years. This requires the detection of many hundreds of GRBs per year (at all redshifts) throughout the mission lifetime and the rapid determination of a redshift to enable pre-selection of the highest priority targets for further study by other follow-up facilities. (2) Perform an unprecedented high-cadence deep monitoring of the X-ray transient Universe in order to identify a wide variety of extragalactic and galactic transients. 

THESEUS would detect $\sim 10$ very-bright low-redshift GRBs per year suitable for the \textsl{Athena} WHIM science requirements (those GRBs capable of producing $\sim 1M$ counts in the \textsl{Athena}/X-IFU). The prompt trigger information and subsequent SXI follow-up data can in principle provide information to help select the best targets. 
While the low-redshift GRB discovery rate required by \textsl{Athena} could in principle be satisfied by current missions, if they are still operational, the current generations of transient monitors, primarily \textsl{Swift}, have only led to the identification of about one high redshift ($z>$6) GRB per 3-4 years on average. This is far below the rate required for \textsl{Athena}. Enabling a statistically reliable study of the period of cosmic dawn requires raising the high-redshift GRB detection rate by more than an order of magnitude (and hence requires raising the total discovery rate). THESEUS was designed to identify $\approx 40-50$ high-redshift GRBs in 4 years, which is well matched to the \textsl{Athena} requirement. Finally, most recent simulations have shown that, thanks to the unique combination of FoV, sensitivity and location accuracy of the SXI, at least 2 SNe shock breakout per year should be discovered and identified by THESEUS. This matches well the number of these events that are required for the \textsl{Athena} science requirement on the BH birth through the SNe channel. 

The two monitors on THESEUS were planned to simultaneously observe the sky over an energy range exceeding five orders of magnitude, and to be particularly sensitive to the softest energies down to 0.3~keV, providing a perfect complement to the \textsl{Athena} observing bandpass. Over this band the sky contains a large variety of transient source classes which vary on timescale from seconds to years. THESEUS could provide real-time triggers and accurate locations (a few arcminutes within a few seconds; $\approx 1$~arcsecond within a few minutes for those also detected by the \textsl{THESEUS/IRT}). These data would provide localization and charaterization of various classes of  Galactic and extragalactic transients, for follow-up by \textsl{Athena} or other facilities e.g.: tens of TDEs per year, many magnetars and soft gamma-ray repeaters, SN shock break-outs, Soft X-ray Transients (SFXTS), thermonuclear bursts from accreting neutron stars, Novae, dwarf novae, stellar flares, AGNs and Blazars. 

With regard to the multi-messenger scenario in 2030, a THESEUS-like mission 
could detect binary mergers than will also be detected by third generation gravitational wave detectors like ET, and can search for electromagnetic (EM) counterparts to off-axis mergers detected by such facilities, thanks to the on-board ability to search for early high-energy or infrared emission. Such identifications could drive follow-up observations by \textit{Athena} and other facilities.

\begin{figure}[!t]
\centering
\includegraphics[angle=-90,width=1.0\textwidth]{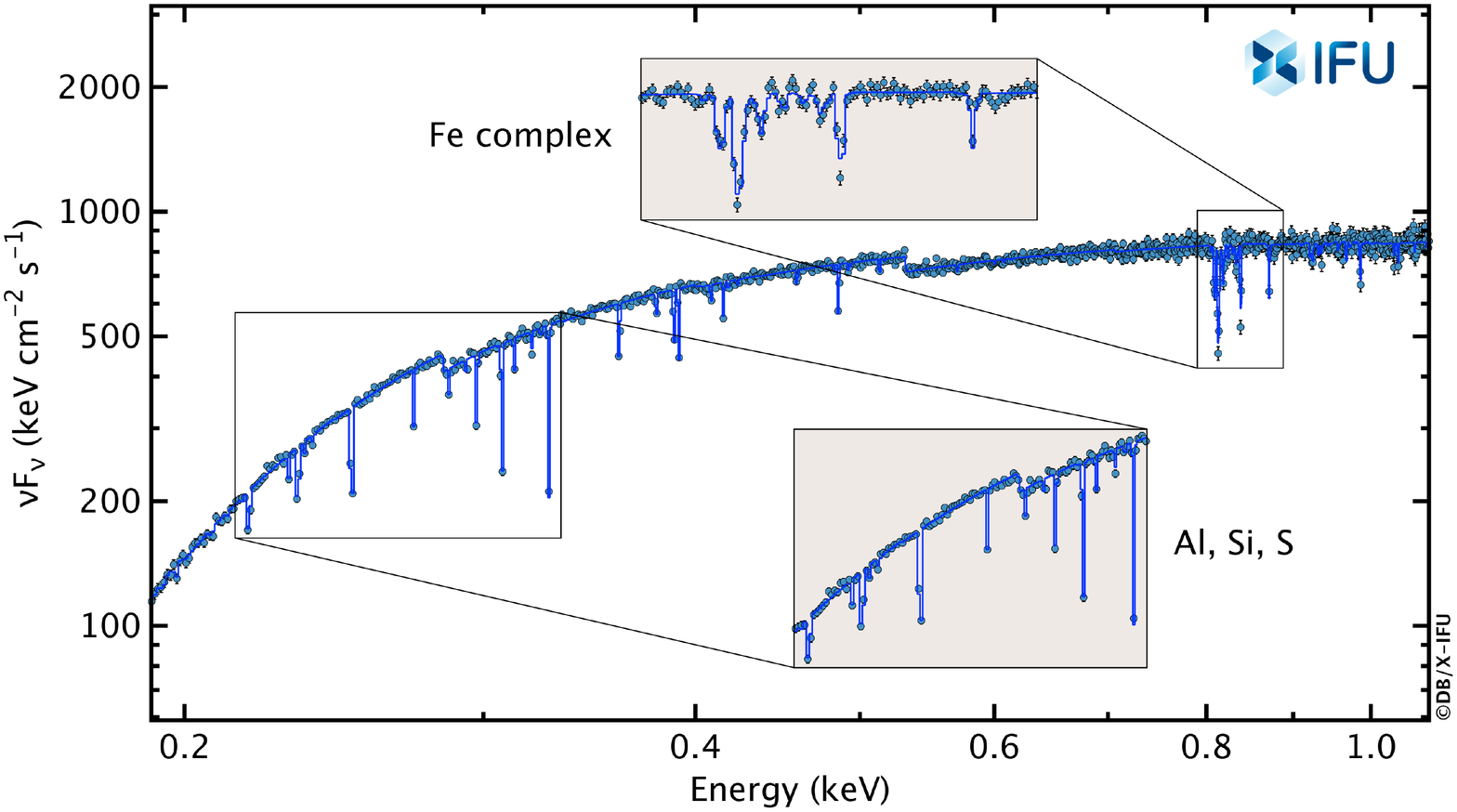}
\caption{Simulated \textsl{Athena}/X-IFU spectrum of a medium bright $z=7$ GRB: fluence $= 4 \times 10^{-7}$ erg cm$^{-2}$, N$_H$ $= 2 \times 10^{22}$ cm$^{-2}$ [Credit: X-IFU Consortium].
}
\label{fig:highz-grb-athena} 
\end{figure}

\section{ Synergy Science Topics}

\subsection{Topic 1: Exploring Cosmic Dawn} 
\label{dawn}
The combination of THESEUS and \textsl{Athena}, together with other multi-wavelength facilities, would enable unprecedented studies of the global star formation history of the Universe up to $z\approx 10$ and possibly beyond.
By locating high-redshift GRBs in numbers far above the rate from current facilities, THESEUS would enable the brightest sources of radiation known to be located rapidly. These GRBs can be used to investigate the re-ionization epoch, the interstellar medium (ISM) and the intergalactic medium (IGM) at all redshifts. Determining the rate of GRBs, where they occur and how much radiation escapes their host galaxies can answer many questions, including: How did re-ionization proceeded as a function of environment? Was radiation from massive stars its primary driver? How did cosmic chemical evolution proceed as a function of time and environment?

The instrument combination on THESEUS could rapidly identify the highest redshift GRBs  from among the many hundreds of others it will detect each year, by exploiting the on-board redshift determination with the IRT (within a matter of few tens of minutes). This information would be communicated down to ground and could be used to re-point \textsl{Athena} at these targets within a few hours. The expected rate of detection would be large enough to allow \textsl{Athena} to build a sample of 25 high-resolution spectra of high-z GRBs.  The newly discovered GRBs would still be bright enough for \textsl{Athena}/X-IFU high-resolution X-ray spectroscopy to study the burst environment.   The medium close to the GRB is expected to be highly ionized, and thus it can be probed only via high-resolution X-ray spectroscopy. A 50~ks exposure using the \textsl{Athena}/X-IFU will typically contain hundreds of thousands of photons from the brightest high-redshift GRBs, allowing to characterize the properties of medium through highly ionized absorption lines (Fig.~\ref{fig:highz-grb-athena}). 
Among this population, \textsl{Athena} may also be able to discover GRBs from the primordial (Pop-III) star population allowing a determination of when the first stars formed, and how the earliest Pop-III and Pop-II stars influencde their environments.

\subsection{Topic 2: Warm Hot Intergalactic Medium} \label{WHIM}

The current census of baryonic matter in the Universe shows a deficit of approximately half --- leading to the so-called ‘missing baryon’ problem. Cosmological hydrodynamic simulations suggest that the missing baryons should be  located  in a hot phase (approximately $10^5$ to $10^7$~K), shock heated following the accretion into dark matter structures.  They could be distributed in a so called warm hot intergalactic medium (WHIM), a gaseous filamentary structure connecting clusters of galaxies, and in a hot Circum-Galactic Medium (CGM) surrounding galaxies or groups. At such temperatures the matter is highly ionized. The only tracers of this important and still elusive baryonic component of the Universe are then highly ionized metals. Such metals are expected to imprint numerous absorption (and emission) lines in high resolution soft X-ray spectra of background quasars and GRBs, analogous to the HI Lyman-$\alpha$ forest seen in the optical spectra of quasars at $z> 2$. The predicted column densities are low, to the extent that the features can only be detected in high signal-to-noise ratio ($S/N$), high resolution spectra \citep{Branchini09}. In the soft X-ray band tentative features have been seen in long exposures with XMM-\textsl{Newton} and \textsl{Chandra} of bright AGN \citep{Nicastro18}. The
unprecedented
effective area of \textsl{Athena}, and its high-resolution X-IFU instrument will revolutionise the study of the WHIM. The associated \textsl{Athena} science requirement is to detect 200 filaments, with about half each coming from observations of bright AGN and bright GRBs. While AGN can only probe relative close systems (due to the paucity of bright enough AGN at high-z), GRBs are particularly important because they provide a backlight at $z \gtrsim 1$, thus allowing to probe also the epoch of the formation of this elusive component.  

THESEUS was designed to
detect large numbers of GRBs per year, including several which will be extremely bright. GRB X-ray spectra are basically of simple power-law form (with some intrinsic absorption) and hence, similar to the use of BL Lacs, provide ideal ‘featureless’ sources for searching for the WHIM.

As with the study of high-redshift GRBs, rapid identification and communication are required to enable \textsl{Athena} to be rapidly re-pointed making maximum use of the precious X-ray light while the GRB is brightest. Taking the brightest GRBs detected by \textsl{Swift} as examples, the \textsl{Athena}/X-IFU could accumulate over a million counts in a 50~ks observation starting within few hours of the GRB trigger for the top 5\% brightest X-ray afterglows. For this sample the \textsl{Athena}/X-IFU  will enable the detection and separation (in velocity space) of multiple  absorbers along a single line of sight, but also for their physical characterization (temperature). \textsl{Athena} will thus allow to routinely detect and characterize the Universe’s missing baryonic mass in the WHIM, and to shed light on the associations between highly ionized metals in the WHIM and the galaxies responsible for their production. This in turn will greatly improve our understanding of the continuous feedback process governing the galaxy-IGM co-evolution throughout cosmic time.

      \subsection{Topic 3: Transient astrophysics} \label{transients}

      THESEUS was designed to detect many hundreds-thousands of transients and variable sources per year arising from the Galaxy and beyond. The monitors were planned to survey many thousands of deg$^2$ per day, repeating such a coverage for several years. This would provide a unique high-cadence survey of the high-energy transient sky. 
      
      Among many extragalactic sources, a THESEUS-like mission can identify flaring AGN and TDEs during its X-ray monitoring survey. X-ray emission from SNe breakout is expected to be detected about twice a year. Some of these may be previously known sources which have entered a bright state or be new transient sources. In either case, THESEUS had both the X-ray capability to identify high-energy transients and the infrared capability to see into galactic centres to confirm nuclear transients. The extraordinary light collecting power of \textsl {Athena} can then be utilized to study the transient properties, including possible X-ray winds from TDEs, massive outflows in AGN, and SNe in close galaxies.

    THESEUS would also be able to localize and identify many Galactic X-ray transients. Thanks to the unprecedented grasp of the THESEUS/SXI, it was expected to discover a large number of still unknown sources and provide for them a preliminary X-ray and IR characterization (beside monitoring the already known recurrent/known transients). Among the most promising classes of objects there are, e.g. the magnetars with their short bursts and the supergiant fast X-ray transients, which are a sub-class of high mass X-ray binaries displaying a peculiarly remarkable variability in X-rays achieving a dynamic range of 10$^5$-10$^6$ on a time scale of few hours. Other classes of Galactic and extragalactic X-ray binaries, as BH candidates and neutron star (NS) thermonuclear bursters, are also clearly at reach for prompt detection with both the SXI and the XGIS. A follow-up of these sources with the high spectroscopic resolution capabilities of \textsl{Athena} will allow the community to get precious insight on any associated accretion/ejection physics, as well as on thermal, non-thermal, and magnetic impulsive emissions. While monitoring with high cadence the relatively long outbursts of, e.g. BH binaries and dwarf novae, 
    a THESEUS-like mission
    would also be able to indicate the best time to carry out deeper \textsl{Athena} follow-up observations to catch the critical stages of timing and/or spectral state transitions of these systems.

        \subsection{Topic 4: Multi-messenger astronomy} \label{multi}

        A THESEUS-like mission will detect the gamma-ray counterpart to on-axis compact binary mergers detected iby the second (2G) and third (3G) generation gravitational wave (GW) detectors 
        as short GRBs for the most distant ones and, for geometrical reasons, may be capable of detecting off-axis jets for the most nearby ($z<$0.1) GRBs. The combination of having a sensitive short GRB detector flying at the same time as the next generation GW detectors are operational will be revolutionary. A synergy between THESEUS and \textsl{Athena} during the 3G  era would enable to build statistical sample on which establish the nature and properties of binary neutron star (BNS) and NS-BH mergers, as well as BNS merger central remnants.

\begin{figure}[!t]
\centering
\includegraphics[width=1.0\textwidth]{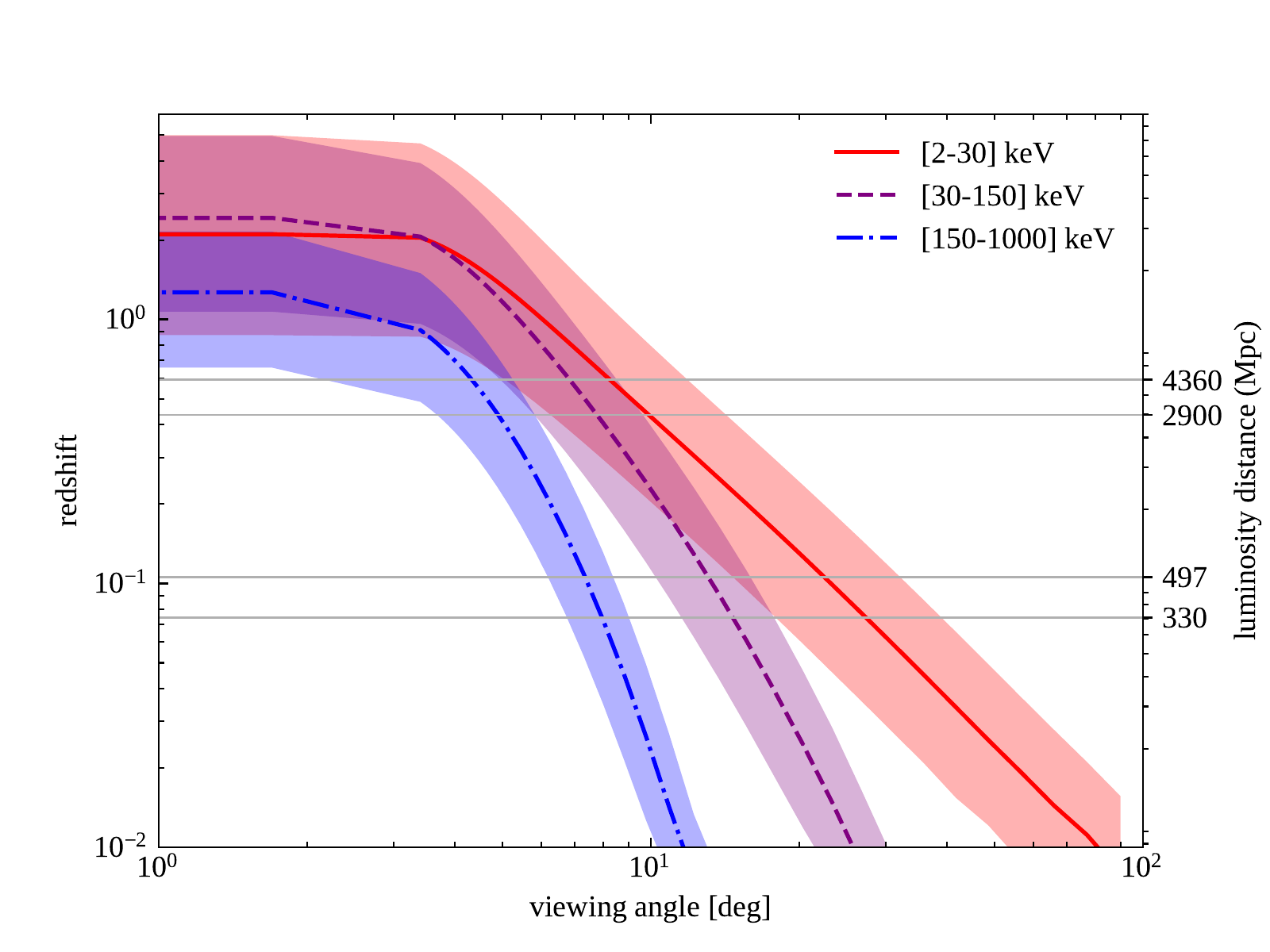}
\caption{Maximum distance/redshift for detecting with THESEUS/XGIS the prompt emission of a GRB like 170817A versus the viewing angle (figure from \citet{2021ExA...tmp..126C})
At the 2G distance reach for NS-NS mergers, THESEUS could detect a GRB up to a viewing angle of $\sim10-40$~deg providing >5-100 times more detections with respect to a viewing angle within the jet core (<4~deg). At distances up to $z\sim$2(4), THESEUS can detect a GRB in the 2-30~keV band up to a viewing angle of $\sim$7(5)~deg thus increasing the on-axis detections by a factor of >3 (1.5). 
}
\label{fig:XGIS_limits_170817} 
\end{figure}

 Huge steps forward in our knowledge on the GRB jet structure were possible due to the off-axis view of a nearby short GRB associated with the gravitational wave source GW170817 \citep{Abbott17gw-gamma}.
    By the 2030s \textsl{Athena} will be the best facility to follow-up the slowly rising, faint X-ray fluxes from the off-axis afterglows, providing great details on their morphology (Sect.~\ref{GWground:science}). Comparison between off-axis short GRBs from NS-NS and NS-BH will also enable to highlights commonalities between these two progenitors. 
    According to our knowledge on compact binary merger electro-magnetic emission, THESEUS was expected to detect a number of off-axis components at early times that would enable \textsl{Athena} to trigger follow-up observations. 
    Indeed, the off-axis view of GRB~170817 enabled us for the first time to quantify how the high-energy prompt emission becomes gradually softer and less energetic as the viewing angle increases with respect to the jet axis. As a result, it has been possible to estimate, for events similar to GRB~170817, the maximum viewing angle at which a given instrument could detect the prompt emission depending on distance 
    (\citet{Salafia19},\citet{Salafia20}, Fig.~\ref{fig:XGIS_limits_170817}). 
    Based on such estimates, the  capabilities of THESEUS offer excellent prospects for detecting the prompt emission from off-axis short GRBs within the relatively small distance reach of GW detectors. For BNS mergers detected by 2G interferometer network, the GRB~170817-like prompt emission would be observable with the XGIS up to 10-40 deg, corresponding to a detection rate increased by a factor of >5-100 with respect to the result for on-axis events only. At the typical distance reached by a 3G detector, such as ET (Fig.~\ref{fig:gw_horizons}), the prompt emission would still be observable by THESEUS up to about 10~deg, potentially increasing the event rate by a factor of >5.
    
\begin{figure}[!t]
\begin{center}
\centering
\includegraphics[width=0.9\textwidth]{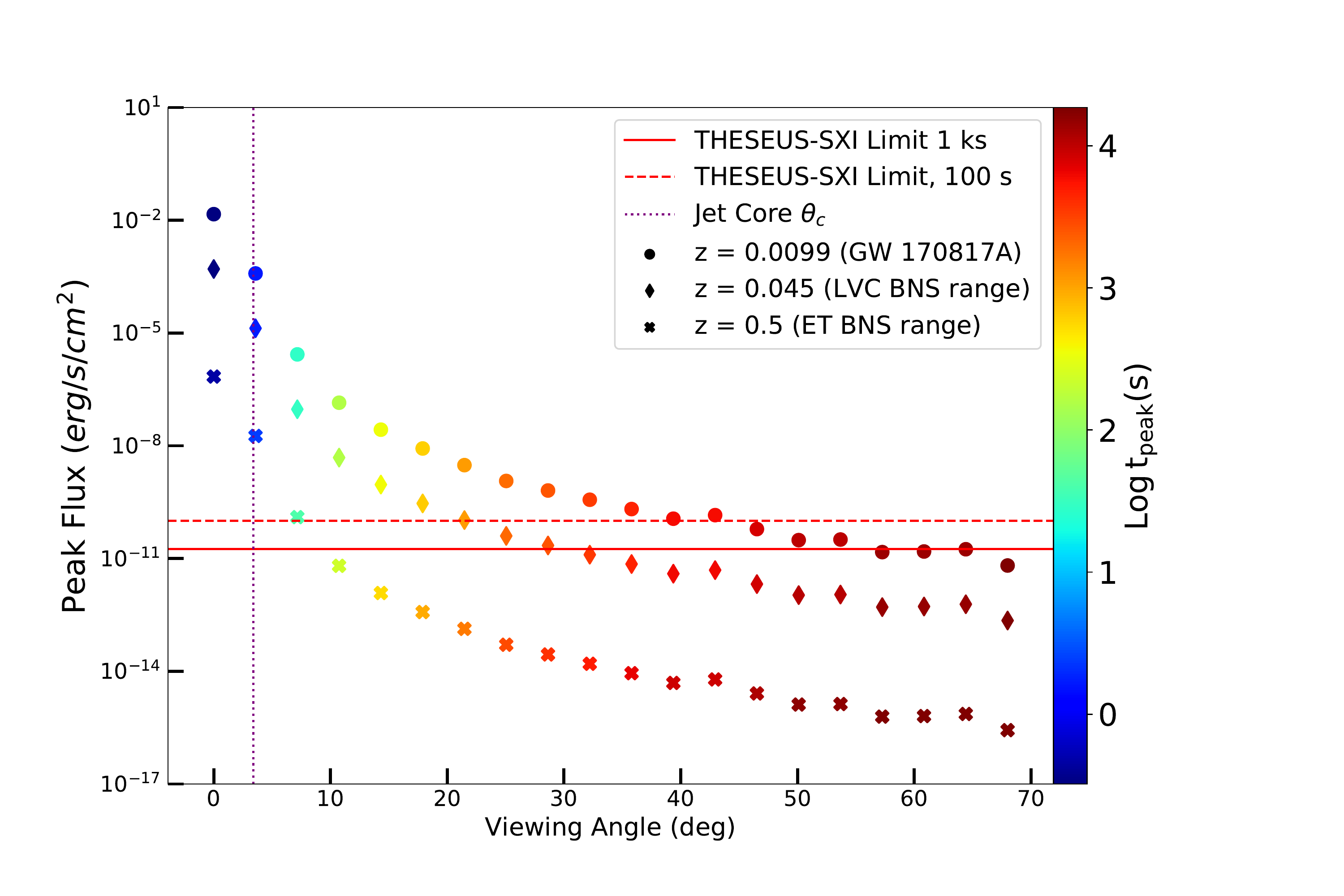}
\caption{Peak flux (in the THESEUS/SXI band 0.3-5~keV) vs viewing angle of the high latitude emission of GRB~170817A assuming the jet structure derived in 
\citet{2020ApJ...893...88O}. Circles, diamonds and crosses represent the same model at three different redshifts. The color code denotes the logarithm of the time of the peak in the observer frame. Red continuous and dashed lines mark the flux limit of THESEUS/SXI integrated in 1~ks and 100~s, respectively. Vertical dotted line marks the inner jet core half-opening angle. [Credit: S. Ascenzi].
}
\label{fig:HLE_170817} 
\end{center}
\end{figure}

    Another potential short GRB-less X-ray emission is predicted by a lateral view of the high latitude emission (HLE) from a structured jet whose energy and bulk Lorentz factor gradually decrease with the angular distance from the jet symmetric axis (\citet{2020ApJ...893...88O}, \citet{2020MNRAS.492.2847B}, \citet{2020A&A...641A..61A}. HLE has been invoked for the first time to interpret the initial steep decay of the X-ray afterglow lightcurve soon after the prompt emission, as the delayed radiation coming from the edges of the innermost, narrow jet core \citep{2000ApJ...541L..51K}. With the same mechanism, a plateau-like feature can be produced with those photons emitted by regions just outside of the jet core in a structured jet. By increasing the viewing angle (i.e. more off-axis lines of sight) the steep-to-plateau transition becomes less evident developing into a shallow rising followed by a power-law decaying X-ray transient that can be detected. 
    Simulations of the expected X-ray  peak flux from a  GRB~170817-like source at different distances show that THESEUS/SXI would be able to detect such component up to 40, 30 and 10 deg for a source at 40~Mpc (GW 170817), z=0.045 (LVC BNS range) and z=0.5 (ET BNS range) (Fig.~\ref{fig:HLE_170817}).  Nearly isotropic X-ray emission is also expected if a magnetar is formed after the merger of two NSs (e.g. \citet{2016ApJ...819...15S}). The predicted luminosities of such a component are very uncertain, ranging from $10^{43}$ to $10^{48}$~erg/s. Detection rates of such off-axis  components strongly depend on several still highly uncertain parameters as the level of collimation, the efficiency in the production of such X-ray emission from NS-NS and NS-BH, the jet structure. 
    The nature of BNS merger remnants has crucial implications on the equation of state (EoS) of NS. So far  no conclusive evidence was found in the post-merger GW signal of GW170817. Short GRB X-ray afterglows can help in answering this important question. Indeed, the formation of a long-lived highly magnetized spinning down NS merger cannot be excluded and actually help in explaining some EM properties of GW170817 \citep{Metzger18,Piro19}. A magnetar scenario has been also invoked in several short GRBs to interpret a peculiar shallow decay observed in the X-ray afterglow light curves (the so-called ``plateau'') on timescales of several minutes up to  hours.  However, alternative scenarios are still viable as for example accretion onto a central black hole or high-latitude emission from structured jets. Joint multi-messenger observations of GW and X-rays from a large sample of BNS will definitively answer to this question. 
    
        The monitors on THESEUS would be able to detect and localize the prompt emission of 10-12 short GRB per year associated with ``face-on'' binary systems (i.e. on-axis short GRBs). The refined arcminute-arcsecond localization (depending if the optical-IR counterpart is detected) can be promptly transmitted to more sensitive X-ray telescopes, such as the one on-board \textsl{Athena}, to fully monitor the X-ray light curves. Since plateaus with low X-ray luminosities seem to last  longer (up to 1 day after the burst) than bright ones,  the huge sensitivity of \textsl{Athena} will enable to observe a large sample of short GRB X-ray plateaus, for which post-merger GW signals will be available. Should a NS remnant become evident from BNS post-merger GW data analysis, the presence/absence of expected pulsating X-ray emission can further characterize the source in terms of alignment of the rotation axis with the line of sight and/or on the presence of surrounding reprocessing matter (e.g. \citet{2017MNRAS.472.1152R}). For the BNSs, \textsl{Athena} will also enable to individuate the presence of extra components  (e.g. flares) that can be produced from a long-lived post-merger compact source, as an accreting BH or a spinning down massive NS \citep{Piro19}. By identifying  the post-merger compact object nature from BNS has relevant consequences on fundamental physics as the EoS of NSs.

\section{Observational strategy}

The feasibility of efficiently observing the different classes of transient/variable sources  with \textsl{Athena} largely depends on the response time of \textsl{Athena} to alerts dispatched by the triggering observatory. The most challenging case is certainly that of the GRBs. It is known that the X-ray flux of the bulk of these objects decays as $t^{-1.3}$ over the first day and thus it is required that \textsl{Athena} responds to the THESEUS alerts within a maximum delay of 4-6~h to make an optimal usage of the largest available number of photons. The EM counterparts of GW sources would also certainly benefit by similarly fast turn around times, although in this case the specific need is still rather uncertain given our currently limited knowledge in the field (Sect.~\ref{sec:GWthemes}). Most of other transient/variable sources mentioned above are characterized by intensity/spectral states that can last significantly longer than the high energy emission from GRBs (up to a week time scale). In all these cases, given the relatively short time scale with which \textsl{Athena} can be re-pointed, it will be the decision loop about possible follow-up observations of the triggering event that will drive the timescale on which \textsl{Athena} will eventually be observing those sources. 

Other science cases requiring ToO observations are listed in Tab.~\ref{tab_too}. The exposure time as after the Athena Mock Observing Plan (MOP). While they constitute just one of the possible instantiations of the \textsl{Athena} science operation plan to demonstrate the feasibility of the core scientific goals of the mission, they can be regraded as an approximate gauge of the time investment that the mission might spend on them, at out current understanding of the \textsl{Athena} science goals and priorities. They do not drive the design of the ground segment and spacecraft response, as well as of the quick-look data processing, because their response time is estimated to be slower than for the GRB case ($\ge$ 12~hours for the fastest cases). We recall here, as the example of a more challenging scenario, the discussion on a possible, complex follow-up strategy of SMBHM triggers identified by LISA in Sect.~\ref{sec:LISA_strategy}, which would require to stretch the response time of the \textsl{Athena} (and LISA!) Science Operation Centers to the requirement to ensure that the error box of the LISA trigger, continuously shrinking as the merging time approaches, can be optimally sampled by \textsl{Athena} to maximize the probability to observe and identify an electro-magnetic counterpart of the GW event in the WFI FoV.
\begin{table}[]
    \centering
    \begin{tabular}{c|c|c|c}
         SCIOBJ & Topic & Number of sources & MOP time (Ms) \\
         251 & Galactic Black Hole Candidate and X-ray Binaries & 20 & 1.72  \\
         252 & Ultra-Luminous X-ray Sources, SgrA$^*$ & 26 & 1.28 \\
         262 & Tidal Disruption Events & 25 & 1.79 \\
         323 & Magnetospheric accretion in low-mass stars & 1 & 0.06 \\
         333 & Accreting White Dwarfs & 2 & 0.25 \\
         334 & Magnetars & 1 & 0.16 \\
         336 & Novae & 1 & 0.21 \\
         338 & Supernovae & 5 & 0.36 \\
    \end{tabular}
    \caption {Other \textsl{Athena} SCIence OBJectives requiring ToO observations, in addition to GRBs and multi-messenger targets, according to the MOP. SCIOBJs with an identifier lower than 300 correspond to the core science.}
    \label{tab_too}
\end{table}

\chapterimage{Images/Athena.pdf} 
\chapter{Summary}


Astronomy has evolved in time from an ensemble of wavelength-specific sectors into a multi-wavelength enterprise, and  is now taking a step further into the multi-messenger era.  This White Paper has highlighted the many synergies between \textsl{Athena} and some of the key multi-messenger facilities that should be operative concurrently with \textsl{Athena}. These facilities include LIGO A+, Advanced Virgo+ and future detectors for ground-based observation of gravitational waves (GW), LISA for space-based observations of GW, IceCube and KM3NeT for neutrino observations, CTA for very high energy  observations. 
A significant part of the sources targeted by multi-messenger facilities is of transient nature. We have thus also discussed the synergy with  wide-field high-energy facilities, taking 
 THESEUS as a case study for transient discovery. This discussion is extended to those \textsl{Athena}  science goals that, while not being strictly multi-messenger, rely on follow-up observations of high-energy transients identified by external observatories. For several science themes, those observations provide independent and complementary information to the multi-messenger observations (e.g. Accretion processes, jet physics, neutro star equation of state -NS EoS-).    A summary of all science themes is presented below.

\section{ Central engine and jet physics in compact binary mergers}
    It is expected that \emph{\textsl{Athena} will observe the location of a binary neutron star (BNS), or a NS- black hole (BH) binary pinpointed by a combination of  GWs  (LIGO A+, Advanced Virgo+ and next generation interferometers, Chapter~\ref{chapter:GW}), 
    GRB prompt emission (e.g THESEUS, Chapter~\ref{chapter:Theseus})} or kilonova observations . In this case, \textsl{Athena} will provide vital input to \emph{clarify the mechanisms of formation, launching and evolution of a jet from the compact remnant created by the merger event} (Chapter~\ref{chapter:GW}), by being able to detect events  up to the redshift probed by third generation GW detectors, and/or to  further off-axis angles, and monitor them at later times. 
    \textsl{Athena} can help determine the outflow geometry  (isotropic vs collimated) from the afterglow evolution at later times and eventually unveil a new population of transients produced by chocked jets. GW selected events are less biased towards on-axis events. Early time observations by \textsl{Athena}  will allow to probe a) the outer structure of the jet,  and the interaction with the merger debris and b) the initial magnetization of the ejecta and the Lorentz factor, which help constrain models for  jet launching. Late time measurements, when the outflow  has turned into sub-relativistic expansion, enable a direct measurement of the kinetic energy of the ejecta.
    
    \emph{It is not obvious if an electromagnetic counterpart of a stellar BH-BH merger exists, but the combined synergy between LISA, ground-based GW interferometers and \textsl{Athena} can  enable the most sensitive  coincident GW-EM detection by taking advantage of the early forecast  by LISA (Chapter~\ref{chapter:LISA}) }
    
    \section{Accretion processes and jet physics in SMBHs}
    \emph{Combined \textsl{Athena}-LISA observations of super massive black holes (SMBHs) in the mass range $10^5-10^7  \msun$ and at $z\lesssim 2$ should probe accretion physics} in extreme regimes and provide an independent measurements of system parameters that drive accretion processes (Chapter~\ref{chapter:LISA}). LISA  will determine masses, spins, inclination and orbital parameters  of the black holes pre- and post- merger, whereas  X-ray observations may provide evidence for the reforming of the active galactic nuclei (AGN) accretion disc and corona,  or by the launch of  powerful relativistic jet, newly formed by the merger. During the inspiral phase X-ray emission could be produced by the accreting and/or shock heated gas in various sites of the system  and be modulated with a frequency commensurate with the GW inspiral signal. \emph{Extreme mass ratio inspiral and intermediate mass ratio inspiral occurring in AGN discs are also an ideal probe for accretion theories.} In the latter case, the inspiralling black hole is massive enough to strongly perturb the surface density of the disc which,  in turn,  is going to affect the intensity and shape of the Fe K reflection line  in a way that depends on the extent of the corona and on the emissivity profile of the disc. 
    
    Furthermore, \emph{\textsl{Athena} will use tidal disruption events, as e.g. identified by THESEUS (Chapter~\ref{chapter:Theseus}), as a probe of the accretion process onto systems hosting black holes at various scales}, including AGNs.  The stellar disruption and the subsequent surge in accretion will probe the dynamics of tidal shearing in the proximity of the event horizon, characterize the orbital and physical evolution of the debris,  and gain insight into the effects of rapid accretion rate changes in AGN systems.
    
    \section{Accretion processes in compact stellar binaries}
    
    \emph{ Observations by \textsl{Athena} of X-ray (BH,NS, white dwarf -WD-) binaries in outburst or in specific luminosity/spectral states, as e.g. identified by X-ray wide field sky monitors like THESEUS (Chapter~\ref{chapter:Theseus}) will allow to study the physics of accretion and outflows} in connection with the nature of the central source. This includes  measuring the accretion and outflow in the different regimes of gravity and magnetic field in BH vs NS binary systems and, for systems hosting accreting WD, to measure their masses, probe the magnetospheric or inner disc interaction, constrain the origin of energy release and study the chemical composition of Novae ejecta. 

Short (< 30~min) period WD binaries are strong UV/X-ray and mHz GW emitters. \emph{Combined \textsl{Athena}-LISA observations of accreting double WD (DWD) binaries would allow us to  study  physical processes, related to accretion physics} and mass transfer stability (Chapter~\ref{chapter:LISA}). For example, the largest uncertainties in predictions of the final fate of DWD binaries (merger versus stable mass transfer) and hence of supernova (SN) Ia rates is the treatment of the onset of mass transfer when the larger WD fills its Roche Lobe and starts to accrete onto its companion. Combined EM-GW data can disentangle the contribution from GW radiation and mass transfer (as derived from X-ray accretion luminosity) from the overall period evolution and study the transport of angular momentum in accreting DWDs.

\section{Improving our understanding of the equation of state in neutron stars} 

In the limit of high densities and low temperatures, quantum chromodynamics (QCD) can only be tested in the extreme astrophysical environment of NS cores. Here QCD predicts a range of rich behaviours depending on assumptions about the way particles interact in this extreme regime. Exotic excitations such as hyperons, or Bose condensates of pions or kaons may appear, as well as phase transition to strange quark matter. \emph{The key “observable” to distinguish between the various models is the equation of state} (EoS), which governs the mechanical equilibrium structure of bound stars and which determines the mass-radius relationship of NS. 

\emph{ Joint \textsl{Athena} and ground based GW observations of a sample of binary NS mergers  and their associated short GRBs afterglows provide a novel approach to determine the EoS} (Chapter~\ref{chapter:GW}). 
GW data constrain the mass of the remnant while the X-ray behaviour  at early and late times can carry the imprints of a (highly magnetized) NS as opposed to a BH. The mass distribution will allow us to pin down the EoS (soft vs stiff), being it  very sensitive to  the maximum mass of a NS. This program will take advantage of the synergy between \textsl{Athena}, ground-based GW interferometers (LIGO A+, Advanced Virgo+, ET and next generation detectors) and wide-field high-energy sky monitors like THESEUS.

Medium and high spectral resolution \emph{observations by \textsl{Athena} of NS will be used to  constrain their masses and radii}, using multiple redundant and complementary diagnostics in a variety of environments and conditions (e.g., isolated stars, binary radio pulsars, quiescent cooling systems, X-ray bursters and accreting binaries). 
Some of these observations are most effective when the source is in \emph{specific intensity or spectral states, that can be identified by  transient facilities such as THESEUS} (Chapter~\ref{chapter:Theseus}).

\section{Cosmic accelerators and origin of cosmic rays}

\emph{ Synergy of \textsl{Athena} with neutrino and very-high-energy gamma-ray observatories (CTA, KM3Net, IceCube and next generation detectors) can provide vital inputs to address fundamental questions on
 what the sources of Galactic and extragalactic cosmic rays (CRs) are,  how these sources are capable of accelerating particles to very high energies, and what fraction of the energy budget goes to the acceleration of particles.}  A unique signature of high energy hadrons, the main constituents of CR, are neutrinos. On the other hand, very-high-energy gamma rays can be produced either by accelerated hadrons or leptons.
 Several classes of X-ray sources are known, or suspected to be, gamma-ray and/or neutrino sources and, thus, candidates for the sites of galactic CRs: supernova remnants (SNR), pulsar wind nebulae, microquasars and star-forming regions, jets and outflows from AGNs and GRBs, shocks in starburst galaxies and clusters of galaxies.
 In order to estimate the energy content of CR one needs to measure the gamma-ray luminosity, determine its nature (leptonic vs hadronic) and model the density and composition of the target material. 
Since the acceleration mechanisms for hadrons and leptons are the same, X-ray observations of the non-thermal synchrotron component are important to probe the active acceleration conditions and, when joined with gamma-ray and neutrino observations, disentangle the hadronic vs leptonic origin. Furthermore, high spectral resolution by \textsl{Athena}  of the thermal X-ray component, if present (like in supernova SNR), helps in assessing the energy content of the hadronic component, by measuring the density and ion temperature, a proxy of CR acceleration efficiency.

\emph{\textsl{Athena} observations of GRB afterglows associated to GW mergers detected by on-ground GW detectors  provide further insight on particle shock acceleration} (Chapter~\ref{chapter:GW}). They enable measurements of the  electron energy distribution throughout the transition from relativistic to non-relativistic stages with unprecedented accuracy, probing shock-acceleration between relativistic and non-relativistic shocks.
 
 \section{ How did intermediate- and high-Z elements form}
 
 SNe are the source of most of the heavy elements in the Universe. SN Ia (SNIa) provide most of iron whereas core collapse SN (SNcc) are the source of intermediate elements (Si, S, and Ca), and are responsible for the enrichment of the early Universe. 
 \emph{ The origin of elements heavier than iron (through r-process) is highly debated, with SNcc and kilonovae being the main contenders.}
In SNe the metal enrichment depends heavily on the explosion mechanism, that is poorly understood.
SNIa are thought to be exploding WDs in a binary system,
while SNcc result from the collapse of the core of a massive star into a NS or BH.  The release of gravitational energy powers the explosion of the rest of the star, but {\emph how the collapse energy drives the explosion is also an issue (neutrino-driven, magnetohydrodynamic-driven),
that affects also the production of r-process elements, expected to be vigorous in the case of a neutrino-wind.}
In addition to  high-resolution spectral observations of SNR, \emph{ \textsl{Athena} is planning to follow-up the prompt discovery of SNe in X-rays by wide-field X-ray facilities like THESEUS, to characterize the properties of the shock break-out, and constrain the explosion models (Chapter~\ref{chapter:Theseus})}. The shock break out marks the first escape of photons related to the explosion, and  takes place on a time scale of minutes - hours, peaking in the soft X-rays. The ensuing luminosity and temperature evolution  determines the radius of the progenitor star as well as the ratio between the explosion energy and the ejecta mass. This will allow for the size of the progenitor star to be determined, thus constraining the explosion models. 

\emph{Neutron rich material from NS merger ejecta (kilonova) has been proposed as the other major source of  of r-process elements.
\textsl{Athena} X-ray observation of kilonovae  (Chapter \ref{chapter:GW}) could measure the ejecta mass}  in various ways, by their shock interaction with the environment, by searching for X-ray features from the  radioactive decay and, possibly, by carrying out deep high resolution X-ray spectroscopy of potential kilonova remnants.

\section{The missing baryons}

The current census of baryonic matter in the Universe shows a deficit of approximately half in the Universe at $z\lesssim$1. According to cosmological simulations, this matter should be  located in a so called warm hot intergalactic medium (WHIM), a gaseous filamentary structure connecting clusters of galaxies, and in a hot circum-galactic medium  around galaxies or groups.  At such temperatures, the matter is  ionized to the extent that  absorption and emission line features can only be detected in soft X-rays via high resolution spectroscopy.
\emph{ \textsl{Athena} will routinely detect hundreds of absorption systems in the spectra of bright AGNs and GRBs, characterizing the location and physical conditions of Universe’s missing baryonic mass} and shedding light on the continuous feedback process governing the galaxy-intergalactic medium co-evolution throughout cosmic time. \emph{GRBs are particularly important because they provide a backlight at $z \gtrsim 1$, thus allowing to probe also the epoch of the formation of this elusive component. This fundamental part of the program will be best exploited in synergy with a GRB monitor} that has the capability to provide the adequate number of bright GRBs to enable fast follow-up observations with \textsl{Athena} (Chapter~\ref{chapter:Theseus}).  
 
\section { Exploring the cosmic dawn with GRBs}
  
GRBs can be used to investigate the re-ionization epoch, and to perform unprecedented studies of the global star formation history of the Universe up to $z= 10$ and possibly beyond. The close environment of the GRB encodes the history of metal enrichment by progenitor stars, so it is expected to be pristine, population-III (Pop-III) or Pop-II enriched. Because of its high ionization it can be probed only via high-resolution X-ray spectroscopy. \emph{ \textsl{Athena} high resolution spectroscopy of high-$z$ GRBs with rapid follow-up, triggered by external facilities such as THESEUS, will allow us to determine when did the first stars form and how did the earliest Pop-III and Pop-II stars influence their environments, the interstellar medium and the intergalactic medium.}


\section { Enhancement of the cosmic distance scale using GW sources as standard sirens} 

\emph{\textsl{Athena} observations of  counterparts of  coalescing binaries observed by LISA or ground-based GW observatories can substantially enhance the measurement of the cosmological parameters} (Chapter \ref{chapter:GW} and \ref{chapter:LISA}).
Coalescing binaries are standard sirens, as the GW signal enables the direct measure of the luminosity distance to the source. If the redshift of the host is derived from the X-ray detection and optical follow-up observations, the resulting  distance-redshift provides a  measurement of the Hubble parameter, that may help to disentangle  tensions between the late- and early universe probes on cosmic expansions from Planck and SNIa. 

\section {Testing General Relativity  and the Standard Model by measuring the speed of multi-messenger carriers} 

The standard model  fails to explain the origin of neutrino mass which is evident by the effect of neutrino flavor oscillations. Physics beyond the standard model (BSM) can  be probed by Violation of Lorentz invariance (LIV), i.e. by different energy-momentum relations of neutrinos,  photons or gravitational waves. This could be observable as time delays between different carriers emitted at the same time from transient sources and arriving at Earth at different times.
\emph{Fast Target of Opportunity  observations with \textsl{Athena} following a neutrino transient source, like a gamma-ray blazar, can produce competitive results on BSM by setting stringent limits on LIV from neutrinos and photons (Chapter~\ref{chapter:HE})}.
In General Relativity, gravitational waves travel with the speed of light (the graviton is massless) and interact very weakly with matter. Alternative theories with a massive graviton predict an additional frequency dependent phase shift of the observed waveform due to dispersion that depends on the graviton’s mass and the distance to the binary.  \emph{Comparing the EM and GW chirp signals from \textsl{Athena} and LISA in SMBHs mergers  will measure the relative propagation speed of photons vs gravitons to  one part in $10^{17}$, allowing a novel test for theories with massive gravity or extra spatial dimensions (Chapter~\ref{chapter:LISA}).}

    
\chapterimage{Images/MergingBH.pdf} 

\chapter*{Acknowledgements}
This research has been  supported by the European Union’s Horizon 2020  programme  under the AHEAD (grant agreement number 654215) and  AHEAD2020 project (grant agreement n. 871158). The authors wish to thank J.M. Torrej\'on for the organization of the \textsl{Athena}-multi-messenger Workshops, held on November  27 - 29, 2018, in Alicante, Spain and  on 5 May 2020 - 6 May 2020 in videoconference.


\chapterimage{Images/Athena.pdf} 
\vspace*{0.5cm}
\renewcommand{\bibname}{References}
\addcontentsline{toc}{chapter}{\textcolor{purple}{References}}
\printbibliography
\let\cleardoublepage\clearpage 
\chapterimage{Images/MergingBH.pdf} 
\chapter*{Acronym List}
\addcontentsline{toc}{chapter}{\textcolor{purple}{Acronym List}}

\begin{tabular}{ll}
 
 AGN \dotfill & active galactic nucleus/nuclei\\
 AMON \dotfill & Astrophysical Multi-messenger Observatory Network \\
 AM CVn \dotfill & accreting white dwarf binary \\
 ARCA \dotfill & \textit{Astroparticle Research with Cosmics in the Abyss}\\
 ASST \dotfill & \textsl{Athena} Science Study Team \\
 BNS \dotfill & binary neutron star\\
 BH \dotfill & black hole\\
 BSM \dotfill & beyond the standard model\\
 CARIBU \dotfill & CAlifornium Rare Isotope Breeder Upgrade\\
 CGM \dotfill & circum-galactic medium \\
 CTA \dotfill & Cherenkov Telescope Array\\
 CV   \dotfill & cataclysmic variable \\
 CR \dotfill & cosmic-ray \\
 EBL \dotfill & extragalactic background light \\
 EM \dotfill &  electromagnetic \\
 EMRI \dotfill & extreme mass ratio inspiral \\
 EoS \dotfill & Equation of State \\
 ESA \dotfill & European Space Agency \\
 ET \dotfill & Einstein Telescope \\
 DWD \dotfill & double white dwarf \\
 \textsl{Fermi} GBM \dotfill & Fermi Gamma-Ray Burst Monitor \\
 FoV \dotfill & Field of View\\
 FRIB \dotfill & Facility for Rare Isotope Beams\\
 GBH \dotfill & galactic black hole\\
 GRB \dotfill & gamma ray burst\\
 GW \dotfill & gravitational wave\\
 IGM \dotfill & intergalactic medium\\
 IMRI \dotfill & intermediate mass ratio inspiral \\
 ISM \dotfill & interstellar medium\\
 IRT \dotfill & infrared telescope \\
 HAWC \dotfill & High Altitude Water Cherenkov \\
 HEW \dotfill & half energy width\\
 HMXB \dotfill & high-mass X-ray binary\\
 $\Lambda$CDM \dotfill & lambda cold dark matter\\
 LGBR \dotfill & long-duration GRB \\
 LMXB \dotfill & low-mass X-ray binary \\
 LST \dotfill & large-sized telescope \\

 \end{tabular}
 
\newpage

\begin{tabular}{ll}

 LVI \dotfill & Lorentz violation invariance \\
 MBH \dotfill & massive black hole \\
 MOS \dotfill & mission observation simulator \\
 MS \dotfill & main sequence \\
 MST \dotfill & medium-sized telescope \\
 NS \dotfill & neutron star \\
 ORC \dotfill & odd radio circle \\
 Pop-II \dotfill & population II\\ 
 Pop-III \dotfill & population III\\
 PWN(e) \dotfill & pulsar wind nebula(e) \\
 QCD \dotfill & quantum chromodynamics \\
 RS \dotfill & reverse shock \\
 RSF \dotfill & resonant shattering flare \\
 SGBR \dotfill & short-duration GRB \\
 SMBH \dotfill & super-massive black hole \\
 SMBHMs \dotfill & super massive blach hole mergers \\
 $S/N$ \dotfill & signal-to-noise ratio \\
 SN(e)  \dotfill & supernova(e) \\
 SNIa \dotfill & Supernova Ia\\
 SNcc \dotfill & core collapse supernova\\
 SNR \dotfill & supernova remnant \\
 SST \dotfill & small-sized telescope \\
 SXI \dotfill & Soft X-ray Imager \\
TDE \dotfill & tidal disruption event\\
 ToO \dotfill & Target of Opportunity\\
 UCXB \dotfill & ultra-compact X-ray binary \\
 ULX \dotfill & ultra-luminous X-ray source\\
 VHE \dotfill & very-high-energy \\
 VLBI \dotfill & Very Long Baseline Interferometry\\
 WD \dotfill & white dwarf \\
 WFI \dotfill & Wide Field Imager \\
 WHIM \dotfill & warm-hot intergalactic medium\\
 X-IFU \dotfill & X-ray Integral Field Unit\\
 XGIS \dotfill & X-Gamma-ray Imaging Spectrometer\\

\end{tabular}

\end{document}